\documentclass[12pt]{article}
\usepackage{amsfonts,amssymb,amsmath,subeqnarray,eqsection,indent}
\usepackage{bm}    
\usepackage{cite}
\usepackage{graphicx}
\usepackage{pstricks,pst-node,pst-text,pst-3d,pst-plot}
\usepackage{url}

\date{October 9, 2012 \\
      Revised November 19, 2012}  

\begin{document}

\title{\vspace*{-1cm} The Hintermann--Merlini--Baxter--Wu  \\
       and the Infinite-Coupling-Limit Ashkin--Teller Models}

\author{
  {\small Yuan Huang}              \\[-2mm]
  {\small\it Hefei National Laboratory for Physical Sciences at Microscale}       \\[-2mm]
  {\small\it and Department of Modern Physics}       \\[-2mm]
  {\small\it University of Science and Technology of China}       \\[-2mm]
  {\small\it Hefei, Anhui 230026, CHINA}       \\[-2mm]
  {\small\tt HUANGY22@MAIL.USTC.EDU.CN}              \\[4mm]
  {\small Youjin Deng}                    \\[-2mm]
  {\small\it Hefei National Laboratory for Physical Sciences at Microscale}       \\[-2mm]
  {\small\it and Department of Modern Physics}       \\[-2mm]
  {\small\it University of Science and Technology of China}       \\[-2mm]
  {\small\it Hefei, Anhui 230026, CHINA}       \\[-2mm]
  {\small\tt YJDENG@USTC.EDU.CN}              \\[4mm]
  {\small Jesper Lykke Jacobsen}              \\[-2mm]
  {\small\it Laboratoire de Physique Th\'eorique}    \\[-2mm]
  {\small\it \'Ecole Normale Sup\'erieure}       \\[-2mm]
  {\small\it 24 rue Lhomond}       \\[-2mm]
  {\small\it 75231 Paris, FRANCE}       \\[-2mm]
  {\small    and} \\[-2mm]
  {\small\it Universit\'e Pierre et Marie Curie} \\[-2mm]
  {\small\it 4 place Jussieu}  \\[-2mm]
  {\small\it 75252 Paris, FRANCE}   \\[-2mm]
  {\small\tt JACOBSEN@LPT.ENS.FR}         \\[4mm]
  {\small Jes\'us Salas}                                    \\[-2mm]
  {\small\it Grupo de Modelizaci\'on, Simulaci\'on Num\'erica y
             Matem\'atica Industrial} \\[-2mm]
  {\small\it Universidad Carlos III de Madrid} \\[-2mm]
  {\small\it Avda.\  de la Universidad, 30}    \\[-2mm]
  {\small\it 28911 Legan\'es, SPAIN}           \\[-2mm]
  {\small    and} \\[-2mm]
  {\small\it Grupo de Teor\'{\i}as de Campos y F\'{\i}sica 
             Estad\'{\i}stica}\\[-2mm]
  {\small\it Instituto Gregorio Mill\'an, Universidad Carlos III de 
             Madrid}\\[-2mm]
  {\small\it Unidad Asociada al IEM-CSIC}\\[-2mm] 
  {\small\it Madrid, SPAIN}           \\[-2mm]
  {\small\tt JSALAS@MATH.UC3M.ES}              \\[4mm]
  {\protect\makebox[5in]{\quad}}  
  \\
}

\maketitle
\thispagestyle{empty}   

\begin{abstract}
We show how the Hintermann--Merlini--Baxter--Wu model (which is a 
generalization of the well-known Baxter--Wu model to a general Eulerian
triangulation) can be mapped onto a particular infinite-coupling-limit 
of the Ashkin--Teller model. We work out some mappings among these models, also including the standard and mixed Ashkin--Teller models.
Finally, we compute the phase diagram of the infinite-coupling-limit 
Ashkin--Teller model on the square, triangular, hexagonal, and kagome lattices. 
\end{abstract}

\bigskip
\noindent
{\bf Key Words:}
Baxter--Wu model; Ashkin--Teller model; 
Hintermann--Merlini--Baxter--Wu model; Infinite-coupling-limit Ashkin--Teller 
model; partial trace transformation; plane Eulerian triangulation.

\clearpage

\newtheorem{theorem}{Theorem}[section]
\newtheorem{proposition}[theorem]{Proposition}
\newtheorem{lemma}[theorem]{Lemma}
\newtheorem{corollary}[theorem]{Corollary}
\newtheorem{definition}[theorem]{Definition}
\newtheorem{conjecture}[theorem]{Conjecture}
\newtheorem{question}[theorem]{Question}
\newtheorem{example}[theorem]{Example}

\renewcommand{\theenumi}{\alph{enumi}}
\renewcommand{\labelenumi}{(\theenumi)}
\def\eop{\hbox{\kern1pt\vrule height6pt width4pt depth1pt\kern1pt}\medskip}
\def\prf{\par\noindent{\bf Proof.\enspace}\rm}
\def\rmk{\par\medskip\noindent{\bf Remark\enspace}\rm}

\newcommand{\be}{\begin{equation}}
\newcommand{\ee}{\end{equation}}
\newcommand{\<}{\langle}
\renewcommand{\>}{\rangle}
\newcommand{\widebar}{\overline}
\def\reff#1{(\protect\ref{#1})}
\def\spose#1{\hbox to 0pt{#1\hss}}
\def\ltapprox{\mathrel{\spose{\lower 3pt\hbox{$\mathchar"218$}}
    \raise 2.0pt\hbox{$\mathchar"13C$}}}
\def\gtapprox{\mathrel{\spose{\lower 3pt\hbox{$\mathchar"218$}}
    \raise 2.0pt\hbox{$\mathchar"13E$}}}
\def\textprime{${}^\prime$}
\def\proof{\par\medskip\noindent{\sc Proof.\ }}
\def\firstproof{\par\medskip\noindent{\sc First Proof.\ }}
\def\secondproof{\par\medskip\noindent{\sc Second Proof.\ }}
\def\thirdproof{\par\medskip\noindent{\sc Third Proof.\ }}
\def\qed{\hbox{\hskip 6pt\vrule width6pt height7pt depth1pt
\hskip1pt}\bigskip}
\def\proofof#1{\bigskip\noindent{\sc Proof of #1.\ }}
\def\firstproofof#1{\bigskip\noindent{\sc First Proof of #1.\ }}
\def\secondproofof#1{\bigskip\noindent{\sc Second Proof of #1.\ }}
\def\thirdproofof#1{\bigskip\noindent{\sc Third Proof of #1.\ }}
\def\altproofof#1{\bigskip\noindent{\sc Alternate Proof of #1.\ }}
\def\half{ {1 \over 2} }
\def\third{ {1 \over 3} }
\def\twothird{ {2 \over 3} }
\def\smfrac#1#2{\textstyle{\frac{#1}{#2}}}
\def\smhalf{ \smfrac{1}{2} }
\newcommand{\real}{\mathop{\rm Re}\nolimits}
\renewcommand{\Re}{\mathop{\rm Re}\nolimits}
\newcommand{\imag}{\mathop{\rm Im}\nolimits}
\renewcommand{\Im}{\mathop{\rm Im}\nolimits}
\newcommand{\sgn}{\mathop{\rm sgn}\nolimits}
\newcommand{\tr}{\mathop{\rm tr}\nolimits}
\newcommand{\supp}{\mathop{\rm supp}\nolimits}
\def\hboxscript#1{ {\hbox{\scriptsize\em #1}} }
\renewcommand{\emptyset}{\varnothing}

\newcommand{\restrict}{\upharpoonright}
\renewcommand{\implies}{\;\Longrightarrow\;}

\newcommand{\scra}{{\mathcal{A}}}
\newcommand{\scrb}{{\mathcal{B}}}
\newcommand{\scrc}{{\mathcal{C}}}
\newcommand{\scrf}{{\mathcal{F}}}
\newcommand{\scrg}{{\mathcal{G}}}
\newcommand{\scrh}{{\mathcal{H}}}
\newcommand{\scrk}{{\mathcal{K}}}
\newcommand{\scrl}{{\mathcal{L}}}
\newcommand{\scrn}{{\mathcal{N}}}
\newcommand{\scro}{{\mathcal{O}}}
\newcommand{\scrp}{{\mathcal{P}}}
\newcommand{\scrr}{{\mathcal{R}}}
\newcommand{\scrs}{{\mathcal{S}}}
\newcommand{\scrt}{{\mathcal{T}}}
\newcommand{\scrv}{{\mathcal{V}}}
\newcommand{\scrw}{{\mathcal{W}}}
\newcommand{\scrz}{{\mathcal{Z}}}

\newcommand{\ahat}{{\widehat{a}}}
\newcommand{\Zhat}{{\widehat{Z}}}
\renewcommand{\k}{{\mathbf{k}}}
\newcommand{\n}{{\mathbf{n}}}
\newcommand{\vv}{{\mathbf{v}}}
\newcommand{\bv}{{\mathbf{v}}}
\newcommand{\br}{{\mathbf{r}}}
\newcommand{\bn}{{\mathbf{n}}}
\newcommand{\w}{{\mathbf{w}}}
\newcommand{\x}{{\mathbf{x}}}
\newcommand{\cc}{{\mathbf{c}}}
\newcommand{\zero}{{\mathbf{0}}}
\newcommand{\one}{{\mathbf{1}}}
\newcommand{\bdelta}{{\boldsymbol{\delta}}}
\newcommand{\bsigma}{{\boldsymbol{\sigma}}}
\newcommand{\bpi}{{\bm{\pi}}}
\newcommand{\B}{{\rm\bm{B}}}

\newcommand{\C}{{\mathbb C}}
\newcommand{\Z}{{\mathbb Z}}
\newcommand{\N}{{\mathbb N}}
\newcommand{\Q}{{\mathbb Q}}
\newcommand{\R}{{\mathbb R}}
\newcommand{\RR}{{\mathbb R}}


\newenvironment{sarray}{
             \textfont0=\scriptfont0
             \scriptfont0=\scriptscriptfont0
             \textfont1=\scriptfont1
             \scriptfont1=\scriptscriptfont1
             \textfont2=\scriptfont2
             \scriptfont2=\scriptscriptfont2
             \textfont3=\scriptfont3
             \scriptfont3=\scriptscriptfont3
           \renewcommand{\arraystretch}{0.7}
           \begin{array}{l}}{\end{array}}

\newenvironment{scarray}{
             \textfont0=\scriptfont0
             \scriptfont0=\scriptscriptfont0
             \textfont1=\scriptfont1
             \scriptfont1=\scriptscriptfont1
             \textfont2=\scriptfont2
             \scriptfont2=\scriptscriptfont2
             \textfont3=\scriptfont3
             \scriptfont3=\scriptscriptfont3
           \renewcommand{\arraystretch}{0.7}
           \begin{array}{c}}{\end{array}}

\clearpage

\section{Introduction}   \label{sec.intro}

In 1944 Onsager \cite{Onsager_44} solved the nearest-neighbor square-lattice
Ising model without magnetic field. This constitutes a major milestone in 
Statistical Mechanics. This achievement opened the door to find other exactly
two-dimensional solvable models using different approaches. Two of such 
lines of research were 
(1) to consider models with multi-spin interactions rather than 
nearest-neighbor couplings, and 
(2) to look for models with larger symmetry groups than $\Z_2$.   

Two different groups obtained positive results along the first line of 
research using 3-spin interactions. In particular, in 1972 Hintermann and 
Merlini \cite{Hintermann_72} solved the Ising model with 3-spin interactions 
on the union-jack lattice. They use a mapping of this model onto the 
8-vertex model \cite{Baxter_book}, and used Baxter's solution of the latter 
\cite{Baxter_book} (another milestone in Statistical Mechanics) to derive 
the critical points of the model. 
Soon after this result, Baxter and Wu \cite{Baxter_Wu_73,Baxter_Wu_74} found 
in 1973 the exact solution of an Ising model on a triangular lattice also 
with pure 3-spin interactions. The method chosen was the Bethe Ansatz 
\cite{Baxter_book}. 
This model is very interesting, as it is believed \cite{Domany_78} 
to belong to the same universality class as the 4-state Potts model
\cite{Potts_52,Baxter_book}, but it 
does not show the logarithmic corrections displayed by the latter model
\cite{Nauenberg_80,Cardy_80,SS_97}.
The field theoretical interpretation of these differing behaviors is the 
following. The dominant part of the action describing the continuum limit of 
both the 4-state Potts model and the Baxter-Wu model is that of a free 
bosonic field. Both models contain further strictly irrelevant operators, 
which are not necessarily identical, since they may depend on the discrete 
symmetries of the spins and of the lattices on which the models are defined. 
But crucially, the 4-state Potts model contains in addition a marginally 
irrelevant operator, which is not present in the Baxter-Wu model. This latter 
operator induces logarithmic corrections under the renormalization-group flow 
that takes the 4-state Potts model to the fixed-point theory. Alternatively, 
one may say that the Baxter-Wu model is designed so that the amplitude of 
the marginally irrelevant operator is set to zero.
The Baxter-Wu universality class is also described via its universal amplitude 
ratios in Ref.~\cite{Shchur_10}.

Both models, the Hintermann--Merlini (HM) and the Baxter--Wu (BW) models,
had many features in common: they were defined on plane triangulations (i.e.,
all faces were triangles),
the dynamical variables were Ising spins, and the Hamiltonian only included
3-spin terms (one for each triangular face). These two models belong to a 
more general family of models, that we will denote as the 
Hintermann--Merlini--Baxter--Wu (HMBW) model. We define this model on 
any plane triangulation $G=(V,E)$ of vertex set $V$ and edge set $E$. Then 
on each vertex $x\in V$ we place an Ising spin $\sigma_x=\pm 1$. These spins
interact through the Hamiltonian:\footnote{
  We could define the HMBW model with face-dependent couplings; but we 
  will not need this generalization in this paper. This comment also applies
  to the other models considered here. 
}
\be
{\mathcal H}_\text{HMBW} \;=\; - J \sum\limits_{\triangle=\{i,j,k\}} 
   \sigma_i \sigma_j \sigma_k \,,
\ee
where the sum is over all triangular faces $\triangle=\{i,j,k\}$ of $G$, and 
$J$ is a coupling constant. 
This kind of models (and generalizations of them) were already considered by 
other authors \cite{Merlini_72,Gruber_77} from a group-theoretical point of 
view. In this paper, we will focus on the particular case of Eulerian 
triangulations: a graph is Eulerian if the degree of all its vertices is
even. This property will play an important role in the following sections. 

The second research line produced (among others) the so-called 
Ashkin--Teller (AT) model \cite{Ashkin_Teller_43}. 
This model was introduced in 1943 and generalized the Ising model to a 
4-state model. It was soon recognized that it could be written as two 
coupled Ising models \cite{Fan_72b}. More precisely, given a graph 
$G=(V,E)$, we place on each vertex $x\in V$ of the graph {\em two} spins 
$\sigma_i,\tau_i$. These spins interact via the Hamiltonian:
\be
{\mathcal H}_\text{AT} \;=\; - \sum\limits_{e=\<ij\>\in E}
  \left[ K_{2} \sigma_i \sigma_j + K'_{2}\tau_i \tau_j + 
         K_{4} \sigma_i \sigma_j \tau_i \tau_j \right] \,, 
\label{def_H_AT}
\ee
where the sum is over all edges of the graph. We are not aware of any  
exact solution to this model, or to any of its particular symmetric case, 
to be reviewed in Section~\ref{sec.models}). 
This AT model contains, as particular cases, the Ising and the 4-state Potts
models. 

Furthermore, in the symmetric case $K'_{2}=K_{2}$, 
there is a curve in the $(K_2,K_4)$-plane that can be mapped to a soluble 
6-vertex model \cite{Baxter_book}. This curve is self-dual, and part of it 
is critical, with critical exponents varying continuously as we move along 
that curve. This phenomenon provides the simplest counter-example to the 
usual notion of universality. 

At first sight, the relation between the HMBW and the AT models is rather 
weak: both models can be defined in terms of Ising spins, and we can find
the 8-vertex model in both the solution of the HM model, and in the 
computation of the self-dual curve of the symmetric AT model. One of the
goals of this paper is to show that the relation is deeper: there is an 
exact mapping between the HMBW model and an infinite-coupling-limit of the AT
model (ICLAT), to be defined below. In particular, certain curves in the 
phase diagram of the ICLAT model on a graph $G$ are equivalent via those
mappings to a HMBW model on a certain triangulation $G'$. 

In order to achieve our main goal, we need to work out a series of 
exact mappings among the HMBW, ICLAT, AT models and some other models
we will introduce in Section~\ref{sec.models}. Some of them are already
known results in the literature, others are alternative versions of 
known results, and finally, some of them are new. 

As an application, our last goal is to compute the phase diagram of the 
ICLAT model on the square, triangular, hexagonal, and kagome lattices, 
which include one or two HMBW models as particular cases. In all cases, we
find a point which corresponds to the point where the self-dual curve of the AT
model on the corresponding lattice hits the ICLAT plane. This point
is denoted B in Section~\ref{sec.phase.diagrams}. 
For the first three lattices, point B belongs to a subspace of the ICLAT 
model that can be mapped onto the HMBW model. So we can approach it  
following two independent paths: either along the HMBW subspace, or by
considering the limit of large couplings of the self-dual curve of the 
corresponding AT model. For these lattices we can make predictions of the
critical exponents along these two paths, and it turns out that they differ. 

The paper is organized as follows. In Section~\ref{sec.graph} we will show the
graph-theoretic set-up. In Section~\ref{sec.models} we will define carefully
the distinct models we will consider in this paper. 
In Section~\ref{sec.mappings} we will work out the mappings between the 
AT and mixed AT models, the mapping between the mixed AT and the ICLAT models,
and finally the partial trace transformation. This latter mapping allows us
to relate the HMBW and the ICLAT models on certain lattices. We conclude this 
section with a nice mapping between the BW model and a loop model. 
The next Section~\ref{sec.phase.diagrams} is devoted to the description of
the ICLAT phase diagrams for several common lattices: square, triangular,
hexagonal, and kagome. In addition we provide the critical exponents for the 
limiting point of the AT self-dual curve when the couplings tend to infinity.
Finally, in Section~\ref{sec.discussion} we summarize our findings. In 
Appendix~\ref{appendix.decimation} we describe the decimation transformation
and how it is related to the partial-trace transformation. 

%
%
\section{Graph-theoretic preliminaries}\label{sec.graph}

In this section we will introduce the main lattices where we will define our
physical models of Section~\ref{sec.models}.

Let us consider a graph $G=(V,E)$ with vertex set $V$, edge set $E$, and 
embedded in the plane (i.e, $G$ is planar). 
We can always define the {\em dual graph\/} $G^*=(V^*,E^*)$ of $G$  
in the standard way as follows:
To each face $f$ in $G$, there corresponds a dual vertex $f^*\in V^*$;
and for every edge $e\in E$, we draw a dual edge $e^* \in E^*$. If the
original edge $e$ lies on the intersection of two faces $f$ and $h$
(possibly $f=h$), then the corresponding dual edge $e^*$ joins the
dual vertices $f^*,h^* \in V^*$. We can draw $G$ and $G^*$ in the plane in such
a way that each edge $e\in E$ intersects its corresponding dual edge
$e^* \in E^*$ exactly once.

A graph $G$ is a {\em plane triangulation} if it is planar and if every  
face of $G$ (including the outer one) is bounded by a triangle.  
If $G$ is a plane triangulation, then $G^*$ is a planar cubic graph (i.e.,
a planar 3--regular graph). 
Therefore, if $G$ is an Eulerian plane triangulation (i.e., all vertices
have even degree), then $G^*$ is a bipartite cubic graph. This implies
the following theorem due to Heawood \cite[Exercise~18, Chapter~6]{Diestel}
(see also \cite{Tsai_West}):%

\begin{theorem}\label{theo.heawood}
A plane triangulation is 3-colorable if and only if it is Eulerian. 
\end{theorem}

\medskip

\noindent
{\bf Remarks.} 
1. In the above theorem, 3-colorable means that we can color the vertices of
$G$ with three colors in such a way that two adjacent vertices are not 
colored alike.

\begin{figure}[htb]
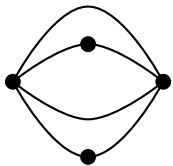

\centering
\psset{xunit=1cm}
\psset{yunit=1cm}
\pspicture(0,0)(2,2) 
\pscurve(0,1)(1,0)(2,1)
\pscurve(0,1)(1,2)(2,1)
\pscurve(0,1)(1,1.5)(2,1)
\pscurve(0,1)(1,0.5)(2,1)
\rput{0}(0,1){\pscircle*{3pt}}
\rput{0}(2,1){\pscircle*{3pt}}
\rput{0}(1,0){\pscircle*{3pt}}
\rput{0}(1,1.5){\pscircle*{3pt}}
\endpspicture
 \caption{\label{fig_G2}
 Graph $G_2$ which is an Eulerian plane triangulation.  
}
\end{figure}

2. This theorem holds for arbitrary Eulerian plane triangulations. In 
particular, we allow for multiple edges. Let us show a simple
example. For any $k\ge 2$, the graph $G_k$ consists of a pair
of vertices connected by $2k$ paths, alternating (as one draws the graph
on a plane) of lengths $1$ and $2$. The smallest example $G_2$ 
is depicted in Figure~\ref{fig_G2}. 

3. Given a Eulerian plane triangulation, then it is {\em uniquely} 
3-colorable (modulo global color permutations).  

\medskip

%
%

As a result, given any Eulerian plane triangulation, there is a {\em unique}\/  
way to split the vertex set into three subsets 
$V=V_1\cup V_2 \cup V_3$, such that they are mutually
disjoint (i.e., $V_i\cap V_j=\emptyset$ for all $j\neq i)$), 
and for each edge $e\in E$, its endpoints belong to different subsets 
(i.e., if $e=\{a,b\}\in E$, then $a\in V_i$ and $b\in V_j$ for $i\neq j$).  
Furthermore, the edge set can also be partitioned into three disjoints sets
$E = E_{12} \cup E_{13} \cup E_{23}$, such that any edge 
$e=\{a,b\}\in E_{ij}$ has one of its endpoints in $V_i$ and the other one in
$V_j$ with $i\neq j$. 

The dual $G^*=(V^*,E^*)$ of an Eulerian plane triangulation $G=(V,E)$ 
is a bipartite cubic graph. Therefore, we can properly color the vertices 
of the dual graph $G^*$ with two colors, or equivalently, we can color 
the faces of $G$ with two colors in such a way that any two neighboring 
faces (i.e., two faces with a common edge) are not colored alike. 
Therefore, this proper face 2-coloring induces
a partition of the (triangular) faces $F=F(G)$ into two disjoint sets  
$F=F_1\cup F_2$, such that all all faces in $F_i$ are colored $i$.  
For a plane triangulation $G$, the following identity 
follows from the handshake lemma on $G^*$: $|E| = 3|F|/2$, which implies
using Euler's formula $|E|-|V|=|F|-2$, that the number of faces $|F|$ is even:
$|F|=2(|V|-2)$. Furthermore, $|F_1|=|F_2| = |V|-2$, because $G^*$ is 3-regular.

\medskip

A graph $G$ is a {\em plane quasi-triangulation} if it is planar and if every
face of $G$ (except the outer one) is bounded by a triangle. This outer
face is bounded by a cycle of length $\ell\ge 4$.
Plane quasi-triangulation are useful when considering a finite
piece of a regular lattice with free boundary conditions: see e.g.,
Figure~\ref{fig_tri_lat}.  
If $G$ is a plane quasi-triangulation with an outer face bounded by a cycle
of length $\ell\ge 4$, we have to make some modifications to the above 
results. On one side, if the degree of all inner vertices is even, then the 
quasi-triangulation is 3-colorable. On the other hand, if there are no
bridges it is also uniquely 3-colorable (modulo global color permutations). 

The number of faces is $|F| = |F_{\rm in}| + 1$, where $|F_{\rm in}|$ is
the number of inner triangular faces. Then the number of edges is given by 
$|E| = (3|F_{\rm in}| +\ell)|/2$. Therefore, $|F_{\rm in}|$ and $\ell$ should
have the same parity, as $|E|$ is an integer. The number of triangular
faces is given by $|F_{\rm in}|=2|V|-\ell-2$. Therefore, we need an outer 
cycle of even length to have an even number of inner triangular faces. 

The dual of a plane quasi-triangulation with all the inner vertices 
of even degree is a graph $G^*=(V^*,E^*)$ constructed as follows: 
the vertex set $V^*=V_{\rm in}^* \cup \{v_{\rm out}^*\}$ contains a vertex
$v^*\in V_{\rm in}^*$ for each inner triangular face in $G$, and an additional
vertex $v_{\rm out}^*$ accounting for the outer face in $G$. The former 
vertices have degree three, and the latter has degree $\ell$. The edge
set $E^*$ is constructed in the usual way. The subgraph induced by
the inner vertices $V_{\rm in}^*$ is indeed bipartite, so the set of inner
faces $F_{\rm in}$ can also be split into two disjoint sets 
$F_{\rm in}=F_1\cup F_2$. However, we cannot ensure that both sets
$F_i$ have the same cardinality, as the dual graph $G^*$ is not regular.  

We are going to study two classes of plane triangulations: the standard triangular lattice (which is a 6-regular graph), and the infinite family of Eulerian
plane triangulations introduced in \cite{Deng_11}. 

\subsection{Triangular lattice} \label{sec.tri.lattice}

%
%
\begin{figure}[htb]
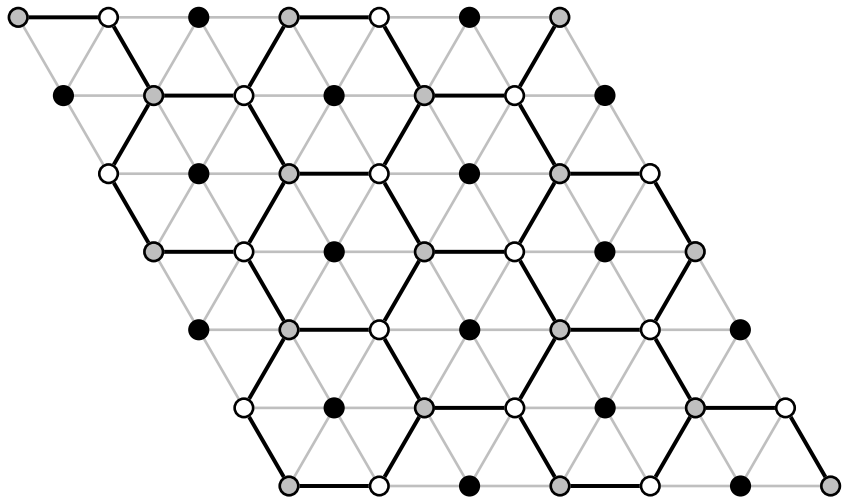

 \centering
 \psset{unit=0.6cm}
 \pspicture(0,-0.5)(18.2,10.5)
 \rput{0}(0,0){ 
   \multirput(-3,5.19615){2}{%
   \multirput(6,0){2}{%
     \psline[linecolor=black,linewidth=1.5pt]( 5,1.73205)( 6,0.000000)
     \psline[linecolor=black,linewidth=1.5pt]( 8,0.00000)( 6,0.000000)
     \psline[linecolor=black,linewidth=1.5pt]( 8,0.00000)( 9,1.732050)
     \psline[linecolor=black,linewidth=1.5pt](11,1.73205)( 9,1.732050)
     \psline[linecolor=black,linewidth=1.5pt]( 5,1.73205)( 6,3.464100)
     \psline[linecolor=black,linewidth=1.5pt]( 9,1.73205)( 8,3.464100)
     \psline[linecolor=black,linewidth=1.5pt]( 6,3.46410)( 8,3.464100)
     \psline[linecolor=black,linewidth=1.5pt]( 6,3.46410)( 5,5.196150)
     \psline[linecolor=black,linewidth=1.5pt]( 8,3.46410)( 9,5.196150)
     \psline[linecolor=lightgray,linewidth=1pt](10,0.00000)( 8,0.000000)
     \psline[linecolor=lightgray,linewidth=1pt](10,0.00000)(12,0.000000)
     \psline[linecolor=lightgray,linewidth=1pt](10,0.00000)( 9,1.732050)
     \psline[linecolor=lightgray,linewidth=1pt](10,0.00000)(11,1.732050)
     \psline[linecolor=lightgray,linewidth=1pt]( 7,1.73205)( 5,1.732050)
     \psline[linecolor=lightgray,linewidth=1pt]( 7,1.73205)( 9,1.732050)
     \psline[linecolor=lightgray,linewidth=1pt]( 7,1.73205)( 6,0.000000)
     \psline[linecolor=lightgray,linewidth=1pt]( 7,1.73205)( 8,0.000000)
     \psline[linecolor=lightgray,linewidth=1pt]( 7,1.73205)( 6,3.464100)
     \psline[linecolor=lightgray,linewidth=1pt]( 7,1.73205)( 8,3.464100)
     \psline[linecolor=lightgray,linewidth=1pt]( 4,3.46410)( 5,1.732050)
     \psline[linecolor=lightgray,linewidth=1pt]( 4,3.46410)( 6,3.464100)
     \psline[linecolor=lightgray,linewidth=1pt]( 4,3.46410)( 5,5.196150)
     \psline[linecolor=lightgray,linewidth=1pt]( 4,3.46410)( 3,5.196150)
     \psline[linecolor=lightgray,linewidth=1pt](10,3.46410)( 9,1.732050)
     \psline[linecolor=lightgray,linewidth=1pt](10,3.46410)( 8,3.464100)
     \psline[linecolor=lightgray,linewidth=1pt]( 7,5.19615)( 6,3.464100)
     \psline[linecolor=lightgray,linewidth=1pt]( 7,5.19615)( 8,3.464100)
     \pscircle*[linecolor=lightgray]         ( 6,0.0){4pt}
     \pscircle[linecolor=black,linewidth=1pt]( 6,0.0){4pt}
     \pscircle*[linecolor=lightgray]         ( 9,1.73205){4pt}
     \pscircle[linecolor=black,linewidth=1pt]( 9,1.73205){4pt}
     \pscircle*[linecolor=lightgray]         ( 6,3.46410){4pt}
     \pscircle[linecolor=black,linewidth=1pt]( 6,3.46410){4pt}
     \pscircle*[linecolor=white]             ( 8,0.0){4pt}
     \pscircle[linecolor=black,linewidth=1pt]( 8,0.0){4pt}
     \pscircle*[linecolor=white]             ( 5,1.73205){4pt}
     \pscircle[linecolor=black,linewidth=1pt]( 5,1.73205){4pt}
     \pscircle*[linecolor=white]             ( 8,3.46410){4pt}
     \pscircle[linecolor=black,linewidth=1pt]( 8,3.46410){4pt}
     \pscircle*[linecolor=black]             (10,0.0){4pt}
     \pscircle*[linecolor=black]             ( 7,1.73205){4pt}
     \pscircle*[linecolor=black]             ( 4,3.46410){4pt}
   }}
   \psline[linecolor=lightgray,linewidth=1pt](18,0)%
                                           (12,10.392300)(0,10.392300)
   \psline[linecolor=black,linewidth=1.5pt](18, 0)       (17, 1.73205)
   \psline[linecolor=black,linewidth=1.5pt](15, 5.19615) (14, 6.9282)
   \psline[linecolor=black,linewidth=1.5pt]( 0,10.392300)( 2,10.392300)
   \psline[linecolor=black,linewidth=1.5pt]( 6,10.392300)( 8,10.392300)
   \pscircle*[linecolor=lightgray]         ( 0,10.392300){4pt}
   \pscircle[linecolor=black,linewidth=1pt]( 0,10.392300){4pt}
   \pscircle*[linecolor=white]             ( 2,10.392300){4pt}
   \pscircle[linecolor=black,linewidth=1pt]( 2,10.392300){4pt}
   \pscircle*[linecolor=black]             ( 4,10.392300){4pt}
   \pscircle*[linecolor=lightgray]         ( 6,10.392300){4pt}
   \pscircle[linecolor=black,linewidth=1pt]( 6,10.392300){4pt}
   \pscircle*[linecolor=white]             ( 8,10.392300){4pt}
   \pscircle[linecolor=black,linewidth=1pt]( 8,10.392300){4pt}
   \pscircle*[linecolor=black]             (10,10.392300){4pt}
   \pscircle*[linecolor=lightgray]         (12,10.392300){4pt}
   \pscircle[linecolor=black,linewidth=1pt](12,10.392300){4pt}
   \pscircle*[linecolor=black]             (13, 8.66025){4pt}
   \pscircle*[linecolor=white]             (14, 6.9282){4pt}
   \pscircle[linecolor=black,linewidth=1pt](14, 6.9282){4pt}
   \pscircle*[linecolor=lightgray]         (15, 5.19615){4pt}
   \pscircle[linecolor=black,linewidth=1pt](15, 5.19615){4pt}
   \pscircle*[linecolor=black]             (16, 3.4641){4pt}
   \pscircle*[linecolor=white]             (17, 1.73205){4pt}
   \pscircle[linecolor=black,linewidth=1pt](17, 1.73205){4pt}
   \pscircle*[linecolor=lightgray]         (18, 0){4pt}
   \pscircle[linecolor=black,linewidth=1pt](18, 0){4pt}
 }
 \endpspicture
 \caption{\label{fig_tri_lat}
  Triangular lattice of size $7\times 7$ with free boundary conditions. 
  The vertex set has a partition into three disjoint sets, depicted as   
  gray, white, and black dots. The edge set is formed by the black and gray 
  edges. The black edges show the subgraph $G_{12} = (V_1\cup V_2,E_{12})$,
  which is a hexagonal lattice.  
}
\end{figure}

An infinite triangular lattice $G=(V,E)$ is a 6-regular 
triangulation of the infinite plane, as shown in Figure~\ref{fig_tri_lat}. 
It is uniquely 3--colorable, therefore we can split the vertex set into 
three disjoint subsets: $V=V_1 \cup V_2 \cup V_3$. Note that 
for each $i\neq j$,  the subgraph $G_{ij}= (V_i  \cup V_j, E_{ij})$ forms 
a hexagonal lattice.

The dual lattice $G_{ij}^* =(V_k,E^*)$ contains as a  vertex set the third 
subset $V_k$ ($k\neq i,j$) of $V$. This also works on the other way around: 
for each $i=1,2,3$, the vertex set $V_i$ can be regarded as the vertex set 
of another triangular lattice (with larger lattice spacing), and whose dual is
precisely the hexagonal lattice $G_{jk}=(V_j\cup V_k,E_{jk})$ with $i\neq j,k$.
For instance, in Figure~\ref{fig_tri_lat} the white and gray dots and
all the black edges form a hexagonal lattice, whose
dual has a vertex set formed by the black dots. 
 
A finite piece of the triangular lattice is a finite subset of the 
infinite lattice described above with linear dimensions $L_x \times L_y$ 
and free boundary conditions. Figure~\ref{fig_tri_lat} shows an example with
$L_x=L_y=7$. This is a plane quasi--triangulation; but it is not 
Eulerian, as there are precisely two vertices with odd degree (the top
right and the bottom left corners). All the inner vertices $i$ have degree
$d_i=6$, but the degree of the vertices belonging to the outer
cycle can be 2, 3, or 4. The length of this outer cycle is
$\ell=2(L_x + L_y -2)$. In order to obtain an Eulerian plane triangulation
from this graph, we  proceed as follows: 
\begin{enumerate}
\item 
   Add an edge $f$ joining the two vertices with odd degree. We obtain 
   an Eulerian graph with all faces being triangles, except two, which 
   are bounded by cycles of length $\ell'=L_x+L_y-1$. 
   
\item If $\ell'$ is not a multiple of $3$, then  
   subdivide the extra edge with $1$ (resp.\/ $2$) vertices when 
   $\ell' \equiv 2 \pmod{3}$ (resp.\/ $\ell' \equiv 1 \pmod{3}$). 
   So we end with two faces bounded by cycles of length multiple of $3$, 
   and all vertices along this cycle have even degree.

\item Take one cycle, and number the vertices $1,2,3$ according to the 
   sublattice $V_i$ the vertex belongs to. Then form triangular 
   faces by adding one edge between two vertices labeled $1$ and $3$, with
   a single vertex labeled $2$ in between. 

\item We obtain one cycle of even length with all vertices of odd degree
   (as in the previous step we added a single edge to each vertex).
   The vertices on this cycle are labeled alternatively $1$ and $3$.
   Now place a new vertex (labeled $2$) inside the face bounded by this
   cycle. Then add edges between this new vertex and any of the vertices 
   defining the cycle. Indeed, the degree of all vertices is now even. 

\item Repeat the last two steps on the other cycle of length multiple of 3. 
\end{enumerate}
In this way, from a finite subset of linear size $L_x \times L_y$ of a
triangular lattice with free boundary conditions, we have produced an 
Eulerian plane triangulation. Indeed, the added vertices and edges are a
negligible fraction of the total number of vertices and edges, when 
$L_x,L_y\to\infty$. Therefore, we expect that their contribution to the 
infinite-volume free energy would be zero.

\subsection{A family of Eulerian plane triangulations} \label{sec.lattices}

%
%
\begin{figure}[htb]
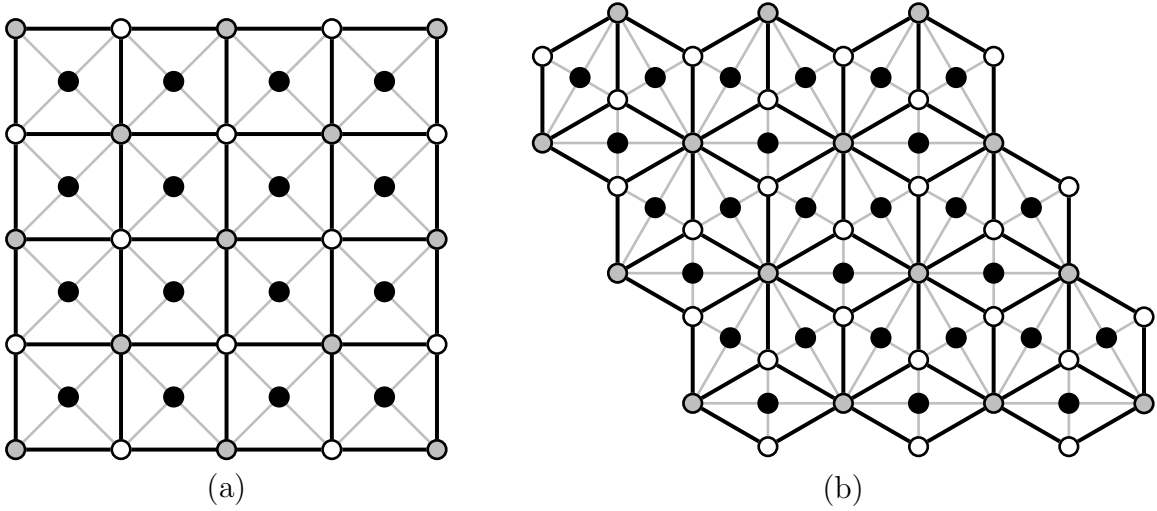

 \centering
 \pspicture(0,-1)(15,6)
%
%
 \psset{unit=0.7cm}
 \rput{0}(0,0){
   \multirput(0,4){2}{%
   \multirput(4,0){2}{%
     \psline[linecolor=black,linewidth=1.5pt](0,0)(2,0)
     \psline[linecolor=black,linewidth=1.5pt](0,0)(0,2)
     \psline[linecolor=black,linewidth=1.5pt](2,0)(4,0)
     \psline[linecolor=black,linewidth=1.5pt](2,0)(2,2)
     \psline[linecolor=black,linewidth=1.5pt](0,2)(2,2)
     \psline[linecolor=black,linewidth=1.5pt](0,2)(0,4)
     \psline[linecolor=black,linewidth=1.5pt](2,2)(4,2)
     \psline[linecolor=black,linewidth=1.5pt](2,2)(2,4)
     \psline[linecolor=lightgray,linewidth=1pt](1,1)(0,0)
     \psline[linecolor=lightgray,linewidth=1pt](1,1)(0,2)
     \psline[linecolor=lightgray,linewidth=1pt](1,1)(2,0)
     \psline[linecolor=lightgray,linewidth=1pt](1,1)(2,2)
     \psline[linecolor=lightgray,linewidth=1pt](3,1)(2,0)
     \psline[linecolor=lightgray,linewidth=1pt](3,1)(2,2)
     \psline[linecolor=lightgray,linewidth=1pt](3,1)(4,0)
     \psline[linecolor=lightgray,linewidth=1pt](3,1)(4,2)
     \psline[linecolor=lightgray,linewidth=1pt](1,3)(0,2)
     \psline[linecolor=lightgray,linewidth=1pt](1,3)(0,4)
     \psline[linecolor=lightgray,linewidth=1pt](1,3)(2,2)
     \psline[linecolor=lightgray,linewidth=1pt](1,3)(2,4)
     \psline[linecolor=lightgray,linewidth=1pt](3,3)(2,2)
     \psline[linecolor=lightgray,linewidth=1pt](3,3)(2,4)
     \psline[linecolor=lightgray,linewidth=1pt](3,3)(4,2)
     \psline[linecolor=lightgray,linewidth=1pt](3,3)(4,4)
     \pscircle*[linecolor=lightgray]         (0,0){4pt}
     \pscircle[linecolor=black,linewidth=1pt](0,0){4pt}
     \pscircle*[linecolor=lightgray]         (2,2){4pt}
     \pscircle[linecolor=black,linewidth=1pt](2,2){4pt}
     \pscircle*[linecolor=white]             (2,0){4pt}
     \pscircle[linecolor=black,linewidth=1pt](2,0){4pt}
     \pscircle*[linecolor=white]             (0,2){4pt}
     \pscircle[linecolor=black,linewidth=1pt](0,2){4pt}
     \pscircle*[linecolor=black]             (1,1){4pt}
     \pscircle*[linecolor=black]             (1,3){4pt}
     \pscircle*[linecolor=black]             (3,1){4pt}
     \pscircle*[linecolor=black]             (3,3){4pt}
   }}
   \psline[linecolor=black,linewidth=1.5pt](8,0)(8,8)
   \psline[linecolor=black,linewidth=1.5pt](0,8)(8,8)
   \multirput(4,0){3}{%
     \pscircle*[linecolor=lightgray]         (0,8){4pt}
     \pscircle[linecolor=black,linewidth=1pt](0,8){4pt}
   }
   \multirput(4,0){2}{%
     \pscircle*[linecolor=white]             (2,8){4pt}
     \pscircle[linecolor=black,linewidth=1pt](2,8){4pt}
   }
   \multirput(0,4){2}{%
     \pscircle*[linecolor=lightgray]         (8,0){4pt}
     \pscircle[linecolor=black,linewidth=1pt](8,0){4pt}
     \pscircle*[linecolor=white]             (8,2){4pt}
     \pscircle[linecolor=black,linewidth=1pt](8,2){4pt}
   }
   \uput[270](4,-0.2){(a)}
 }
%
%
 \psset{unit=0.5cm}
 \rput{0}(11,-0.5){
   \multirput(-2.0,3.46410){3}{%
   \multirput(4,0){3}{%
     \psline[linecolor=black,linewidth=1.5pt](7.0,1.73205)(9.0,0.57735)
     \psline[linecolor=black,linewidth=1.5pt](7.0,1.73205)(9.0,2.88675)
     \psline[linecolor=black,linewidth=1.5pt](9.0,2.88675)(11.0,1.732050)
     \psline[linecolor=black,linewidth=1.5pt](9.0,0.57735)(11.0,1.732050)
     \psline[linecolor=black,linewidth=1.5pt](9.0,2.88675)(9.0,5.19615)
     \psline[linecolor=black,linewidth=1.5pt](7.0,1.73205)(7.0,4.04145)
     \psline[linecolor=lightgray,linewidth=1pt](7.0,1.732050)(11.0,1.732050)
     \psline[linecolor=lightgray,linewidth=1pt](9.0,0.57735)(9.0,2.88675)
     \psline[linecolor=lightgray,linewidth=1pt](9.0,2.88675)(8.0,3.47410)
     \psline[linecolor=lightgray,linewidth=1pt](9.0,5.19615)(7.0,1.73205)
     \psline[linecolor=lightgray,linewidth=1pt](9.0,5.19615)(11.0,1.732050)
     \psline[linecolor=lightgray,linewidth=1pt](8.0,3.47410)(7.0,4.04145)
     \psline[linecolor=lightgray,linewidth=1pt](9.0,2.88675)(10.0,3.47410)
     \psline[linecolor=lightgray,linewidth=1pt](11.0,4.06145)(10.0,3.47410)
   }}
   \multirput(3,10.9597)(4,0){3}{%
     \psline[linecolor=black,linewidth=1.5pt](0,0)(2,1.1547)(4,0)
   }
   \multirput(15,8.6503)(2,-3.46410){2}{%
     \psline[linecolor=black,linewidth=1.5pt](0,0)(2,-1.1547)
   }
   \multirput(15,10.9597)(2,-3.46410){3}{%
     \psline[linecolor=black,linewidth=1.5pt](0,0)(0,-2.3094)
   }
   \multirput(-2.0,3.46410){3}{%
   \multirput(4,0){3}{%
     \pscircle*[linecolor=lightgray]         (7.0,1.73205){4pt}
     \pscircle[linecolor=black,linewidth=1pt](7.0,1.73205){4pt}
     \pscircle*[linecolor=black]             (9.0,1.73205){4pt}
     \pscircle*[linecolor=white]             (9.0,2.88675){4pt}
     \pscircle[linecolor=black,linewidth=1pt](9.0,2.88675){4pt}
     \pscircle*[linecolor=white]             (9.0,0.57735){4pt}
     \pscircle[linecolor=black,linewidth=1pt](9.0,0.57735){4pt}
     \pscircle*[linecolor=black]             (8.0,3.47410){4pt}
     \pscircle*[linecolor=black]             (10.0,3.47410){4pt}
   }}
   \multirput(3,10.9597)(4,0){3}{%
     \pscircle*[linecolor=white]             (0,0){4pt}
     \pscircle[linecolor=black,linewidth=1pt](0,0){4pt}
     \pscircle*[linecolor=lightgray]         (2,1.1547){4pt}
     \pscircle[linecolor=black,linewidth=1pt](2,1.1547){4pt}
   }
   \multirput(15,10.9597)(2,-3.46410){3}{%
     \pscircle*[linecolor=white]             (0,0){4pt}
     \pscircle[linecolor=black,linewidth=1pt](0,0){4pt}
     \pscircle*[linecolor=lightgray]         (0,-2.3094){4pt}
     \pscircle[linecolor=black,linewidth=1pt](0,-2.3094){4pt}
   }
   \uput[270](11,0.2){(b)}
 }
 \endpspicture
 \caption{\label{fig_lattices}
  Two pieces of planar triangulations with free boundary conditions: 
  (a) the union-jack lattice of size $5\times 5$, and (b) the 
  bisected hexagonal lattice of size $4\times 4$ (measured in units of the 
  underlying triangular Bravais lattice i.e., the gray dots). 
  The vertex set has a partition into three disjoint sets, depicted as
  gray, white, and black dots. The gray and white dots and the black solid
  lines form the lattice $\widehat{G}$: square (a) and diced (b). The new
  vertices and edges added to $\widehat{G}$ to get $\widetilde{G}$ are 
  depicted as black dots and solid gray lines, respectively.
} 
\end{figure}

In this section, we follow Ref.~\cite{Deng_11}.
Let us construct this family of Eulerian planar triangulation by starting from
a connected planar graph $G=(V,E)$. We can then define its dual $G^*=(V^*,E^*)$.
The next step is to build the graph $\widehat{G}=(V\cup V^*,\widehat{E})$
with vertex set $V\cup V^*$ and edges $ij$ whenever $i\in V$ lies on the 
boundary of the face of $G$ that contains $j\in V^*$. The graph 
$\widehat{G}$ is a plane quadrangulation: on each face of $\widehat{G}$,
one pair of diametrically opposite vertices corresponds to an edge $e\in E$,
and the other pair corresponds to an edge $e^*\in E^*$. In other words,
$\widehat{G}$ is the dual of the medial graph 
$\mathcal{M}(G)=\mathcal{M}(G^*)$. 
Conversely, every quadrangulation $\widehat{G}$ arises via this construction 
from some pair $G,G^*$. (See \cite[Figure~1]{Deng_11}.) 
The vertices of $G$ (resp.\/ $G^*$) are depicted gray (resp.\/ white),
and the edges in $\widehat{E}$ are depicted as solid black lines 
in Figure~\ref{fig_lattices}. 

Let us now define the graph $\widetilde{G}$ constructed from  
$\widehat{G}$ by adjoining a new vertex in each face of $\widehat{G}$,
and four new edges connecting this new vertex to the four corners of the
face. The new vertex is depicted as a black dot, and the new edges as
solid gray lines in Figure~\ref{fig_lattices}. The graph 
$\widetilde{G}$ is an Eulerian plane triangulation, with vertex tripartition 
$V=V \cup V^* \cup V_3$, where $V_3$ is the set consisting of the new degree-4
vertices. Conversely, every Eulerian plane triangulation in which one 
sublattice consists of degree-4 vertices arises in this way. Indeed, this 
does not cover the triangular lattice.   

If $G=G^*=\widehat{G}=$ square lattice, then $\widetilde{G}= $ union-jack 
lattice, displayed in Figure~\ref{fig_lattices}(a). If $G= $ triangular
lattice, $G^*=$ hexagonal lattice, then $\widehat{G}=$ diced lattice, and
$\widetilde{G}= $ bisected hexagonal lattice, displayed in 
Figure~\ref{fig_lattices}(b). According to the standard notation
\cite{tilings}, the union-jack lattice is the $[4\cdot 8^2]$ Leaves tiling,
and the bisected hexagonal lattice corresponds to the Leaves tiling 
$[4\cdot 6 \cdot 12]$.

\medskip

The infinite union-jack lattice is not regular since $d_i=8$ for all 
$i\in V\cup V^*$, while $d_i=4$ for all $i\in V_3$. This lattice can be 
regarded as a square Bravais lattice with a two-point basis.   
Notice that the union-jack lattice satisfies the following properties:

\begin{itemize}
\item The graph $\widehat{G}=(V\cup V^*,\widehat{E})$ has a square-lattice dual
      $\widehat{G}^* = (V_3, \widehat{E}^*)$.  

\item The subgraph $G_{13} = (V\cup V_3, E_{13})$, 
      where $E_{13}$ are the set of edges joining vertices $e=\{ij\}$ with 
      $i\in V$ and $j\in V_3$, can be regarded as a decorated square lattice 
      tilted 45$^\text{o}$ with respect the original lattice. Its dual 
      is another tilted square lattice $G_{13}^* = (V^*, E_{13}^*)$, 
      where every edge in $G_{13}^*$ is doubled. The same occurs to the 
      subgraph  $G_{23} = (V^*\cup V_3, E_{23})$.  
\end{itemize}

\medskip

The bisected hexagonal lattice is also non-regular: the degrees of the vertices
in $V$ (gray dots), $V^*$ (white dots), and $V_3$ (black dots) are 
respectively 12, 6, and 4 (see Figure~\ref{fig_lattices}(b)).   
This lattice can be regarded as a triangular Bravais lattice (formed by the 
vertices in $V$) with a 6-point basis. 
Notice that the bisected hexagonal lattice satisfies the following properties:

\begin{itemize}
\item The graph $\widehat{G}=(V\cup V^*,\widehat{E})$ is a diced lattice, 
      and its dual is $\widehat{G}^* = (V_3, \widehat{E}^*)$ which is a 
      kagome lattice. 

\item The subgraph $G_{13} = (V\cup V_3, E_{13})$, where $E_{13}$ are the set 
      of edges joining vertices $e=\{ij\}$ with $i\in V$ and $j\in V_3$, 
      can be regarded as a decorated triangular lattice. Its dual
      is the hexagonal lattice $G_{13}^* = (V^*, E_{13}^*)$, where
      every edge in $G_{13}^*$ is doubled. 
 
\item The subgraph $G_{23} = (V^*\cup V_3, E_{23})$, where $E_{23}$ are the set 
      of edges joining vertices $e=\{ij\}$ with $i\in V^*$ and $j\in V_3$, 
      can be regarded as a decorated hexagonal lattice. Its dual
      is the triangular lattice $G_{13}^* = (V, E_{23}^*)$, where
      every edge in $G_{23}^*$ is doubled. 
\end{itemize}

\medskip

The union-jack and bisected-hexagonal lattices with free boundary conditions
shown in  Figure~\ref{fig_lattices} are quasi triangulations. But in both cases
it is very easy to modify them slightly so they become Eulerian triangulations:
if suffices to add an outer vertex belonging to $V_3$ (black dot) and join 
this extra vertex to every vertex on the outer cycle of each graph. As the
degree of all of them was odd, with the new edge, it becomes even, and we 
obtain the desired Eulerian plane triangulation. 

%
%
\section{Models to be studied} \label{sec.models} 

In this section we will introduce the models we are going to use in the 
following sections, and study their main properties. 
We cover a well-known model (the AT model \cite{Ashkin_Teller_43}), 
a model known in the literature in a different context 
(the HMBW model \cite{Hintermann_72,Merlini_72,Gruber_77}), a model used as 
a mere technical step (but we think it has some interest in its own right: 
the mixed AT model \cite{Baxter_book}), and a {\em new} model 
(the infinite-coupling limit AT model).

%
%
\subsection{HMBW model on an Eulerian plane triangulation}
\label{sec.HMBW}

Given an Eulerian plane triangulation $G=(V,E)$ with vertex set $V$, edge set 
$E$, and the set $F(G)$ of all the {\em triangular} faces,
we define the {\em Hintermann--Merlini--Baxter--Wu model} (HMBW) on $G$ 
as follows: on each vertex $i\in V$, we place an Ising spin $\sigma_i=\pm 1$, 
and these spins interact via the Hamiltonian:
\be
\mathcal{H}_\text{HMBW} \;=\; 
- J \sum\limits_{ \{i,j,k\} = t \in F(G)} \sigma_i \sigma_j \sigma_k \,,
\label{def_H_HMBW}
\ee
where the sum is over all the triangular faces $t=\{i,j,k\}$ of $G$  
bounded by the vertices $i,j,k$, and the corresponding 3--spin interaction 
has a coupling constant $J$.  
The partition function is 
\be
Z_\text{HMBW}(G;J) \;=\; \sum\limits_{\{\sigma\}}   
   e^{- \mathcal{H}_\text{HMBW}}\,.
\label{def_Z_HMBW}
\ee

In the ferromagnetic regime $J\ge 0$, if $G$ is a plane Eulerian triangulation
the HMBW model has exactly four ground states: one in which all spins take
the same value $\sigma=+1$, and three states in which one spin takes the
value $\sigma=+1$, and the other two take the opposite value.

One important property of the ferromagnetic HMBW model on an Eulerian 
plane triangulation is that this model is self-dual \cite{Gruber_77} with 
dual coupling satisfying the identity
\be
v \, v^* \;=\; 2 \,,
\label{def_dual_v}
\ee
where the temperature-like variable $v$ is defined as 
\be
v \;=\; e^{2 J} - 1 \,.
\label{def_v_HMBW}
\ee
This self-dual point satisfies:
\begin{subeqnarray}
\slabel{def_vc_HMBW}
v_c       &=& \sqrt{2} \\
e^{2J_c} &=& 1 + \sqrt{2} 
\slabel{def_Jc_HMBW}
\label{def_Tc_HMBW}
\end{subeqnarray}
In the BW and HM models this self-dual point \reff{def_Tc_HMBW}
constitutes the {\em unique} critical point of the model, as the exact 
solution shows \cite{Hintermann_72,Baxter_Wu_73,Baxter_Wu_74}.
Notice that the critical coupling \reff{def_Jc_HMBW} is  the same as the 
critical coupling for the square--lattice Ising model \cite{Onsager_44}. 

%
%
\subsection{The Ashkin--Teller model}
\label{sec.AT}

Let us consider an arbitrary graph $G=(V,E)$, with vertex set $V$ and 
edge set $E$. The {\em Ashkin--Teller model}\/ (AT) is defined on $G$ 
as follows: 
on each vertex $i\in V$ we place two Ising spins
$\sigma_i,\tau_i=\pm 1$, and these spins interact via the Hamiltonian
\reff{def_H_AT} \cite{Ashkin_Teller_43}.
Therefore, we can regard this system as two copies of the graph $G$, with the
$\sigma$ spins living on one copy, and the $\tau$ spins living on the other.
The $\sigma$ (resp.\ $\tau$) spins interact among themselves via a 
nearest-neighbor coupling $K_{2}$ (resp.\ $K'_{2}$),
and both copies interact via a 4--spin coupling $K_{4}$. 
The partition function for this system is 
\be
Z_\text{AT}(G;K_2,K'_2,K_4) \;=\; \sum\limits_{\{\sigma,\tau\}} 
   e^{-\mathcal{H}_\text{AT}} \,.
\label{def_Z_AT}
\ee 

The Boltzmann weight for each edge $e=\<ij\>$ can be read from 
\reff{def_H_AT}/\reff{def_Z_AT}, and depends only on the products 
$\sigma_i\sigma_j,\tau_i\tau_j=\pm 1$:  
\be
   \omega_{ij}(\sigma_i\sigma_j,\tau_i\tau_j)
   \;=\;
   \exp\left[ K_{2} \, \sigma_i \sigma_j \,+\,
              K'_{2}\, \tau_i   \tau_j   \,+\,
              K_{4} \, \sigma_i \sigma_j \tau_i \tau_j
       \right] \,.
   \label{def_weight_AT_general}
\ee
Indeed, the weights $\omega_{ij}$ are the same for all edges $\<ij\>\in E$.
For each edge $\<ij\>\in E$, there are four possible spin configurations.
Their corresponding weights are denoted by $\omega_{k}$ with $k=0,1,2,3$,
and they are given in the second column of 
Table~\ref{table_weights_AT_general}. We use the standard definitions
(see e.g., Ref.~\cite{Wu_77}):
\begin{subeqnarray}
\omega_{0}&=&\omega_{ij}(+1,+1)\\
\omega_{1}&=&\omega_{ij}(+1,-1)\\
\omega_{2}&=&\omega_{ij}(-1,+1)\\
\omega_{3}&=&\omega_{ij}(-1,-1)
\label{def_omega_AT}
\end{subeqnarray}
We can also use these weights \reff{def_omega_AT} to define the corresponding
AT model:
\be
Z_\text{AT}(G;K_2,K'_2,K_4) \;\equiv \; 
Z_\text{AT}(G;\omega_0, \omega_1, \omega_2,\omega_4)\,.
\label{def_Z_AT_bis}
\ee 

The values of the couplings $K_2,K'_2,K_4$ in terms of 
the weights $\omega_0,\omega_1,\omega_2,\omega_3$
can be obtained from the equations given in the second column of 
Table~\ref{table_weights_AT_general}:
\begin{subeqnarray}
e^{4K_{0}} &=& \omega_{0} \, \omega_{1}\,  \omega_{2}\,  \omega_{3} \\[2mm] 
e^{4K_{2}} &=&  \frac{\omega_{0}\,  \omega_{1}}
                     {\omega_{2}\,  \omega_{3}}\\[2mm] 
e^{4K_{2}'} &=& \frac{\omega_{0}\, \omega_{2}}
                     {\omega_{1}\, \omega_{3}}\\[2mm] 
e^{4K_{4}} &=&  \frac{\omega_{0}\, \omega_{3}}
                     {\omega_{1}\, \omega_{2}}
\label{def_omega_couplings_AT_vs_omega}
\end{subeqnarray}
where $K_{0}$ is an arbitrary constant fixing the zero of energy. 
In Eq.~\reff{def_H_AT} we choose $K_{0}=0$ for simplicity; therefore
the weights $\omega_{k}$ satisfy: 
\be
\omega_{0} \, \omega_{1} \, \omega_{2}\,  \omega_{3} \;=\; 1 \,.
\label{def_convention_K0_AT}
\ee
 
%
%
\begin{table}[htb]
\centering
\begin{tabular}{ccc}
\hline\hline
Configuration &  Weight & (Normalized) Weight \\
\hline \\[-2mm]
$\sigma_i\sigma_j=+1\,, \tau_i\tau_j=+1$& 
$\omega_{0}=e^{K_{0}+K_{2}+K'_{2}+K_{4}}$ & $1$\\[2mm]
$\sigma_i\sigma_j=+1\,, \tau_i\tau_j=-1$& 
$\omega_{1}=e^{K_{0}+K_{2}-K'_{2}-K_{4}}$ & $e^{-2(K'_{2}+K_{4})}$\\[2mm] 
$\sigma_i\sigma_j=-1\,, \tau_i\tau_j=+1$& 
$\omega_{2}=e^{K_{0}-K_{2}+K'_{2}-K_{4}}$ & $e^{-2(K_{2}+K_{4})}$\\[2mm]
$\sigma_i\sigma_j=-1\,, \tau_i\tau_j=-1$& 
$\omega_{3}=e^{K_{0}-K_{2}-K'_{2}+K_{4}}$ & $e^{-2(K_{2}+K'_{2})}$\\[2mm] 
\hline\hline
\end{tabular}
\caption{\label{table_weights_AT_general}
Boltzmann weights for the AT model \reff{def_H_AT}/\reff{def_Z_AT}.
For each edge $\<ij\>\in E$, we first give the configuration of the 
corresponding $\sigma,\tau$ variables, we then quote the weight
read off from \reff{def_weight_AT_general} [`` Weight''],
and the normalized weight obtained by making the first one equal to $1$
[``(Normalized) Weight'']. $K_{0}$ is an arbitrary constant fixing the
zero of energy for the contribution of each edge $\<ij\>\in E$. 
In \protect\reff{def_H_AT} we choose $K_{0}=0$ for simplicity. 
}
\end{table}

For our purposes, it is also interesting to compute the normalized 
weights obtained by making the weight corresponding to the 
configuration $\sigma_i=\sigma_j$, $\tau_i=\tau_j$ equal to one
(i.e., $\omega_{0}=1$). 
These normalized weights are also displayed in the third column of  
Table~\ref{table_weights_AT_general}.

The AT model \reff{def_H_AT}/\reff{def_Z_AT} 
contains two important particular cases: when $K_4=0$, we
obtain two decoupled Ising models with couplings $K_2$ and $K_2'$. 
At the other extreme, the limit $K_4\to+\infty$ corresponds to a single
Ising model ($\sigma=\tau)$ with coupling $K_2+K_2'$. Finally, the line 
$K_2=K_2'=K_4$, corresponds to the 4--state Potts model with 
$J_{\rm Potts}=4K_2$. 

The Hamiltonian \reff{def_H_AT} is invariant under any permutation $\pi$ of 
the coupling constants $(K_2,K_2',K_4)$, as the fields 
$\sigma,\tau,\sigma\tau$ play symmetric roles in the model: 
\be
(K_2,K_2',K_4) \;\to\; \pi (K_2,K_2',K_4)\,, \qquad \pi \in S_3  \,. 
\label{def_symmetries_AT}
\ee
If the graph $G$ is {\em bipartite}, then the vertex set can be split into
two disjoints sets $V=V_1 \cup V_2$, such that each edge $e=\<ij\>$ satisfies
that $i\in V_i$ and $j=V_j$ with $i\neq j$. In this case, there are additional 
symmetries because we can flip $\sigma$, $\tau$, or both, on any of two vertex
subsets $V_i$. Then, the uniform AT model on a bipartite graph $G$ is 
invariant under the transformations:
\begin{subeqnarray}
(K_2,K_2',K_4) &\to& (-K_2,K'_2,-K_4) \\
(K_2,K_2',K_4) &\to& ( K_2,-K'_2,-K_4) \\
(K_2,K_2',K_4) &\to& (-K_2,-K'_2,K_4) 
\label{def_symmetries_AT_bipartite}
\end{subeqnarray} 
When $K_2 = K_2'$ we obtain the {\em symmetric Ashkin--Teller model}: 
\be
\mathcal{H}_\text{sAT} \;=\; - \sum\limits_{\<ij\>\in E}
\left[ K_{2} (\sigma_{i}\sigma_{j} \,+\,
              \tau_i \tau_j) \,+\,
       K_{4} \sigma_{i}\sigma_{j} \tau_i \tau_j \right] \,.
\label{def_H_sAT}
\ee
%

%
%
\subsection{The Infinite-Coupling Limit Ashkin--Teller model}
\label{sec.ICLAT}

As far as we can tell this model is new in the literature.\footnote{
   After the completion of this work, we learned that a particular case 
   of this model had been previously considered by Ikhlef and Rajabpour 
   \cite{Ikhlef_12}. Their findings have been summarized in a remark 
   in Section~\ref{sec.sq.ICLAT}. We thank the referee for bringing this
   paper to our attention.
} 
Its motivation is rather simple: if we look at the normalized weights in 
Table~\ref{table_weights_AT_general}, 
we see that in the limit $K_{2}$, $K'_{2}$, $-K_{4} \rightarrow +\infty$, 
with $K_{2}+K_{4}=L_{2}$ and $K'_{2}+K_{4}=L'_{2}$ 
kept finite, one has that the normalized weight 
for $\sigma_i\sigma_j=\tau_i\tau_j=-1$ tends to zero, while the 
other configurations keep nonzero normalized weights. These weights 
correspond to the {\em infinite-coupling-limit Ashkin--Teller model}\/
(ICLAT). The Hamiltonian of this model can be alternatively written as:
\be
\mathcal{H}_\text{ICLAT} \;=\; - \sum\limits_{\<ij\>\in E} 
\left[ (L_{2}-L) \sigma_{i}\sigma_{j} \,+\, 
       (L'_{2}-L)\tau_i \tau_j \,+\, 
       L \sigma_{i}\sigma_{j} \tau_i \tau_j \right] \,,
\label{def_H_ICLAT}
\ee 
where we take the limit $L\to-\infty$, with the couplings 
$L_{2},L'_{2}$ kept finite. The corresponding partition function is
\be
Z_\text{ICLAT}(G;L_2,L'_2) \;=\; \lim_{L\to-\infty}
\sum\limits_{\{\sigma,\tau\}} 
   e^{-\mathcal{H}_\text{ICLAT}} \,.
\label{def_Z_ICLAT}
\ee 
The Boltzmann weight for a given edge $e=\<ij\>$ is given by
\reff{def_H_ICLAT}/\reff{def_Z_ICLAT} and it reads:
\be
   \omega(\sigma_i\sigma_j,\tau_i\tau_j)
   \;=\;
   \exp\left[ (L_{2}-L) \, \sigma_i \sigma_j \,+\,
              (L'_{2}-L)\, \tau_i   \tau_j   \,+\,
              L \sigma_i \sigma_j \tau_i \tau_j
       \right] \,.
   \label{def_weight_ICLAT_general}
\ee
The values of this weight for the four possible spin configuration are 
given in Table~\ref{table_weights_ICLAT_general}. Indeed, if we normalize
these weights so that the normalized weight
for $\sigma_i\sigma_j=\tau_i\tau_j=1$ is equal to one (see the third 
column in Table~\ref{table_weights_ICLAT_general}), we obtain the same
weights as those obtained by taking the appropriate infinite--coupling 
limit in the normalized weights for the AT model (displayed in the third
column in Table~\ref{table_weights_AT_general}). 

%
%
\begin{table}[htb]
\centering
\begin{tabular}{ccc}
\hline\hline
Configuration &  Weight & (Normalized) Weight\\
\hline \\[-2mm]
$\sigma_i\sigma_j=+1\,,    \tau_i\tau_j=+1$& 
$\omega_{0}=e^{L_{0}+L_{2} + L'_{2}-L}$ & $1$ \\[2mm]
$\sigma_i\sigma_j=+1\,,    \tau_i\tau_j=-1$& 
$\omega_{1}=e^{L_{0}+L_{2} - L'_{2}-L}$   & $e^{-2L'_{2}}$ \\[2mm]
$\sigma_i\sigma_j=-1\,, \tau_i\tau_j=+1$& 
$\omega_{2}=e^{L_{0}-L_{2} + L'_{2}-L}$  & $e^{-2L_{2}}$ \\[2mm]
$\sigma_i\sigma_j=-1\,, \tau_i\tau_j=-1$& 
$\omega_{3}=e^{L_{0}+3L-L_{2}-L'_{2}}$&$e^{2(2L-L_{2}-L'_{2})}\to 0$ \\[2mm]
\hline\hline
\end{tabular}
\caption{\label{table_weights_ICLAT_general}
Boltzmann weights for the ICLAT model \reff{def_H_ICLAT}/\reff{def_Z_ICLAT}.
For each edge $\<ij\>\in E$, we first give the configuration of the
corresponding $\sigma,\tau$ variables, we then quote the weight
read off from \reff{def_weight_ICLAT_general} [`` Weight''],
and the normalized weight obtained by making the first one equal to $1$
[``(Normalized) Weight''] in the limit $L\to-\infty$.  
$L_0$ is an arbitrary constant fixing the
zero of energy for the contribution of each edge $\<ij\>\in E$.
In \protect\reff{def_H_ICLAT} we choose $L_0=0$ for simplicity.
}
\end{table}

When $L_2\to+\infty$, then the $\sigma$ spins are all equal; therefore, the
model reduces to an Ising model on the $\tau$ spins with coupling $L_2'$. 
Indeed, when $L_2'\to+\infty$, the $\tau$ spins are all equal, and 
the model reduces to an Ising model on the $\sigma$ spins with coupling $L_2$.  

In the ICLAT model \reff{def_H_ICLAT} $\sigma$ and $\tau$ play a 
symmetric role, therefore, the model is invariant under the transformation
\be
(L_2,L_2') \;\to\; (L_2',L_2) \,.
\label{def_symmetries_ICLAT}
\ee

If the graph $G$ is {\em bipartite}, we can define a new spin 
$\rho_i = \sigma_i \tau_i = \pm 1$, so that the Hamiltonian \reff{def_H_ICLAT}
reads:
\be
\mathcal{H}_\text{ICLAT} \;=\; - \sum\limits_{\<ij\>\in E}
\left[ (L_{2}-L) \sigma_{i}\sigma_{j} \,+\,
       (L'_{2}-L)\rho_i \rho_j \sigma_i \sigma_j \,+\,
       L \rho_{i}\rho_{j} \right] \,,
\label{def_H_ICLAT_bis}
\ee
when $L \to -\infty$. If we flip the spins $\rho_i$ on one of two vertex
subsets $V_i$, then the model is invariant under the transformation  
$(L_2-L,L'_2-L,L)\to (L_2-L,L-L_2',-L)$. We obtain
\be
\mathcal{H}_\text{ICLAT} \;=\; - \sum\limits_{\<ij\>\in E}
\left[ (L_{2}-L) \sigma_{i}\sigma_{j} \,-\,
       (L'_{2}-L)\rho_i \rho_j \sigma_i \sigma_j \,-\,
       L \rho_{i}\rho_{j}  \right] \,,
\label{def_H_ICLAT_tris}
\ee
in the limit $L \to -\infty$. 
By comparing \reff{def_H_ICLAT}/\reff{def_H_ICLAT_tris}, we see that, if we
redefine $L \to L + L_2' \to -\infty$, we obtain the equivalent 
Hamiltonian  
\be
\mathcal{H}_\text{ICLAT} \;=\; - \sum\limits_{\<ij\>\in E}
\left[ (L_{2}-L_2'-L) \sigma_{i}\sigma_{j} \,+\,
       (-L_2'-L) \rho_{i}\rho_{j}  \,+\,
       L \rho_i \rho_j \sigma_i \sigma_j \right]\,, 
\label{def_H_ICLAT_4}
\ee
in the limit $L \to -\infty$. Therefore, the ICLAT model on a bipartite 
graph is invariant under the transformations:
\begin{subeqnarray}
(L_2,L_2') &\to& (L_2 - L_2',-L_2') \\
(L_2,L_2') &\to& (-L_2, L_2' - L_2) 
\label{def_symmetries_ICLAT_bipartite}
\end{subeqnarray} 
where the second equation comes from using the variables $(\tau,\rho)$ 
instead of $(\sigma,\rho)$. 

If we take $L_2=L_2'=0$ in \reff{def_H_ICLAT}, then it reduces to 
\be
\mathcal{H}_\text{ICLAT} \;=\; -\sum\limits_{\<ij\>\in E}
\left[ - L \sigma_{i}\sigma_{j} \,-\,
       L\tau_i \tau_j \,+\,
       L \sigma_{i}\sigma_{j} \tau_i \tau_j \right] \,,
\label{def_H_uICLAT_zero}
\ee
in the limit $L\to -\infty$. If the lattice $G$ is bipartite, then 
we can flip the $\tau$ and $\sigma$ spins on one of the vertex subsets $V_i$, 
so that all the couplings become equal to $L$. So it corresponds to the
zero-temperature limit of the antiferromagnetic 4--state Potts model.  

%
%
\subsection{Mixed Ashkin--Teller model}

This model appears in a natural way when one consider the duality 
transformation of an AT model. 
Let us now suppose that $G=(V,E)$ is an arbitrary planar graph; hence, it has 
a planar dual pair  $G^* = (V^*,E^*)$, Indeed, one can define an AT model on
both $G$ or $G^*$ as in the preceding section. However, we can 
define a {\em mixed Ashkin--Teller model} (mAT) with one set of spins
living on $G$, and the other set on $G^*$, and a clever coupling of both
Ising models. This coupling is based on the following fact 
about dual graphs: the number of edges is the same in both $G$ and its dual 
$G^*$, and there is natural bijection between these two edge sets: we 
can always draw $G$ and $G^*$ in the plane in such a way that 
each edge $e=\<ij\>\in E$ intersects its corresponding dual edge 
$e^*=\<i^*j^*\>\in E^*$ exactly once. 

The formal definition of the mixed AT model is as follows. 
We place at each vertex 
$i\in V$ (resp.\/ at each dual vertex $i^*\in V^*$) 
of the graph $G$ (resp.\/ of the dual graph $G^*$) an Ising spin 
$\sigma_i=\pm 1$ (resp.\/ $\tau_{i^*}=\pm 1$), and these spins interact 
through a nearest-neighbor coupling $K_{2}$ 
(resp.\/ $K'_{2}$). These two
Ising models are coupled via the four-spin interaction 
$K_{4}\sigma_i\sigma_j \tau_{i^*}\tau_{j^*}$, where $e^*=\<i^*j^*\>\in E^*$ 
is the unique dual edge associated to the edge $e=\<ij\>\in E$.
Therefore, the Hamiltonian for this model is: 
\be
\mathcal{H}_\text{mAT} \;=\; -  
  K_2  \sum\limits_{\<ij\>\in E}        \sigma_{i}\sigma_{j} \,-\, 
  K'_2 \sum\limits_{\<i^*j^*\>\in E^*}  \tau_{i^*}\tau_{j^*} \,-\, 
  K_4  \sum\limits_{\<ij\>\in E}  \sigma_{i}\sigma_{j} \tau_{i^*} \tau_{j^*} 
\,.
\label{def_H_mAT}
\ee
The partition function for this mixed model is given by:
\be
Z_\text{mAT}(G,G^*; K_2,K'_2,K_4) \;=\; 
\sum\limits_{\begin{scarray} 
            \{\sigma_i\} \\
            i\in V 
            \end{scarray}}
\sum\limits_{\begin{scarray} 
            \{\tau_{i^*}\} \\
            i^*\in V^* 
            \end{scarray}}
   e^{-\mathcal{H}_\text{mAT}} \,.
\label{def_Z_mAT}
\ee
It is important to note that the spins living on $G$ couple through the 
coupling $K_2$, while those living on its dual $G^*$, couple via $K'_2$. 

The notation in \reff{def_H_mAT} can be lightened by noting again that the
correspondence $\<ij\> \to \<i^*j^*\>$ is bijective. Thus, 
we can loosely use the labels $ij$ (instead of $i^*j^*$) for the $\tau$ spins 
(but one should remember that in the mixed AT model, the $\tau$ spins live 
on the dual vertices $V^*$, while the $\sigma$ spins live on $V$). We 
can loosely write the Hamiltonian for this model as
in  Eq.~\reff{def_H_AT}.

Finally, we can also represent the partition function of the mixed AT model
\reff{def_Z_mAT} in terms of the weights $\{w_k\}$ with $k=0,1,2,3$
[cf.\ \reff{def_omega_AT}] as in Eq.~\reff{def_Z_AT_bis}:  
\be
Z_\text{mAT}(G,G^*;K_2,K'_2,K_4) \;\equiv \;
Z_\text{mAT}(G,G^*;\omega_0, \omega_1, \omega_2,\omega_3)\,.
\label{def_Z_mAT_bis}
\ee
It is important to recall that, if we interchange $G\leftrightarrow G^*$, 
then we should interchange the couplings 
$K_2 \leftrightarrow K'_2$, or equivalently, the weights 
$\omega_1 \leftrightarrow \omega_2$:  
\begin{subeqnarray}
Z_\text{mAT}(G,G^*;K_2, K'_2,K_4) &=& 
Z_\text{mAT}(G^*,G;K'_2,K_2, K_4) \\[2mm]
&=& 
Z_\text{mAT}(G^*,G;\omega_0, \omega_2, \omega_1,\omega_3)
\label{def_Z_mAT_symmetry}
\end{subeqnarray} 

\medskip

\noindent
{\bf Remark.} Note that, contrary to the standard AT model, the number 
of $\sigma$ spins ($=|V|$) is in general different from the number of $\tau$
spins $(=|V^*|$).

\medskip

If $G=(V,E)$ is an Eulerian plane triangulation, then it is 3--colorable,
and there is a natural tripartition of the vertex set $V=V_1\cup V_2 \cup V_3$,
and edge set $E=E_{12}\cup E_{13} \cup E_{23}$. If we consider the
subgraph $G_{12}=(V_1 \cup V_2, E_{12})$, then its dual subgraph
is given by $G_{12}^*=(V_3, E_{12}^*)$, where the dual
edge set $E_{12}^*$ is built in the standard way. Then, for each
edge $e\in E_{12}$, there corresponds a dual edge $e^*\in E_{12}^*$
such that it crosses $e$ once. Then the Hamiltonian \reff{def_H_mAT} 
can be (loosely) rewritten as
\be
\label{def_H_mAT2}
\mathcal{H}_\text{mAT} \;=\; 
- \sum\limits_{\<ij\>\in E_{12}}  \left[ 
K_{2} \sigma_{i}\sigma_{j} \,+\,
K'_{2} \tau_{i}\tau_{j} \,+\,
K_{4}\sigma_{i}\sigma_{j} \tau_{i} \tau_{j} \right] \,, 
\ee
and the corresponding partition function \reff{def_Z_mAT} as 
\be
Z_\text{mAT}(G_{12},G_{12}^*;K_2,K'_2,K_4)\;=\; 
\sum\limits_{\begin{scarray}
            \{\sigma_i\} \\
            i\in V_1 \cup V_2
            \end{scarray}}
\sum\limits_{\begin{scarray}
            \{\tau_{i^*}\} \\
            i^*\in V_3  
            \end{scarray}}
   e^{-\mathcal{H}_\text{mAT}} \,. 
\label{def_Z_mAT2}
\ee

We can also take the same infinite-coupling limit as in the previous
section, namely $K_{2},K'_{2},-K_{4}\to +\infty$, while 
$L_{2}=K_{2}+K_{4}$ and $L'_{2}=K'_{2}+K_{4}$ are kept
finite. Using the above-discussed bijection between $E$ and $E^*$,
this model can be written loosely as in  Eq.~\reff{def_H_ICLAT}, but
again, keeping in mind that the $\sigma$ (resp.\/ $\tau$) spins live on $V$ 
(resp.\/ $V^*$).  

%
%
\section{Mappings among models} \label{sec.mappings} 

%
%
\subsection{Mapping between the AT and the mixed AT models}
  
In this section we will consider a general AT model with 
couplings $K_2,K_2',K_4$ on a planar graph $G=(V,E)$, not necessarily Eulerian.
We assume that either $K'_{2}\ge |K_{4}|$ or $K_{2}\ge |K_{4}|$.
Without loss of generality, let us assume that 
$K'_{2}\ge |K_{4}|$.  
The relation between the AT and mixed AT models arises when trying to derive
the duality relation for the former model. We will follow the ideas of 
Ref.~\cite{Wu_77}; but instead of looking for a relation with the 
8-vertex model, we will seek the relation with the mixed AT model 
\reff{def_H_mAT}/\reff{def_Z_mAT}. 
(See also \cite{Fan_72,Wu_Wang_76,Baxter_book}.)

Let us start with an AT model on a graph $G=(V,E)$ 
[cf.\ \reff{def_H_AT}/\reff{def_Z_AT}]:  
\be
Z_{\rm AT}(G;K_2,K'_2,K_4) \;=\; \sum\limits_{\{\sigma \}} \prod_{\<ij\>\in E}
e^{ K_{2} \sigma_i \sigma_j} \,  
\sum\limits_{\{t\}}  \prod_{\<ij\>\in E}
e^{ \widetilde{K}_{2ij} t_i t_j} \,,
\label{def_Z_AT_start}
\ee
where we have split the sum over the spin configurations into two
terms: one over the $\sigma$ spins, and the other over the $t$ 
spins (we use the letter $t$ to denote the second set of Ising spins in 
this case). The effective {\em edge-dependent} two--spin coupling between the 
$t$ spins is given by
\be
\widetilde{K}_{2ij}  \;=\; K'_{2} + K_{4}\, \sigma_i  \sigma_j \,.
\label{def_K2tilde}
\ee
Notice that the condition $K'_{2}\ge |K_{4}|$ implies that
$\widetilde{K}_{2ij} \ge 0$ for all edges $\<ij\>\in E$. 
   
We now do a high-temperature expansion using the fact that
\be
e^{\widetilde{K}_{2ij} t_i t_j} \;=\; 
    \cosh(\widetilde{K}_{2ij}) \left[  
1 + t_i t_j \, \tanh( \widetilde{K}_{2ij}) \right] \,. 
\label{trick1_new}
\ee
When we expand the product over the edges $e \in E$, we get a sum
over spanning subgraphs $E'\subseteq E$, where $E'$ is the subset of
edges contributing with the factor 
$t_i t_j \, \tanh( \widetilde{K}_{2ij})$. 
When we sum over all possible configurations of the $t$ spins, 
only those subgraphs $E'$ with an {\em even}\/
number of incident occupied edges on every vertex $x \in V$ survive. 
In summary, the partition sum reads
\begin{eqnarray}
Z_{\rm AT}(G;K_2,K'_2,K_4) &=& 2^{|V|} \, 
\sum\limits_{\{\sigma \}}
\prod_{\<ij\>\in E}
\left[ e^{ K_{2} \sigma_i \sigma_j} \, \cosh(\widetilde{K}_{2ij}) \right] 
    \nonumber \\
     & & \qquad \times  
    \sum\limits_{\begin{scarray}
                  E'\subseteq E \\
                  \hbox{\scriptsize $E'$ Eulerian}
                  \end{scarray}
                 }
    \prod_{\<ij\> \in E'} \tanh( \widetilde{K}_{2ij}) \,,  
\label{def_Z_AT_1}
\end{eqnarray}
where $\widetilde{K}_{2ij}$ is given by \reff{def_K2tilde}.

\begin{figure}[htb]
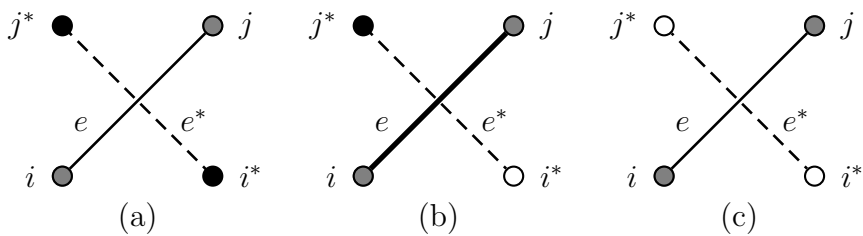

 \centering
 \psset{unit=1cm}
 \psset{labelsep=10pt}
 \pspicture(-1,-1)(12,3)
 \rput{0}(0,0){%
    \psline[linecolor=black,linewidth=1pt](0,0)(2,2)
    \psline[linecolor=black,linewidth=1pt,linestyle=dashed](2,0)(0,2)
    \pscircle*[linecolor=gray](0,0){4pt}
    \pscircle[linecolor=black] (0,0){4pt}
    \pscircle*[linecolor=gray](2,2){4pt}
    \pscircle[linecolor=black] (2,2){4pt}
    \pscircle*[linecolor=black](0,2){4pt}
    \pscircle*[linecolor=black](2,0){4pt}
    \uput[270](1,0){(a)}
    \uput[180](0,0){$i$}
    \uput[180](0,2){$j^*$}
    \uput[0](2,0){$i^*$}
    \uput[0](2,2){$j$}
    \uput[90](0.25,0.25){$e$}
    \uput[90](1.75,0.25){$e^*$}
 }
 \rput{0}(4,0){%
    \psline[linecolor=black,linewidth=2pt](0,0)(2,2)
    \psline[linecolor=black,linewidth=1pt,linestyle=dashed](2,0)(0,2)
    \pscircle*[linecolor=gray](0,0){4pt}
    \pscircle[linecolor=black] (0,0){4pt}
    \pscircle*[linecolor=gray](2,2){4pt}
    \pscircle[linecolor=black] (2,2){4pt}
    \pscircle*[linecolor=black](0,2){4pt}
    \pscircle*[linecolor=white](2,0){4pt}
    \pscircle[linecolor=black](2,0){4pt}
    \uput[270](1,0){(b)}
    \uput[180](0,0){$i$}
    \uput[180](0,2){$j^*$}
    \uput[0](2,0){$i^*$}
    \uput[0](2,2){$j$}
    \uput[90](0.25,0.25){$e$}
    \uput[90](1.75,0.25){$e^*$}
 }
 \rput{0}(8,0){%
    \psline[linecolor=black,linewidth=1pt](0,0)(2,2)
    \psline[linecolor=black,linewidth=1pt,linestyle=dashed](2,0)(0,2)
    \pscircle*[linecolor=gray](0,0){4pt}
    \pscircle[linecolor=black] (0,0){4pt}
    \pscircle*[linecolor=gray](2,2){4pt}
    \pscircle[linecolor=black] (2,2){4pt}
    \pscircle*[linecolor=white](0,2){4pt}
    \pscircle[linecolor=black](0,2){4pt}
    \pscircle*[linecolor=white](2,0){4pt}
    \pscircle[linecolor=black](2,0){4pt}
    \uput[270](1,0){(c)}
    \uput[180](0,0){$i$}
    \uput[180](0,2){$j^*$}
    \uput[0](2,0){$i^*$}
    \uput[0](2,2){$j$}
    \uput[90](0.25,0.25){$e$}
    \uput[90](1.75,0.25){$e^*$}
 }
 \endpspicture
\caption{ \label{fig.def.tau}
  (a) Dual edge $e^*=\<i^*j^*\>\in E^*$ corresponding to an edge 
      $e=\<ij\>\in E$. The vertices $i,j$ belong to $V$, and
      the dual ones $i^*,j^*$ to $V^*$. 
  (b) When $e$ belongs to the spanning subgraph $(V,E')$ (depicted
      as a thick line), then 
      $\tau_{i^*}\tau_{j^*}=-1$ (depicted as dots of different colors).
  (c) When $e$ does not belong to the spanning subgraph $(V,E')$, 
      then $\tau_{i^*}\tau_{j^*}=1$ (depicted as dots of equal color).
}
\end{figure}

We now introduce the new Ising variables $\tau_i= \pm1$ living on the
dual vertex set $V^*$. They are assigned such that for any Eulerian 
spanning subgraph $(V,E')$ of $G=(V,E)$:
\be
e=\<ij\> \in E' \;\Longleftrightarrow \; \tau_{i^*} \tau_{j^*} \;=\; -1 \,,
\label{def_tau_rule}
\ee
where $i^*,j^*$ are the vertices of the dual edge $e^*$ corresponding to the
edge $e$. (See Figure~\ref{fig.def.tau}.)
In words, given an edge $e\in E$, if this edge belongs 
(resp.\ does not belong) to the subset $E'\subseteq E$, then the 
$\tau$ variables associated to this edge have different (resp.\  equal) signs. 

%
%

%
%
\begin{figure}[htb]
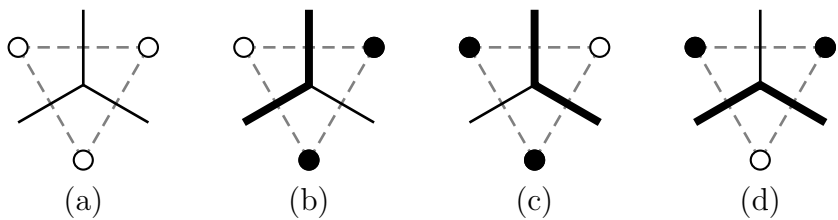

 \centering
 \psset{unit=1cm}
 \psset{labelsep=10pt}
 \pspicture(-0.5,-1)(11,2)
 \rput{0}(0,0){
  \psline[linecolor=gray,linewidth=1pt,linestyle=dashed](0.866025,0)(0,1.5)
  \psline[linecolor=gray,linewidth=1pt,linestyle=dashed]%
                                                  (0.866025,0)(1.73205,1.5)
  \psline[linecolor=gray,linewidth=1pt,linestyle=dashed](1.73205,1.5)(0,1.5) 
  \psline[linecolor=black,linewidth=1pt](0,0.5)(0.866025,1)(1.73205,0.5)
  \psline[linecolor=black,linewidth=1pt](0.866025,1)(0.866025,2)
  \pscircle*[linecolor=white](0.866025,0){4pt}
  \pscircle[linecolor=black] (0.866025,0){4pt}
  \pscircle*[linecolor=white](0,1.5){4pt}
  \pscircle[linecolor=black] (0,1.5){4pt}
  \pscircle*[linecolor=white](1.73205,1.5){4pt}
  \pscircle[linecolor=black] (1.73205,1.5){4pt}
  \uput[270](0.866025,0){(a)}
 }
 \rput{0}(3,0){
  \psline[linecolor=gray,linewidth=1pt,linestyle=dashed](0.866025,0)(0,1.5)
  \psline[linecolor=gray,linewidth=1pt,linestyle=dashed]%
                                                  (0.866025,0)(1.73205,1.5)
  \psline[linecolor=gray,linewidth=1pt,linestyle=dashed](1.73205,1.5)(0,1.5) 
  \psline[linecolor=black,linewidth=3pt](0,0.5)(0.866025,1)(0.866025,2)
  \psline[linecolor=black,linewidth=1pt](0.866025,1)(1.73205,0.5)
  \pscircle*[linecolor=black](0.866025,0){4pt}
  \pscircle[linecolor=black] (0.866025,0){4pt}
  \pscircle*[linecolor=white](0,1.5){4pt}
  \pscircle[linecolor=black] (0,1.5){4pt}
  \pscircle*[linecolor=black](1.73205,1.5){4pt}
  \pscircle[linecolor=black] (1.73205,1.5){4pt}
  \uput[270](0.866025,0){(b)}
 }
 \rput{0}(6,0){
  \psline[linecolor=gray,linewidth=1pt,linestyle=dashed](0.866025,0)(0,1.5)
  \psline[linecolor=gray,linewidth=1pt,linestyle=dashed]%
                                                  (0.866025,0)(1.73205,1.5)
  \psline[linecolor=gray,linewidth=1pt,linestyle=dashed](1.73205,1.5)(0,1.5) 
  \psline[linecolor=black,linewidth=3pt](1.73205,0.5)(0.866025,1)(0.866025,2)
  \psline[linecolor=black,linewidth=1pt](0.866025,1)(0,0.5)
  \pscircle*[linecolor=black](0.866025,0){4pt}
  \pscircle[linecolor=black] (0.866025,0){4pt}
  \pscircle*[linecolor=black](0,1.5){4pt}
  \pscircle[linecolor=black] (0,1.5){4pt}
  \pscircle*[linecolor=white](1.73205,1.5){4pt}
  \pscircle[linecolor=black] (1.73205,1.5){4pt}
  \uput[270](0.866025,0){(c)}
 }
 \rput{0}(9,0){
  \psline[linecolor=gray,linewidth=1pt,linestyle=dashed](0.866025,0)(0,1.5)
  \psline[linecolor=gray,linewidth=1pt,linestyle=dashed]%
                                                  (0.866025,0)(1.73205,1.5)
  \psline[linecolor=gray,linewidth=1pt,linestyle=dashed](1.73205,1.5)(0,1.5) 
  \psline[linecolor=black,linewidth=3pt](1.73205,0.5)(0.866025,1)(0,0.5)
  \psline[linecolor=black,linewidth=1pt](0.866025,1)(0.866025,2)
  \pscircle*[linecolor=white](0.866025,0){4pt}
  \pscircle[linecolor=black] (0.866025,0){4pt}
  \pscircle*[linecolor=black](0,1.5){4pt}
  \pscircle[linecolor=black] (0,1.5){4pt}
  \pscircle*[linecolor=black](1.73205,1.5){4pt}
  \pscircle[linecolor=black] (1.73205,1.5){4pt}
  \uput[270](0.866025,0){(d)}
 }
 \endpspicture
 \caption{\label{fig_Eulerian}
 Mapping \protect\reff{def_tau_rule} 
 between Eulerian spanning subgraphs on a vertex of degree 3.  
 The solid black (resp.\ dashed gray) edges belong to $E$ 
 (resp.\ $E^*$), and the dots correspond to vertices of $V^*$.
 If $e\in E$ belongs (resp.\ does not belong) to the spanning 
 subgraph $(V,E')$, the corresponding edge is depicted as a thick 
 (resp.\ thin) solid black line.  
 In panels (a)--(d) we show the 
 four different configurations: no edge in $E'$ (i.e., all Ising spins
 have the same value), and two edges in $E'$ (i.e., two Ising spins
 take the same value, and the other one takes the other value). 
}
\end{figure}

%
%
\begin{figure}[htb]
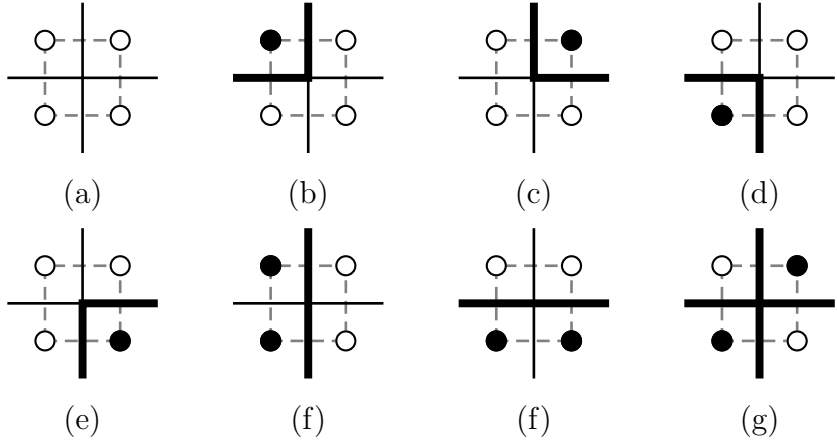

 \centering
 \psset{unit=1cm}
 \psset{labelsep=10pt}
 \pspicture(-0.5,-1)(11.5,5)
 \rput{0}(0,3){
  \psline[linecolor=gray,linewidth=1pt,linestyle=dashed](0.5,0.5)(0.5,1.5)%
         (1.5,1.5)(1.5,0.5)(0.5,0.5)
  \psline[linecolor=black,linewidth=1pt](0,1)(2,1)
  \psline[linecolor=black,linewidth=1pt](1,0)(1,2)
  \pscircle*[linecolor=white](0.5,0.5){4pt}
  \pscircle[linecolor=black] (0.5,0.5){4pt}
  \pscircle*[linecolor=white](0.5,1.5){4pt}
  \pscircle[linecolor=black] (0.5,1.5){4pt}
  \pscircle*[linecolor=white](1.5,0.5){4pt}
  \pscircle[linecolor=black] (1.5,0.5){4pt}
  \pscircle*[linecolor=white](1.5,1.5){4pt}
  \pscircle[linecolor=black] (1.5,1.5){4pt}
  \uput[270](1,0){(a)}
 }
 \rput{0}(3,3){
  \psline[linecolor=gray,linewidth=1pt,linestyle=dashed](0.5,0.5)(0.5,1.5)%
         (1.5,1.5)(1.5,0.5)(0.5,0.5)
  \psline[linecolor=black,linewidth=3pt](0,1)(1,1)(1,2)
  \psline[linecolor=black,linewidth=1pt](1,0)(1,1)(2,1)
  \pscircle*[linecolor=white](0.5,0.5){4pt}
  \pscircle[linecolor=black] (0.5,0.5){4pt}
  \pscircle*[linecolor=black](0.5,1.5){4pt}
  \pscircle[linecolor=black] (0.5,1.5){4pt}
  \pscircle*[linecolor=white](1.5,0.5){4pt}
  \pscircle[linecolor=black] (1.5,0.5){4pt}
  \pscircle*[linecolor=white](1.5,1.5){4pt}
  \pscircle[linecolor=black] (1.5,1.5){4pt}
  \uput[270](1,0){(b)}
 }
 \rput{0}(6,3){
  \psline[linecolor=gray,linewidth=1pt,linestyle=dashed](0.5,0.5)(0.5,1.5)%
         (1.5,1.5)(1.5,0.5)(0.5,0.5)
  \psline[linecolor=black,linewidth=3pt](1,2)(1,1)(2,1)
  \psline[linecolor=black,linewidth=1pt](1,0)(1,1)(0,1)
  \pscircle*[linecolor=white](0.5,0.5){4pt}
  \pscircle[linecolor=black] (0.5,0.5){4pt}
  \pscircle*[linecolor=white](0.5,1.5){4pt}
  \pscircle[linecolor=black] (0.5,1.5){4pt}
  \pscircle*[linecolor=white](1.5,0.5){4pt}
  \pscircle[linecolor=black] (1.5,0.5){4pt}
  \pscircle*[linecolor=black](1.5,1.5){4pt}
  \pscircle[linecolor=black] (1.5,1.5){4pt}
  \uput[270](1,0){(c)}
 }
 \rput{0}(9,3){
  \psline[linecolor=gray,linewidth=1pt,linestyle=dashed](0.5,0.5)(0.5,1.5)%
         (1.5,1.5)(1.5,0.5)(0.5,0.5)
  \psline[linecolor=black,linewidth=3pt](0,1)(1,1)(1,0)
  \psline[linecolor=black,linewidth=1pt](1,2)(1,1)(2,1)
  \pscircle*[linecolor=black](0.5,0.5){4pt}
  \pscircle[linecolor=black] (0.5,0.5){4pt}
  \pscircle*[linecolor=white](0.5,1.5){4pt}
  \pscircle[linecolor=black] (0.5,1.5){4pt}
  \pscircle*[linecolor=white](1.5,0.5){4pt}
  \pscircle[linecolor=black] (1.5,0.5){4pt}
  \pscircle*[linecolor=white](1.5,1.5){4pt}
  \pscircle[linecolor=black] (1.5,1.5){4pt}
  \uput[270](1,0){(d)}
 }
 \rput{0}(0,0){
  \psline[linecolor=gray,linewidth=1pt,linestyle=dashed](0.5,0.5)(0.5,1.5)%
         (1.5,1.5)(1.5,0.5)(0.5,0.5)
  \psline[linecolor=black,linewidth=3pt](1,0)(1,1)(2,1)
  \psline[linecolor=black,linewidth=1pt](0,1)(1,1)(1,2)
  \pscircle*[linecolor=white](0.5,0.5){4pt}
  \pscircle[linecolor=black] (0.5,0.5){4pt}
  \pscircle*[linecolor=white](0.5,1.5){4pt}
  \pscircle[linecolor=black] (0.5,1.5){4pt}
  \pscircle*[linecolor=black](1.5,0.5){4pt}
  \pscircle[linecolor=black] (1.5,0.5){4pt}
  \pscircle*[linecolor=white](1.5,1.5){4pt}
  \pscircle[linecolor=black] (1.5,1.5){4pt}
  \uput[270](1,0){(e)}
 }
 \rput{0}(3,0){
  \psline[linecolor=gray,linewidth=1pt,linestyle=dashed](0.5,0.5)(0.5,1.5)%
         (1.5,1.5)(1.5,0.5)(0.5,0.5)
  \psline[linecolor=black,linewidth=3pt](1,0)(1,2)
  \psline[linecolor=black,linewidth=1pt](0,1)(2,1)
  \pscircle*[linecolor=black](0.5,0.5){4pt}
  \pscircle[linecolor=black] (0.5,0.5){4pt}
  \pscircle*[linecolor=black](0.5,1.5){4pt}
  \pscircle[linecolor=black] (0.5,1.5){4pt}
  \pscircle*[linecolor=white](1.5,0.5){4pt}
  \pscircle[linecolor=black] (1.5,0.5){4pt}
  \pscircle*[linecolor=white](1.5,1.5){4pt}
  \pscircle[linecolor=black] (1.5,1.5){4pt}
  \uput[270](1,0){(f)}
 }
 \rput{0}(6,0){
  \psline[linecolor=gray,linewidth=1pt,linestyle=dashed](0.5,0.5)(0.5,1.5)%
         (1.5,1.5)(1.5,0.5)(0.5,0.5)
  \psline[linecolor=black,linewidth=1pt](1,0)(1,2)
  \psline[linecolor=black,linewidth=3pt](0,1)(2,1)
  \pscircle*[linecolor=black](0.5,0.5){4pt}
  \pscircle[linecolor=black] (0.5,0.5){4pt}
  \pscircle*[linecolor=white](0.5,1.5){4pt}
  \pscircle[linecolor=black] (0.5,1.5){4pt}
  \pscircle*[linecolor=black](1.5,0.5){4pt}
  \pscircle[linecolor=black] (1.5,0.5){4pt}
  \pscircle*[linecolor=white](1.5,1.5){4pt}
  \pscircle[linecolor=black] (1.5,1.5){4pt}
  \uput[270](1,0){(f)}
 }
 \rput{0}(9,0){
  \psline[linecolor=gray,linewidth=1pt,linestyle=dashed](0.5,0.5)(0.5,1.5)%
         (1.5,1.5)(1.5,0.5)(0.5,0.5)
  \psline[linecolor=black,linewidth=3pt](1,0)(1,2)
  \psline[linecolor=black,linewidth=3pt](0,1)(2,1)
  \pscircle*[linecolor=black](0.5,0.5){4pt}
  \pscircle[linecolor=black] (0.5,0.5){4pt}
  \pscircle*[linecolor=white](0.5,1.5){4pt}
  \pscircle[linecolor=black] (0.5,1.5){4pt}
  \pscircle*[linecolor=white](1.5,0.5){4pt}
  \pscircle[linecolor=black] (1.5,0.5){4pt}
  \pscircle*[linecolor=black](1.5,1.5){4pt}
  \pscircle[linecolor=black] (1.5,1.5){4pt}
  \uput[270](1,0){(g)}
 }
 \endpspicture
 \caption{\label{fig_Eulerian2}
 Mapping \protect\reff{def_tau_rule} between Eulerian spanning subgraphs i
 on a vertex of degree 4.  In panels (a)--(g) we show the  
 eight different configurations: no edge in $E'$ (i.e., all Ising spins
 have the same value), two edges in $E'$ (i.e., two consecutive Ising spins
 take the same value, and the other two take the other value; or one spin
 takes one value, and the other three spins take the opposite value), and 
 four edges in $E'$ (the spin values alternate as we move around).  
}
\end{figure}

For instance, the relation between edge configurations and $\tau$ spin
configurations for vertices of degree $3$ and $4$ are given, respectively 
in Figures~\ref{fig_Eulerian} and \ref{fig_Eulerian2}. 
For a vertex $v\in V$ of degree $d$, 
the number of possible Eulerian configurations (i.e., an even number of 
edges incident to $v$) is given by 
\be
\sum\limits_{n\ge 0} \binom{d}{2n} \;=\; 2^{d-1} \,,
\ee
which is exactly the same number of $\tau$-spin configurations of the 
$d$ neighboring vertices (modulo a global reversal of the spin values). 
Indeed, there is a one-to-two correspondence between edges subsets $E'$ 
and $\{\tau\}$ configurations: for each subset $E'$ there are two 
equivalent $\tau$ configurations differing by a global change of sign.  

Note that the $\sigma$ (resp.\/ $\tau$) spins live on the vertices
of $V$ (resp.\/ $V^*$). Thus, we arrive at a mixed AT model defined on $G$ and
$G^*$:   
\begin{subeqnarray}
Z_{\rm AT}(G;K_2,K'_2,K_4) &=& 2^{|V|-1} \, 
\sum\limits_{\begin{scarray}
                  \{\sigma_i \} \\
                  i\in V
                  \end{scarray}}
\sum\limits_{\begin{scarray}
                  \{\tau_i \} \\
                  i\in V^*
                  \end{scarray}}
\prod_{\<ij\>\in E}
\left[ e^{ K_{2} \sigma_i \sigma_j} \, 
       \cosh(\widetilde{K}_{2ij}) \right] 
    \nonumber \\
     & & \; \times  
    \prod_{\<ij\> \in E} \left[ 
    \frac{1+\tau_{i^*}\tau_{j^*}}{2} + 
    \frac{1-\tau_{i^*}\tau_{j^*}}{2}  
    \tanh(\widetilde{K}_{2ij})\right] \\  
  &=& 2^{|V|-1} \, \sum\limits_{\begin{scarray}
                  \{\sigma_i \} \\
                  i\in V
                  \end{scarray}}
\sum\limits_{\begin{scarray}
                  \{\tau_i \} \\
                  i\in V^*
                  \end{scarray}}
\prod_{\<ij\>\in E}
\widehat{\omega}_{ij}(\sigma_i\sigma_j,\tau_{i^*}\tau_{j^*}) 
\label{def_Z_AT_2}
\end{subeqnarray}
where $\widetilde{K}_{2ij}$ is given by \reff{def_K2tilde}, and 
to each edge $e=\<ij\>\in E$ there corresponds a unique dual
edge $e^*=\<i^*j^*\>\in E^*$, and vice versa. The extra factor $2^{-1}$  
comes from the above-mentioned two-to-one
relation between $\tau$ configurations and Eulerian subgraphs 
$E'\subseteq E$. 
The second term in the product is just a way to express the fact that
when $e\in E'$ (so $\tau_{i^*}=-\tau_{j^*}$), there is a contribution
$\tanh(\widetilde{K}_{2ij})$; and when $e\not\in E'$,
(so $\tau_{i^*}=\tau_{j^*}$), then the contribution is~$1$.
The Boltzmann weights $\widehat{\omega}_{ij}$ 
associated to the above representation
\reff{def_Z_AT_2} are given in Table~\ref{table_weights2}.

%
%
\begin{table}[htb]
\centering
\begin{tabular}{lcc}
\hline\hline
Configuration & Weight & (Normalized) Weight \\
\hline \\[-2mm]
$\sigma_i\sigma_j=+1\,,    \tau_i\tau_j=+1$& 
$\widehat{\omega}_{0}=e^{K_{2}} \cosh(K'_{2} + K_{4})$ & 
$1$ \\[3mm]
$\sigma_i\sigma_j=+1\,,    \tau_i\tau_j=-1$& 
$\widehat{\omega}_{1}=e^{K_{2}} \sinh(K'_{2} + K_{4})$ & 
$\tanh(K'_{2} + K_{4})$ \\[3mm]
$\sigma_i\sigma_j=-1\,, \tau_i\tau_j=+1$& 
$\widehat{\omega}_{2}=e^{-K_{2}} \cosh(K'_{2} - K_{4})$ &
$e^{-2K_{2ij}} \frac{\cosh(K'_{2} - K_{4})}
                    {\cosh(K'_{2} + K_{4})}$ \\[3mm]
$\sigma_i\sigma_j=-1\,, \tau_i\tau_j=-1$& 
$\widehat{\omega}_{3}=e^{-K_{2}} \sinh(K'_{2} - K_{4})$ &
$e^{-2K_{2}} \frac{\sinh(K'_{2} - K_{4})}
                  {\cosh(K'_{2} + K_{4})}$ \\[3mm]
\hline\hline
\end{tabular}
\caption{\label{table_weights2}
Boltzmann weights for the model \protect\reff{def_Z_AT_2}.
For each configuration of the $\sigma,\tau$ variables, we first quote 
the weight (up to an unimportant global factor) read off from 
\protect\reff{def_Z_AT_2} [`` Weight''],
and then the normalized weight obtained by making the first one equal to $1$
[``(Normalized) Weight'']. 
}
\end{table}

The relation between the weights $\widehat{\omega}_k$ of the 
transformed model \reff{def_Z_AT_2} and the original ones $\omega_k$
is given for every edge $\<ij\>\in E$ by 
\begin{subeqnarray}
\widehat{\omega}_{0} &=& \frac{1}{2} (\omega_{0}+\omega_{1}) \\[2mm]
\widehat{\omega}_{1} &=& \frac{1}{2} (\omega_{0}-\omega_{1}) \\[2mm]
\widehat{\omega}_{2} &=& \frac{1}{2} (\omega_{2}+\omega_{3}) \\[2mm]
\widehat{\omega}_{3} &=& \frac{1}{2} (\omega_{2}-\omega_{3}) 
\label{def_omegahat}
\end{subeqnarray}
The above results can be summarized in the following

\begin{theorem} \label{theo.AT_vs_mAT}
Let $G=(V,E)$ be a planar graph. Then the Ashkin--Teller model 
\reff{def_H_AT}/\reff{def_Z_AT} on $G$, with $K'_{2}\ge |K_{4}|$, 
is equivalent to a mixed Ashkin--Teller model \reff{def_H_mAT}/\reff{def_Z_mAT}
on $G$ and $G^*$ such that
\be
Z_{\rm AT}(G;\omega_0,\omega_1,\omega_2,\omega_3)
\;=\; 2^{|V|-1} \, 
Z_{\rm mAT}(G,G^*;\widehat{\omega}_0,\widehat{\omega}_1,
                  \widehat{\omega}_2,\widehat{\omega}_3) \,,
\label{def_Z_AT_vs Z_mAT}
\ee
where the weights $\widehat{\omega}_k$ are given in terms of the
weights $\omega_k$ by \reff{def_omegahat}. 
\end{theorem}

\medskip

\noindent
{\bf Remark}. Let us remind that in the mixed AT model in 
Theorem~\ref{theo.AT_vs_mAT} the $\sigma$ (resp.\/ $\tau$) spins live on 
$V$ (resp.\/ $V^*$). 
The $\sigma$  (resp.\/ $\tau$) spins have a nearest-neighbor coupling 
$K_2$ (resp.\/ $K'_2$); and both $\sigma,\tau$ spins interact via a 
4-spin coupling $K_4$.  

\medskip

The duality transformation for the AT model can be easily obtained from 
Theorem~\ref{theo.AT_vs_mAT} by playing the same game. We want to 
express the partition function of the original AT model on $G$ in terms of 
the partition function of another AT model on the dual graph $G^*$. 
This is easily done as follows:
\begin{subeqnarray}
2^{-|V|+1}\, Z_{\rm AT}(G;\omega_0,\omega_1,\omega_2,\omega_3) &=& 
Z_{\rm mAT}(G,G^*;\widehat{\omega}_0,\widehat{\omega}_1,
                  \widehat{\omega}_2,\widehat{\omega}_3) \\[2mm]
&=& Z_{\rm mAT}(G^*,G;\widehat{\omega}_0,\widehat{\omega}_2,
                  \widehat{\omega}_1,\widehat{\omega}_3) \\[2mm]
&=& 2^{-|V|^*+1}Z_{\rm AT}(G^*;\omega^*_0,\omega^*_1,\omega^*_2,\omega^*_3)
\label{def_relations_ATs}
\end{subeqnarray}
where the first equality is just Theorem~\ref{theo.AT_vs_mAT},
the second line comes from using the symmetry relation \reff{def_Z_mAT_symmetry}
for the mixed AT model, and the last one comes from using 
Theorem~\ref{theo.AT_vs_mAT} on the dual graph $G^*$. The expression of
the new weights $\omega_k^*$ in terms of the original ones is
\begin{subeqnarray}
\omega_{0}^* &=& \widehat{\omega}_{0}+\widehat{\omega}_{2} 
\;=\; \frac{1}{2}(\omega_{0}+\omega_{1}+\omega_{2}+\omega_{3})\\[2mm]
\omega_{1}^* &=& \widehat{\omega}_{0}-\widehat{\omega}_{2} 
\;=\; \frac{1}{2}(\omega_{0}+\omega_{1}-\omega_{2}-\omega_{3})\\[2mm]
\omega_{2}^* &=& \widehat{\omega}_{1}+\widehat{\omega}_{3} 
\;=\; \frac{1}{2}(\omega_{0}+\omega_{2}-\omega_{1}-\omega_{3})\\[2mm]
\omega_{3}^* &=& \widehat{\omega}_{1}-\widehat{\omega}_{3}
\;=\; \frac{1}{2}(\omega_{0}+\omega_{3}-\omega_{1}-\omega_{2})
\label{def_omega_dual}
\end{subeqnarray}
The above results can be summarized in the following

\begin{theorem}[Wu \protect\cite{Wu_77}] \label{theo.duality_AT}
Let $G=(V,E)$ be a planar graph. Then the Ashkin--Teller model
\reff{def_H_AT}/\reff{def_Z_AT} on $G$ is equivalent to another 
Ashkin--Teller model on the dual graph $G^*$ such that
\be
2^{-|V|}\, 
Z_{\rm AT}(G;\omega_0,\omega_1,\omega_2,\omega_3) \;=\; 2^{-|V^*|} \,
Z_{\rm AT}(G^*;\omega^*_0,\omega^*_1,\omega^*_2,\omega^*_3) \,,
\label{def_Z_AT_vs Z_mAT_theo}
\ee
where the weights $\omega^*_k$ are given in terms of the
weights $\omega_k$ by \reff{def_omega_dual}.
\end{theorem}

\medskip

If $G$ is self-dual, then the corresponding AT model is self-dual when
the following condition holds: 
\be
\omega_{0} \;=\; \omega_{1} + \omega_{2} + \omega_{3} \,.
\label{def_AT_selfdual}
\ee
This situation occurs for instance for the infinite square lattice.

%
%
\subsection{Mapping between the mixed AT and the ICLAT models}

Let us now explore a little further the relation between the mixed 
AT on $G$ and $G^*$, and the AT model on $G^*$. We know from
\reff{def_relations_ATs}/\reff{def_omega_dual} that:
\be
Z_{\rm mAT}(G,G^*;\omega_0,\omega_1,\omega_2,\omega_3)
\;=\; 2^{-|V^*|+1} \, 
Z_{\rm AT}(G^*;\omega^*_0,\omega^*_1,\omega^*_2,\omega^*_3)
\ee
where the new weights are given by
\begin{subeqnarray}
\omega_{0}^* &=& \omega_{0}+\omega_{2} \\ 
\omega_{1}^* &=& \omega_{0}-\omega_{2} \\ 
\omega_{2}^* &=& \omega_{1}+\omega_{3} \\ 
\omega_{3}^* &=& \omega_{1}-\omega_{3} 
\label{def_omega_star}
\end{subeqnarray}
It is easy to see that $\omega_{3}^*=0$ if and only if 
$\omega_{1}=\omega_{3}$. And the last equality holds true if and
only if $K_{2}=K_{4}$. In this particular case, we have 
\begin{subeqnarray}
2^{|V^*|-1}\,
Z_{\rm mAT}(G,G^*;\omega_0,\omega_1,\omega_2,\omega_1) 
&=& 
Z_{\rm AT}(G^*;\omega^*_0,\omega^*_1,\omega^*_2,0)\qquad\qquad\qquad \\[2mm]
&=& 
Z_{\rm AT}\left(G^*;1,\frac{\omega^*_1}{\omega^*_0},
                      \frac{\omega^*_2}{\omega^*_0}, 0\right)
\prod_{\<ij\>\in E} \omega_{0}^* 
\end{subeqnarray}
But the ratios of these weights are precisely the couplings of the 
ICLAT model:
\begin{subeqnarray}
\frac{\omega^*_{1}}{\omega^*_{0}} &=& e^{-2K^{*'}_{2}-2K^*_{4}} \;=\; 
                                      e^{-2L_{2}^{*'}} \\[2mm]
\frac{\omega^*_{2}}{\omega^*_{0}} &=& e^{-2K^*_{2}-2K^*_{4}} \;=\; 
                                      e^{-2L_{2}^*}
\end{subeqnarray}
In terms of the original couplings, we get that the couplings of the ICLAT
model are given by the expressions:
\begin{subeqnarray}
e^{-2L_2^{*'}} &=& \frac{\omega_{0}-\omega_{2}}{\omega_{0}+\omega_{2}} \;=\; 
\tanh(2K_{2}) \\[2mm]
e^{-2L_2^{*}} &=& \frac{2\omega_{1}}
                       {\omega_{0}+\omega_{2}} \;=\; 
\frac{e^{-2K'_{2}}}{\cosh(2K_{2})} 
\label{def_L_ICLAT}
\end{subeqnarray}
The above considerations prove the following

\begin{theorem} \label{theo.mAT_ICLAT}
Let $G=(V,E)$ be a planar graph. Then the mixed Ashkin--Teller model
\reff{def_H_mAT}/\reff{def_Z_mAT} on $G$ and $G^*$ with equal
couplings $K_{2}=K_{4}$, is equivalent to the infinite--coupling--limit 
Ashkin--Teller model on $G^*$, such that  
\be
Z_{\rm mAT}(G,G^*;\omega_0,\omega_1,\omega_2,\omega_1)
\;=\; 2^{-|V^*|+1} \left[ \prod_{\<ij\>\in E} \omega_{0}^* \right] 
Z_{\rm ICLAT}(G^*;L^*_2,L_2^{*'})  
\label{def_Z_mAT_vs Z_ICLAT_theo}
\ee
where the weights couplings $L^*_2,L_2^{*'}$ are given  
in terms of the couplings $K^*_2,K_2^{*'}$ by 
\reff{def_L_ICLAT}.  
\end{theorem}

\medskip

If we prefer to write the partition function of the mixed AT model as
an ICLAT model on $G$, the procedure is similar: in this case we have
to use \reff{def_omegahat} to define the couplings of the AT model
$\omega_k$ in terms of those of the mixed AT model 
$\widehat{\omega}_k$. We obtain (compare to \reff{def_omega_star}) 
\begin{subeqnarray}
\omega_{0} &=& \widehat{\omega}_{0}+\widehat{\omega}_{1} \\
\omega_{1} &=& \widehat{\omega}_{0}-\widehat{\omega}_{1} \\
\omega_{2} &=& \widehat{\omega}_{2}+\widehat{\omega}_{3} \\
\omega_{3} &=& \widehat{\omega}_{2}-\widehat{\omega}_{3}
\label{def_omegahat_bis}
\end{subeqnarray}
so that
\be
Z_{\rm mAT}(G,G^*;\widehat{\omega}_0,\widehat{\omega}_1,
                  \widehat{\omega}_2,\widehat{\omega}_3)
\;=\; 
Z_{\rm AT}(G; \omega_0,\omega_1,\omega_2,\omega_3) 
\ee
We then notice that $\omega_{3}=0$ if and only if 
$\widehat{\omega}_{2}=\widehat{\omega}_{3}$, which is in turn 
equivalent to $\widehat{K}'_{2}=\widehat{K}_{4}$.  

But the ratios of weights for this AT mode are precisely the couplings of the 
ICLAT model:
\begin{subeqnarray}
\frac{\omega_{1}}{\omega_{0}} &=& e^{-2K'_{2}-2K_{4}} \;=\; 
                                          e^{-2L'_{2}} \\[2mm]
\frac{\omega_{2}}{\omega_{0}} &=& e^{-2K_{2}-2K_{4}} \;=\; 
                                          e^{-2L_{2}}
\end{subeqnarray}
In terms of the mixed AT couplings we get that the couplings of the ICLAT
model are given by the expressions:
\begin{subeqnarray}
e^{-2L'_2} &=& \frac{\widehat{\omega}_{0}-\widehat{\omega}_{1}}
                    {\widehat{\omega}_{0}+\widehat{\omega}_{1}} \;=\; 
\tanh(2\widehat{K}'_{2}) \\[2mm]
e^{-2L_2} &=& \frac{2\widehat{\omega}_{2}}
                   {\widehat{\omega}_{0}+\widehat{\omega}_{1}} \;=\; 
\frac{e^{-2\widehat{K}_{2}}}{\cosh(2\widehat{K}'_{2})} 
\label{def_L_ICLAT_bis}
\end{subeqnarray}
The above considerations prove the following

\begin{theorem} \label{theo.mAT_ICLAT_bis}
Let $G=(V,E)$ be a planar graph. Then the mixed Ashkin--Teller model
\reff{def_H_mAT}/\reff{def_Z_mAT} on $G$ and $G^*$ with equal
couplings $K'_{2}=K_{4}$, is equivalent to the infinite--coupling--limit 
Ashkin--Teller model on $G$, such that  
\be
 Z_{\rm mAT}(G,G^*;\widehat{\omega}_0,\widehat{\omega}_1,
                   \widehat{\omega}_2,\widehat{\omega}_2)
 \;=\; 2^{-|V|+1} \left[ \prod_{\<ij\>\in E} \omega_{0} \right] 
 Z_{\rm ICLAT}(G;L_2,L'_2)  
 \label{def_Z_mAT_vs Z_ICLAT_theo_bis}
\ee
where the weights couplings $L_2,L'_2$ are given  
in terms of the couplings $\widehat{K}_2,\widehat{K}'_2$ by 
\reff{def_L_ICLAT_bis}. 
\end{theorem}

%
%
\subsection{Partial trace transformation} 
\label{subsec.partial.trace}

The goal of this section is to show that the HMBW model on various Eulerian
plane triangulations can be mapped onto ICLAT models defined on the most 
common lattices (square, hexagonal, triangular, and kagome).
Let us start with the following lemma: 

\begin{lemma} \label{lemma.partial.trace}
Let $\sigma_1,\ldots,\sigma_{2k}$ take the values $\pm 1$,
and define $\sigma_{2k+1} = \sigma_1$.
Then, if $1\le k \le 3$,
\be
   \cosh \left[ J \sum_{i=1}^{2k} \sigma_i \sigma_{i+1} \right]
   \;=\;
   \cosh \left[ J \sum_{i=1}^{k} \sigma_{2i} \,+\,
           J \sum_{j=1}^{k} \sigma_{2j-1} \sigma_{2j} \sigma_{2j+1} \right]
   \;.
\label{eq_map}
\ee
Moreover, when $k\ge 4$ this equality does not hold. 
\end{lemma}

\proof
The statement \reff{eq_map} is equivalent to prove that the 
following quantity $A(k)$ vanishes when all $\sigma_i^2 = 1$,
$\sigma_{2k+1}=\sigma_1$, and $1\le k \le 3$: 
\be
   A(k) \;=\; \frac{1}{2}\left[ 
      \left( \sum_{i=1}^{2k} \sigma_i \sigma_{i+1} \right)^2
   \,-\,
   \left( \sum_{i=1}^{2k} \sigma_{2i} 
          \,+\,
          \sum_{j=1}^{k} \sigma_{2j-1} \sigma_{2j} \sigma_{2j+1} \right)^2
   \right]\,. 
\label{lemma_def_A}
\ee

The most direct way to show that $A(k)=0$ for $k=1,2,3$ when 
$\sigma_i^2=1$ and $\sigma_{2k+1}=\sigma_1$ is to write down the above 
expression with $t_i=1-\sigma_i^2$ and taking into account the latter
condition ($\sigma_{2k+1}=\sigma_1$). We arrive easily at the expressions:
\begin{subeqnarray}
A(1) &=& t_1^2 (t_2-1) \,, \\ 
A(2) &=& -t_1 t_3 (\sigma_2+\sigma_4)^2\,,  \\
A(3) &=& -\sum\limits_{j=1}^3 t_{2j-1} \left[ 
   t_{2j+1} t_{2j} - 2 \sigma_{2j}\sigma_{2j-2} 
   +2 \sigma_{2j}\sigma_{2j+1}\sigma_{2j-2}\sigma_{2j-3} \right]\,, 
\end{subeqnarray}
where we have explicitly used the identifications $\sigma_{2k+1}=\sigma_1$
and $\sigma_{2k}=\sigma_0$. 
It is obvious that when $t_i=0$, then $A(k)=0$ for $k=1,2,3$. 

For $k\ge 4$ \reff{lemma_def_A} cannot be true as the first term of the 
r.h.s.\ of Eq.~\reff{lemma_def_A} contains products of up to four spins, 
while the second term of the r.h.s.\ of Eq.~\reff{lemma_def_A} contains 
products of six spins, and none of them can be simplified using the boundary
conditions $\sigma_{2k+1}=\sigma_1$.  \hfill \qed

\medskip

\noindent
{\bf Remark}. Eq.~\reff{eq_map} can be rewritten as a sum over a spin 
$\sigma_0$:  
\be
   \sum\limits_{\sigma_0=\pm 1} 
        e^{ \displaystyle J \sum_{i=1}^{2k} \sigma_0 \sigma_i \sigma_{i+1} }
   \;=\;
   \sum\limits_{\sigma_0=\pm 1}  \exp \left[ 
           J \sum_{i=1}^{k} \sigma_0 \sigma_{2i} \,+\,
           J \sum_{j=1}^{k} \sigma_0 \sigma_{2j-1} \sigma_{2j} \sigma_{2j+1} 
                                  \right]
   \;.
\label{eq_map_bis}
\ee
Therefore, a 3-spin Ising model can be written as a certain Ising model with
2- and 4-spin interactions.

\bigskip

%
%
\begin{figure}[htb]
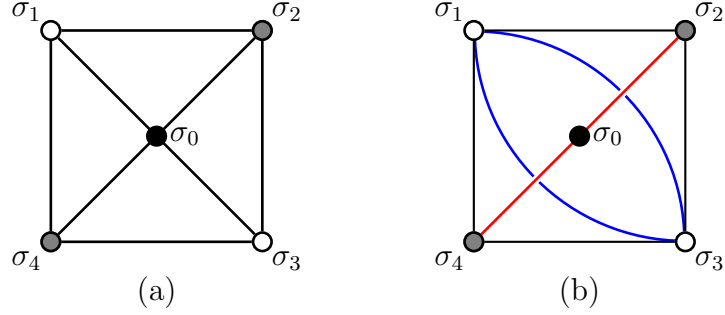

 \centering 
 \psset{xunit=40pt}
 \psset{yunit=40pt}
 \pspicture(-0.5,-0.7)(6.5,2.5)
 \rput{0}(0,0){
  %
  %
  \psline[linecolor=black,linewidth=1pt](0,0)(2,0)(2,2)(0,2)(0,0)
  \psline[linecolor=black,linewidth=1pt](0,0)(2,2)
  \psline[linecolor=black,linewidth=1pt](2,0)(0,2)
  \pscircle*[linecolor=gray] (0,0){4pt}
  \pscircle[linewidth=1pt]   (0,0){4pt}
  \pscircle*[linecolor=gray] (2,2){4pt}
  \pscircle[linewidth=1pt]   (2,2){4pt}
  \pscircle*[linecolor=white](2,0){4pt}
  \pscircle[linewidth=1pt]   (2,0){4pt}
  \pscircle*[linecolor=white](0,2){4pt}
  \pscircle[linewidth=1pt]   (0,2){4pt}
  \pscircle*[linecolor=black](1,1){4pt}
  \uput[225](0,0){$\sigma_4$}
  \uput[45] (2,2){$\sigma_2$}
  \uput[315](2,0){$\sigma_3$}
  \uput[135](0,2){$\sigma_1$}
  \uput[0]  (1,1){$\sigma_0$}
  \uput[270](1.0,-0.2){(a)}
 }

 \rput{0}(4,0){
  %
  %
  \psarc[linewidth=1pt,linecolor=blue](0,0){2.8}{0}{90}
  \psarc[linewidth=1pt,linecolor=blue](2,2){2.8}{180}{270}
  \psline[linecolor=black](0,0)(2,0)(2,2)(0,2)(0,0)
  \psline[linecolor=red,linewidth=1pt,border=1pt](0,0)(2,2)
  \pscircle*[linecolor=gray] (0,0){4pt}
  \pscircle[linewidth=1pt]   (0,0){4pt}
  \pscircle*[linecolor=gray] (2,2){4pt}
  \pscircle[linewidth=1pt]   (2,2){4pt}
  \pscircle*[linecolor=white](2,0){4pt}
  \pscircle[linewidth=1pt]   (2,0){4pt}
  \pscircle*[linecolor=white](0,2){4pt}
  \pscircle[linewidth=1pt]   (0,2){4pt}
  \pscircle*[linecolor=black](1,1){4pt}
  \uput[225](0,0){$\sigma_4$}
  \uput[45] (2,2){$\sigma_2$}
  \uput[315](2,0){$\sigma_3$}
  \uput[135](0,2){$\sigma_1$}
  \uput[0]  (1,1){$\sigma_0$}
  \uput[270](1.0,-0.2){(b)}
 }
\endpspicture
\caption{\label{fig_map_sqa_NEW}
(a) The Ising model with uniform 3-spin interaction
    $J (\sigma_0\sigma_1\sigma_2 + \sigma_0\sigma_2\sigma_3 
       +\sigma_0\sigma_3\sigma_4 + \sigma_0\sigma_4\sigma_1)$. 
(b) The Ising model with 2-spin interaction
    $J(\sigma_0\sigma_2 + \sigma_0\sigma_4)$, and 4-spin interaction  
    $J(\sigma_0\sigma_1\sigma_2\sigma_3
      +\sigma_0\sigma_3\sigma_4\sigma_1)$ obtained from (a)
    after applying Lemma~\ref{lemma.partial.trace} with $k=2$.
}
\end{figure}

The first non-trivial application corresponds to $k=2$ (see 
Figure~\ref{fig_map_sqa_NEW}). Let us suppose we have an
Eulerian plane triangulation $G=(V,E)$, such that $V=V_1 \cup V_2 \cup V_3$, 
and such that each vertex $i\in V_3$ has degree $\Delta_3=4$. 
(This happens for all the lattices belonging to the family described in 
Section~\ref{sec.lattices}.)
If we perform the transformation of Lemma~\ref{lemma.partial.trace}
to the vertices on $V_3$, then the HMBW model on $G$ with 
coupling constant $J$ \reff{def_H_HMBW}/\reff{def_Z_HMBW} can be written as
a mixed AT model on $G_{13}=(V_1\cup V_3,E_{13})$, whose dual is 
$G_{13}^* = (V_2,E_{13}^*)$ [or on $G_{23}=(V_2\cup V_3,E_{23})$ whose dual is
$G_{23}^* = (V_1,E_{23}^*)$]. 
Both the two--spin coupling $\sigma_i\sigma_j$, and the 
4--spin coupling $\sigma_i\sigma_j \sigma_{i^*}\sigma_{j^*}$ (for any
$\<ij\>\in E_{13}$) take the same value $J$; but the coupling between 
dual spins is zero. Again, the edge 
$e^*=\<i^*j^*\>\in E_{13}^*$ is the one {\em uniquely}\/ associated to the
edge $e\in E_{13}$.  
Notice that the vertices in $V_3$ (depicted as black dots in
Figure~\ref{fig_map_sqa_NEW}) subdivide the edge joining the
two vertices in $V_1$ (depicted as gray dots in Figure~\ref{fig_map_sqa_NEW}).
Therefore, there are two dual edges (depicted as blue curves on
Figure~\ref{fig_map_sqa_NEW}(b)) 
joining the two dual vertices in $V_2$
(depicted as white dots in Figure~\ref{fig_map_sqa_NEW}).
The Hamiltonian of the transformed system is therefore
\begin{subeqnarray}
\mathcal{H} &=&  
-J \sum_{\<ij\>\in E_{13}} \sigma_i \sigma_j
-J \sum_{\<ij\>\in E_{13}} \sigma_i \sigma_j \sigma_{i^*} \sigma_{j^*} \\
 &=& -J \sum_{\<ij\>\in E_{13}} \left(1 \,+\, \sigma_{i^*} \sigma_{j^*}
       \right) \, \sigma_i \sigma_j 
\label{def_H_Is4s0}
\end{subeqnarray}
The relation between partition functions is clear: given an 
Eulerian plane triangulation 
$G=(G,V)$, such that $V=V_1\cup V_2 \cup V_3$, 
$E= E_{12}\cup E_{13}\cup E_{23}$, and the degree of all vertices in $V_3$
is $\Delta_3=4$, then 
\be
Z_{\rm HMBW}(G;J) \;=\; Z_{\rm mAT}(G_{13},G_{13}^*; J,0,J) \,,
\label{def_Z_Is4s0_OK}
\ee
where $G_{13}=(V_1\cup V_3,E_{13})$, and its dual graph is
$G_{13}^* = (V_2,E_{13}^*)$. 

%
%
\begin{figure}[htb]
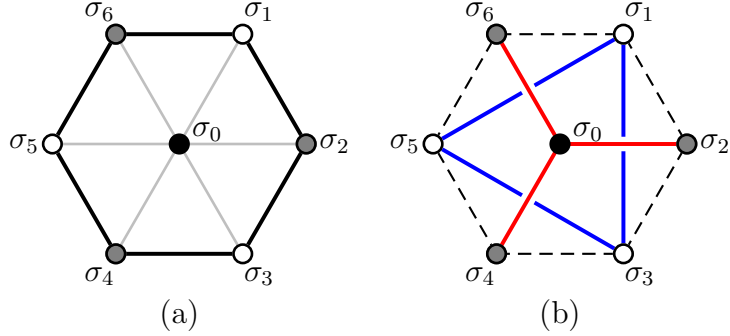

 \centering
 \psset{xunit=24pt}
 \psset{yunit=24pt}
 \pspicture(-1.8,-1.5)(10,4.5)
 \rput{0}(0,0){
 %
 %
  \psline[linecolor=black,linewidth=1.5pt](0,0)(2,0)(3,1.73205)%
                             (2,3.46410)(0,3.46410)(-1,1.73205)(0,0)
  \psline[linecolor=lightgray,linewidth=1pt]( 0,0)(2,3.46410)
  \psline[linecolor=lightgray,linewidth=1pt]( 2,0)(0,3.46410)
  \psline[linecolor=lightgray,linewidth=1pt](-1,1.73205)(3,1.73205)
  \pscircle*[linecolor=gray] (0,0){4pt}
  \pscircle[linewidth=1pt]   (0,0){4pt}
  \pscircle*[linecolor=gray] (3,1.73205){4pt}
  \pscircle[linewidth=1pt]   (3,1.73205){4pt}
  \pscircle*[linecolor=gray] (0,3.46410){4pt}
  \pscircle[linewidth=1pt]   (0,3.46410){4pt}
  \pscircle*[linecolor=white](2,0){4pt}
  \pscircle[linewidth=1pt]   (2,0){4pt}
  \pscircle*[linecolor=white](-1,1.73205){4pt}
  \pscircle[linewidth=1pt]   (-1,1.73205){4pt}
  \pscircle*[linecolor=white](2,3.46410){4pt}
  \pscircle[linewidth=1pt]   (2,3.46410){4pt}
  \pscircle*[linecolor=black](1,1.73205){4pt}
  \uput[240](0,0){$\sigma_4$}
  \uput[300](2,0){$\sigma_3$}
  \uput[  0](3.0,1.73205){$\sigma_2$}
  \uput[ 60](2.0,3.46410){$\sigma_1$}
  \uput[120](0.0,3.46410){$\sigma_6$}
  \uput[180](-1.0,1.73205){$\sigma_5$}
  \uput[30] (1,1.73205){$\sigma_0$}
  \uput[270](1.0,-0.5){(a)}
 }

 \rput{0}(0,0){ 
 %
 %
  \psline[linecolor=black,linestyle=dashed](6,0)(8,0)(9,1.73205)%
                               (8,3.46410)(6,3.46410)(5,1.73205)(6,0)
  \psline[linecolor=blue,linewidth=1.5pt](5,1.73205)(8,0)
  \psline[linecolor=blue,linewidth=1.5pt](8,3.46410)(8,0)
  \psline[linecolor=blue,linewidth=1.5pt](5,1.73205)(8,3.46410)
  \psline[linecolor=red,linewidth=1.5pt,border=2pt](7,1.73205)(9,1.73205)
  \psline[linecolor=red,linewidth=1.5pt,border=2pt](7,1.73205)(6,3.46410)
  \psline[linecolor=red,linewidth=1.5pt,border=2pt](7,1.73205)(6,0)
  \pscircle*[linecolor=gray] (6,0){4pt}
  \pscircle[linewidth=1pt]   (6,0){4pt}
  \pscircle*[linecolor=gray] (9,1.73205){4pt}
  \pscircle[linewidth=1pt]   (9,1.73205){4pt}
  \pscircle*[linecolor=gray] (6,3.46410){4pt}
  \pscircle[linewidth=1pt]   (6,3.46410){4pt}
  \pscircle*[linecolor=white](8,0){4pt}
  \pscircle[linewidth=1pt]   (8,0){4pt}
  \pscircle*[linecolor=white](5,1.73205){4pt}
  \pscircle[linewidth=1pt]   (5,1.73205){4pt}
  \pscircle*[linecolor=white](8,3.46410){4pt}
  \pscircle[linewidth=1pt]   (8,3.46410){4pt}
  \pscircle*[linecolor=black](7,1.73205){4pt}
  \uput[240](6,0){$\sigma_4$}
  \uput[300](8,0){$\sigma_3$}
  \uput[  0](9.0,1.73205){$\sigma_2$}
  \uput[ 60](8.0,3.46410){$\sigma_1$}
  \uput[120](6.0,3.46410){$\sigma_6$}
  \uput[180](5,1.73205){$\sigma_5$}
  \uput[30] (7,1.73205){$\sigma_0$}
  \uput[270](7,-0.5){(b)}
}
 \endpspicture
\caption{\label{fig_map_tri_NEW}
(a) The Ising model with uniform 3-spin interaction
    $J (\sigma_0\sigma_1\sigma_2 + \sigma_0\sigma_2\sigma_3 
       +\sigma_0\sigma_3\sigma_4 + \sigma_0\sigma_4\sigma_5
       +\sigma_0\sigma_5\sigma_6 + \sigma_0\sigma_6\sigma_1)$.
(b) The Ising model with 2-spin interaction
    $J(\sigma_0\sigma_2 + \sigma_0\sigma_4 + \sigma_0\sigma_6)$, and 
     4-spin interaction
    $J(\sigma_0\sigma_1\sigma_2\sigma_3
      +\sigma_0\sigma_3\sigma_4\sigma_5
      +\sigma_0\sigma_5\sigma_6\sigma_1)$ obtained from (a)
    after applying Lemma~\ref{lemma.partial.trace} with $k=3$.
} 
\end{figure}

The $k=3$ case is shown in Figure~\ref{fig_map_tri_NEW}. 
Let us suppose we have an Eulerian plane triangulation $G=(V,E)$, such 
that $V=V_1 \cup V_2 \cup V_3$, and
such that each vertex $i\in V_3$ has degree $\Delta_3=6$.
(This happens for the triangular and bisected-hexagonal lattices.)
If we perform the transformation of Lemma~\ref{lemma.partial.trace}
to the vertices on $V_3$, then the HMBW model on $G$ with
coupling constant $J$ \reff{def_H_HMBW}/\reff{def_Z_HMBW} can be written as
a mixed AT model on $G_{13}=(V_1\cup V_3,E_{13})$ [or on
$G_{23}=(V_2\cup V_3,E_{23})$], whose dual graph is
$G_{13}^* = (V_2,E_{13}^*)$; and again all couplings are 
equal to $J$, except the couplings between dual vertices, which are 
zero. The model can be written as in \reff{def_H_Is4s0}/\reff{def_Z_Is4s0_OK}. 

We can gather the above results in the following 

\begin{theorem} \label{theo.partial.trace}
Let $G=(V,E)$ be an Eulerian plane triangulation with vertex set 
$V=V_1\cup V_2 \cup V_3$, edge set $E= E_{12}\cup E_{13}\cup E_{23})$,
and such that the degree of all vertices in $V_3$ is $\Delta_3=4$ or $6$.
Then the Hintermann--Merlini--Baxter--Wu model on $G$ 
\reff{def_H_HMBW}/\reff{def_Z_HMBW} with coupling constant $J$ 
is equivalent to a mixed Ashkin--Teller model 
\reff{def_H_mAT}/\reff{def_Z_mAT} with uniform coupling 
constant $K_2=K_4=J$ and $K'_2=0$ on $G_{13}=(V_1\cup V_3,E_{13})$ 
[with $G_{13}^* = (V_2,E_{13}^*)$], such that  
\be
Z_{\rm HMBW}(G;J) \;=\; Z_{\rm mAT}(G_{13},G_{13}^*; J,0,J) \,.
\label{def_Z_HMBW_vs_Z_mAT.theo}
\ee
(We can equally define the mixed AT model on $G_{23}=(V_2\cup V_3,E_{23})$,
and the formulas are the same after interchanging $1\leftrightarrow 2$.)
\end{theorem}   
Using Theorem~\ref{theo.mAT_ICLAT}, we can obtain the following
\begin{corollary} \label{coro.partial.trace}
Let $G=(V,E)$ be an Eulerian plane triangulation with vertex set 
$V=V_1\cup V_2 \cup V_3$, edge set $E= E_{12}\cup E_{13}\cup E_{23})$,
and such that the degree of all vertices in $V_3$ is $\Delta_3=4$ or $6$.
Then the uniform Hintermann--Merlini--Baxter--Wu model on $G$ 
\reff{def_H_HMBW}/\reff{def_Z_HMBW} with coupling constant $J$ 
is equivalent to an infinite--coupling--limit Ashkin--Teller model  
\reff{def_H_ICLAT}/\reff{def_Z_ICLAT} on $G_{13}^*=(V_2,E_{13}^*)$, where
$G_{13}=(V_1\cup V_3,E_{13})$, and such that 
\be
Z_{\rm HMBW}(G;J) \;=\; 2^{-|V_2|+1}\, Z_{\rm ICLAT}(G_{13}^*;L_2,L_2') \,,
\label{def_Z_HMBW_vs_Z_ICLAT.coro}
\ee
where 
\begin{subeqnarray}
e^{-2L_2}  &=& \frac{1}{\cosh(2J)} \\[2mm]
e^{-2L'_2} &=& \tanh(2J) 
\label{def_L.coro}
\end{subeqnarray}
These couplings satisfy 
\be
\left(e^{-2L_2\phantom{'}}\right)^2 +\left(e^{-2L_2'}\right)^2 \;=\; 1 \,.
\label{rel_ICLAT_tri}
\ee
\end{corollary}   

We can apply this corollary to the main lattices in this paper:

\begin{itemize}

\item[(a)] 
      If $G$ is the triangular lattice and we choose $G_{13}$ to be one of its 
      hexagonal sublattices, then $G_{13}^*$ is the other triangular 
      sublattice.  
      Then, the triangular-lattice HMBW model with coupling $J$ 
      is equivalent to a triangular-lattice ICLAT model with 
      couplings $(L_2,L_2')$ satisfying the condition \reff{rel_ICLAT_tri}.

\item[(b)]
      If $G$ is the union-jack lattice and we choose $G_{13}$ to be 
      a decorated square lattice (formed by the degree--4 black vertices
      and the degree--8 gray vertices in Figure~\ref{fig_lattices}(a)), 
      then $G_{13}^*$ is also a square lattice, tilted $45^\circ$, 
      formed by the degree--8 white dots in Figure~\ref{fig_lattices}(a),
      and with double edges. Then, the union--jack--lattice HMBW 
      model with coupling $J$ is equivalent to
      a square--lattice ICLAT model with couplings $(L_2,L_2')$ 
      satisfying condition 
\be
e^{-2L'_2} + e^{-2L_2'} \;=\; 1 \,,
\label{rel_ICLAT_square}
\ee
      because of the double edges. 
\end{itemize}

If $G$ is the bisected-hexagonal lattice, then we have several
choices for $G_{13}$:
 
\begin{itemize}
\item[(c)]
      If $G_{13}$ is the decorated triangular lattice formed by the degree--4
      and degree--12 vertices in $G$ (black and gray dots, respectively in
      Figure~\ref{fig_lattices}(b)), then $G_{13}^*$ is a hexagonal 
      lattice formed by the degree--6 vertices in $G$ (white dots) and 
      double edges. 
      Then, the bisected-hexagonal--lattice HMBW model 
      with coupling $J$ 
      is equivalent to a hexagonal-lattice ICLAT model with 
      couplings $(L_2,L_2')$ satisfying the condition \reff{rel_ICLAT_square}.

\item[(d)]
      If $G_{13}$ is the  hexagonal lattice formed by the degree--4 
      and degree--6 vertices in $G$ (black and white dots, respectively in 
      Figure~\ref{fig_lattices}(b)), then $G_{13}^*$ is a triangular 
      lattice formed by the degree--12 vertices in $G$ (gray dots), and 
      double edges.
      Then, the bisected-hexagonal--lattice HMBW model with coupling $J$ 
      is equivalent to a triangular-lattice ICLAT model with 
      couplings $(L_2,L_2')$ satisfying the condition \reff{rel_ICLAT_square}.

\item[(e)]
      If $G_{13}$ is the diced lattice formed by the degree--6 
      and degree--12 vertices in $G$ (white and gray dots, respectively in 
      Figure~\ref{fig_lattices}(b)), then $G_{13}^*$ is a kagome 
      lattice formed by the degree--4 vertices in $G$ (black dots).
      Then, the bisected-hexagonal--lattice HMBW model with coupling $J$ 
      is equivalent to a kagome-lattice ICLAT model with 
      couplings $(L_2,L_2')$ satisfying the condition \reff{rel_ICLAT_tri}.
\end{itemize}

In conclusion, the HMBW model on several Eulerian triangulations can be 
exactly mapped along certain curves (\reff{rel_ICLAT_tri} or 
\reff{rel_ICLAT_square}) on the ICLAT model defined on other lattices.

Some of these results can be obtained using the decimation transformation 
(see Appendix~\ref{appendix.decimation}). This latter transformation 
can be applied
in the case $k=2$ with general couplings $J_i$; but reduce to some of the 
above results in the homogeneous case $J_i=J$.  

%
%
\subsection{Mapping between the BW and two-color non-overlapping 
           Eulerian-bond models} \label{sec.BW.2CNEB}

Let us start with a BW model: that is, a HMBW model 
\reff{def_H_HMBW}/\reff{def_Z_HMBW} model defined  
on the triangular lattice $G=(V,E)$. Then, both the vertex and the edge
sets can be partitioned into three disjoint sets 
$V=V_1\cup V_2 \cup V_3$, and $E=E_{12}\cup E_{13}\cup E_{23}$, so that
for all $i\neq j$, $E_{ij}$ contains all edges $e=\<xy\>$ with $x\in V_i$
and $y\in V_j$. Each triangular face $t=\{x,y,z\}$ contains one edge from
each set $E_{ij}$; therefore, we can rewrite the Hamiltonian 
\reff{def_H_HMBW} as
\be
\mathcal{H}_\text{BW} \;=\;  
- J \sum\limits_{ \{i,j\} \in E_{23}} \sigma_i \sigma_j (\sigma_p + \sigma_q)\,,
\label{def_H_HMBW_bis} 
\ee
where $\{i,j,p\}$ and $\{i,j,q\}$ are only two triangular faces sharing the
common edge $e=\<ij\>\in E_{23}$. Notice that the subgraph 
$(V_2\cup V_3,E_{23})$ is a hexagonal lattice. 

The Boltzmann weight associated to an arbitrary edge $e=\<ij\>\in E_{23}$ can
be rewritten as
\be
e^{J\sigma_i \sigma_j (\sigma_p + \sigma_q)} \;=\; 
\cosh\left[J(\sigma_p + \sigma_q)\right] \left\{ 
1 + \sigma_i \sigma_j \tanh \left[J(\sigma_p + \sigma_q)\right] \right\}\,.
\label{map.trick1}
\ee
If we now use the trivial identities:
\begin{subeqnarray}
\cosh\left[J(\sigma_p + \sigma_q)\right] &=& 
  \frac{\cosh 2J }{(\cosh 2J)^{\delta_{\sigma_p,-\sigma_q}}} \;=\; 
  \begin{cases} 
    \cosh 2J & \quad \text{if $\sigma_p = \sigma_q$} \\
    1        & \quad \text{if $\sigma_p \neq \sigma_q$} \end{cases}
\\[2mm]
\tanh\left[J(\sigma_p + \sigma_q)\right] &=& \sigma_p 
  \delta_{\sigma_p,\sigma_q} \tanh 2J \;=\; 
  \begin{cases}
    \sigma_p\tanh 2J & \, \text{if $\sigma_p = \sigma_q$} \\
    0                & \, \text{if $\sigma_p \neq \sigma_q$} \end{cases}
\label{map.trick2}
\end{subeqnarray}
we can perform a high--temperature expansion of the partition function
\reff{def_Z_HMBW}/\reff{def_H_HMBW_bis}: 
\begin{eqnarray}
Z_{\rm BW}(G;J) &=& \left( \cosh 2J\right)^{|E_{23}|} \, 
                    2^{|V_2\cup V_3|} \, 
\sum\limits_{\begin{scarray}
                  \{\sigma_p \} \\
                  p\in V_1
                  \end{scarray}} 
\sum\limits_{\begin{scarray}
              E'\subseteq E_{23}\\
              E' \, \hbox{ \scriptsize Eulerian } 
             \end{scarray}} 
\prod_{e\in E_{23}} 
\left( \frac{1}{\cosh 2J} \right)^{\delta_{\sigma_p,-\sigma_q}} 
\nonumber \\[2mm]
& & \qquad \times  
\prod_{e\in E'} \sigma_p \delta_{\sigma_p,\sigma_q} \tanh 2J \,, 
\label{map.trick3}
\end{eqnarray}
with $E'$ being the subset of edges in $E_{23}$ contributing with
a factor proportional to $\tanh 2J$. Notice that 
\begin{enumerate}
\item  The sum over subsets $E'$ only contains Eulerian subgraphs  
       $(V_2\cup V_3,E')$. Because $(V_2\cup V_3,E_{23})$ is a hexagonal
       lattice of maximum degree $\Delta=3$, the subgraphs that contribute
       to the partition function \reff{map.trick3} are precisely the class
       of non--intersecting Eulerian loops.  

\item The spins that appear in  \reff{map.trick3} live on the sublattice $V_1$.
      Because of the factor $\sigma_p \delta_{\sigma_p,\sigma_q}$, all spins
      on any boundary of a loop should be equal. 

\item Because the hexagonal lattice is bipartite, all loops have even length.
      Therefore, the factors $\sigma_p$ in \reff{map.trick3} cancel out,
      and always give a contribution equal to $1$. 
\end{enumerate}

With these considerations \reff{map.trick3} can be rewritten as
\begin{eqnarray}
Z_{\rm BW}(G;J) &=& \left( \cosh 2J\right)^{|E_{23}|} \, 
                    2^{|V_2\cup V_3|} \, 
\sum\limits_{\begin{scarray}
              E'\subseteq E_{23}\\
              E' \, \hbox{ \scriptsize Eulerian } 
             \end{scarray}} 
             \left( \tanh 2J \right)^{|E'|} \, 
\nonumber \\[2mm]
& & \qquad \times  
\sum\limits_{\begin{scarray}
                  \{\sigma_p \} \\
                  p\in V_1
                  \end{scarray}} 
\prod_{e\in E_{23}}
\left( \frac{1}{\cosh 2J} \right)^{\delta_{\sigma_p,-\sigma_q}} 
\prod_{e\in E'} \delta_{\sigma_p,\sigma_q} \,, 
\label{map.trick4}
\end{eqnarray}

\begin{figure}[htb]
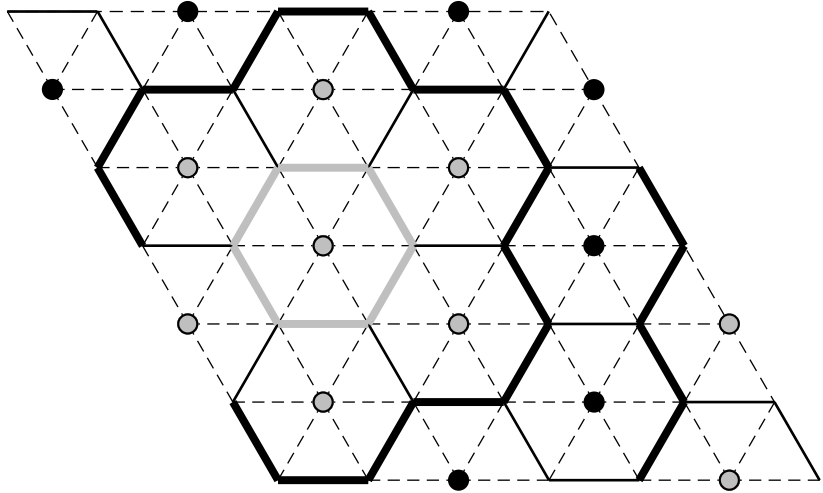

 \centering
 \psset{unit=0.6cm}
 \pspicture(0,-0.5)(18.2,10.5)
 \rput{0}(0,0){ 
   \multirput(-3,5.19615){2}{%
   \multirput(6,0){2}{%
     \psline[linewidth=1.0pt]( 5,1.73205)( 6,0.000000)
     \psline[linewidth=1.0pt]( 8,0.00000)( 6,0.000000)
     \psline[linewidth=1.0pt]( 8,0.00000)( 9,1.732050)
     \psline[linewidth=1.0pt](11,1.73205)( 9,1.732050)
     \psline[linewidth=1.0pt]( 5,1.73205)( 6,3.464100)
     \psline[linewidth=1.0pt]( 9,1.73205)( 8,3.464100)
     \psline[linewidth=1.0pt]( 6,3.46410)( 8,3.464100)
     \psline[linewidth=1.0pt]( 6,3.46410)( 5,5.196150)
     \psline[linewidth=1.0pt]( 8,3.46410)( 9,5.196150)
     \psline[linewidth=0.5pt,linestyle=dashed](10,0.00000)( 8,0.000000)
     \psline[linewidth=0.5pt,linestyle=dashed](10,0.00000)(12,0.000000)
     \psline[linewidth=0.5pt,linestyle=dashed](10,0.00000)( 9,1.732050)
     \psline[linewidth=0.5pt,linestyle=dashed](10,0.00000)(11,1.732050)
     \psline[linewidth=0.5pt,linestyle=dashed]( 7,1.73205)( 5,1.732050)
     \psline[linewidth=0.5pt,linestyle=dashed]( 7,1.73205)( 9,1.732050)
     \psline[linewidth=0.5pt,linestyle=dashed]( 7,1.73205)( 6,0.000000)
     \psline[linewidth=0.5pt,linestyle=dashed]( 7,1.73205)( 8,0.000000)
     \psline[linewidth=0.5pt,linestyle=dashed]( 7,1.73205)( 6,3.464100)
     \psline[linewidth=0.5pt,linestyle=dashed]( 7,1.73205)( 8,3.464100)
     \psline[linewidth=0.5pt,linestyle=dashed]( 4,3.46410)( 5,1.732050)
     \psline[linewidth=0.5pt,linestyle=dashed]( 4,3.46410)( 6,3.464100)
     \psline[linewidth=0.5pt,linestyle=dashed]( 4,3.46410)( 5,5.196150)
     \psline[linewidth=0.5pt,linestyle=dashed]( 4,3.46410)( 3,5.196150)
     \psline[linewidth=0.5pt,linestyle=dashed](10,3.46410)( 9,1.732050)
     \psline[linewidth=0.5pt,linestyle=dashed](10,3.46410)( 8,3.464100)
     \psline[linewidth=0.5pt,linestyle=dashed]( 7,5.19615)( 6,3.464100)
     \psline[linewidth=0.5pt,linestyle=dashed]( 7,5.19615)( 8,3.464100)
     \pscircle*[linecolor=black]             (10,0.0){4pt}
     \pscircle*[linecolor=black]             ( 7,1.73205){4pt}
     \pscircle*[linecolor=black]             ( 4,3.46410){4pt}
   }}
   \psline[linewidth=0.5pt,linestyle=dashed](18,0)%
                                           (12,10.392300)(0,10.392300)
   \psline[linewidth=1.0pt](18, 0)       (17, 1.73205)
   \psline[linewidth=1.0pt](15, 5.19615) (14, 6.9282)
   \psline[linewidth=1.0pt]( 0,10.392300)( 2,10.392300)
   \psline[linewidth=1.0pt]( 6,10.392300)( 8,10.392300)

   \psline[linewidth=3.0pt,linecolor=lightgray]( 6,3.46410)( 8,3.464100)
   \psline[linewidth=3.0pt,linecolor=lightgray]( 6,3.46410)( 5,5.196150)
   \psline[linewidth=3.0pt,linecolor=lightgray]( 8,3.46410)( 9,5.196150)
   \psline[linewidth=3.0pt,linecolor=lightgray]( 5,5.19615)( 6,6.92820)
   \psline[linewidth=3.0pt,linecolor=lightgray]( 9,5.19615)( 8,6.92820)
   \psline[linewidth=3.0pt,linecolor=lightgray]( 6,6.92820)( 8,6.92820)

   \psline[linewidth=3.0pt](3,5.19615)(2,6.92820)
   \psline[linewidth=3.0pt](2,6.92820)(3,8.66025)
   \psline[linewidth=3.0pt](3,8.66025)(5,8.66025)
   \psline[linewidth=3.0pt](5,8.66025)(6,10.3923)
   \psline[linewidth=3.0pt](6,10.3923)(8,10.3923)
   \psline[linewidth=3.0pt](8,10.3923)(9,8.66025)
   \psline[linewidth=3.0pt](9,8.66025)(11,8.66025)
   \psline[linewidth=3.0pt](11,8.66025)(12,6.92820)
   \psline[linewidth=3.0pt](12,6.92820)(11,5.19615)
   \psline[linewidth=3.0pt](11,5.19615)(12,3.4641)
   \psline[linewidth=3.0pt](12,3.4641)(11,1.73205)
   \psline[linewidth=3.0pt](11,1.73205)(9,1.73205)
   \psline[linewidth=3.0pt](9,1.73205)(8,0)
   \psline[linewidth=3.0pt](8,0)(6,0)
   \psline[linewidth=3.0pt](6,0)(5,1.73205 )
   \psline[linewidth=3.0pt](14,0)(15,1.73205)
   \psline[linewidth=3.0pt](15,1.73205)(14,3.4641)
   \psline[linewidth=3.0pt](14,3.4641)(15,5.19615)
   \psline[linewidth=3.0pt](15,5.19615)(14,6.92820)

   \pscircle*[linecolor=black]             ( 4,10.392300){4pt}
   \pscircle*[linecolor=black]             (10,10.392300){4pt}
   \pscircle*[linecolor=black]             (13, 8.66025){4pt}

   \pscircle*[linecolor=lightgray](16, 3.4641){4pt}
   \pscircle                      (16, 3.4641){4pt}
   \pscircle*[linecolor=lightgray](16, 0){4pt}
   \pscircle                      (16, 0){4pt}
   \pscircle*[linecolor=lightgray](10, 3.4641){4pt}
   \pscircle                      (10, 3.4641){4pt}
   \pscircle*[linecolor=lightgray](10, 6.9282){4pt}
   \pscircle                      (10, 6.9282){4pt}
   \pscircle*[linecolor=lightgray](7,  8.66025){4pt}
   \pscircle                      (7,  8.66025){4pt}
   \pscircle*[linecolor=lightgray](7,  5.19615){4pt}
   \pscircle                      (7,  5.19615){4pt}
   \pscircle*[linecolor=lightgray](7,  1.73205){4pt}
   \pscircle                      (7,  1.73205){4pt}
   \pscircle*[linecolor=lightgray](4,  3.4641){4pt}
   \pscircle                      (4,  3.4641){4pt}
   \pscircle*[linecolor=lightgray](4,  6.9282){4pt}
   \pscircle                      (4,  6.9282){4pt}

 }
 \endpspicture
 \caption{\label{fig_tri_lat_loop}
  The BW model as a two--color 
  non-overlapping loop model on the hexagonal lattice. We show a finite
  subset of a triangular lattice $G=(V,E)$. The hexagonal sublattice
  is formed by the subgraph $(V_2\cup V_3,E_{23})$; the vertices of this
  sublattice are not depicted, while the edges are depicted as solid lines.
  The vertices belonging to the subset $V_1$ are depicted
  as dots: a black (resp.\/ gray) dot corresponds to an associated 
  spin taking the value $+1$ (resp.\/ $-1$). The edges not in $E_{23}$ are
  depicted as dashed thin lines.  
  The low--temperature loops (depicted as thick black lines) separate 
  regions with opposite spin values. The high--temperature
  loops (depicted as thick solid gray lines) are surrounded by spins all 
  taking the same value. As this is a finite piece of the triangular 
  lattice, we can only show some parts of the loops.  
}
\end{figure}

The inverse powers of $\cosh 2J$ can be interpreted as contributions from
domain--wall loops: i.e., low--temperature loops which separate regions
with all spins equal to $+1$ from regions with all spins equal to $-1$.
Notice that $(\cosh 2J)^{-\delta_{\sigma_p,-\sigma_q}}$ only gives a
non-trivial contribution when $\sigma_p\neq \sigma_q$. These low--temperature
loops are also non-intersecting, and in addition, they cannot cross any 
high--temperature loop. See Figure~\ref{fig_tri_lat_loop} for an 
example.  

%
%

In summary, the partition function of the BW model 
\reff{def_H_HMBW}/\reff{def_H_HMBW_bis} is equivalent to a loop gas with 
the following two conditions: (1) the loops are non-intersecting, and (2)
there are two species of loops: the high--temperature loops with weight
$\tanh 2J$ per edge, and the low--temperature loops with weight 
$\cosh^{-1} 2J$ per edge. The partition function can be finally written
as
\be
Z_{\rm BW}(G;J) \;=\; \left( 2\cosh 2J\right)^{|E_{23}|} \, 
\sum\limits_{k\ge 0}
\sum\limits_{\begin{scarray}
                  \hbox{\scriptsize loops with $k$} \\
                  \hbox{\scriptsize components} 
                  \end{scarray}} 
\prod_{j=1}^k 
\left\{ \left( \tanh 2J\right)^{\ell_j} + 
       \left(\frac{1}{\cosh 2J}\right)^{\ell_j} 
\right\} \,, 
\label{def_H_sHMBW_tris}
\ee
where $\ell_j$ is the length of the $j$-th loop. 
 
Because $\tanh^2 2J + \cosh^{-2} 2J = 1$, we can parametrize the weights
with a single parameter $\theta$:
\be
\sin\theta \;=\; \frac{1}{\cosh 2J}\,, \qquad  
\cos\theta \;=\; \tanh 2J\,, 
\label{def_theta}
\ee
so that
\be
Z_{\rm BW}(G;J) \;=\; \left( 2\cosh 2J\right)^{|E_{23}|} \,
\sum\limits_{k\ge 0}
\sum\limits_{\begin{scarray}
                  \hbox{\scriptsize loops with $k$} \\
                  \hbox{\scriptsize components}
                  \end{scarray}}
\prod_{j=1}^k
\left\{ \sin^{\ell_j} \theta + \cos^{\ell_j} \theta   
\right\} \,.
\label{def_H_sHMBW_4}
\ee

If we perform a duality transformation in the HMBW model with 
$J\to J^*$ ($v \to 2/v^*$) \cite{Gruber_77}, the loop weights are interchanged:
$\tanh 2J \to \cosh^{-1} 2J^*$, and $\cosh^{-1} 2J\to \tanh 2J^*$. This 
agrees with the interpretation of those loops as high--temperature and
low--temperature loops. The uniform HMBW model is self-dual when $J=J^*$.
In terms of the loop model, we obtain that \reff{def_H_sHMBW_4} is self-dual
at
\be
\sin \theta_c \;=\; \cos \theta_c \;=\; \frac{1}{\sqrt{2}} \,.
\label{def_theta_selfdual}
\ee 

Notice that the loop model \reff{def_H_sHMBW_4} is an inhomogeneous
$O(2)$ model \cite{Nienhuis_82} 
\be
Z_{O(n)}(\beta_1,\ldots,\beta_n) \;=\; \sum\limits_{k\ge 0} 
\sum\limits_{\begin{scarray}
                  \hbox{\scriptsize loops with $k$} \\
                  \hbox{\scriptsize components}
                  \end{scarray}}
\prod_{j=1}^k \left( \sum\limits_{\alpha=1}^n \beta_\alpha^{\ell_j} \right)  
\label{def_Z_On}
\ee
with weights
\be
\beta_1 \;=\; \tanh 2J \,, \qquad \beta_2 \;=\; \cosh^{-1} 2J \,.
\label{def_weights_On}
\ee
This model becomes homogeneous at the self-dual point 
\reff{def_theta_selfdual}, and this self-dual point corresponds to the
critical point for the $O(2)$ model \cite{Nienhuis_82} on the 
hexagonal lattice:
\be
\beta_c(n) \;=\; \frac{1}{\sqrt{2 + \sqrt{2-n}}} \,,
\label{def_betac_On}
\ee
as
\be
\lim_{n\to 2} \beta_c(n) \;=\; \frac{1}{\sqrt{2}} \;=\; \tanh 2J_c \,.
\ee
This is indeed the critical value Baxter and Wu found for the 
BW model [cf.\/ \reff{def_Jc_HMBW}]. 

\medskip

\noindent
{\bf Remarks}. 1. It is worth noticing that when 
$\beta_1\to\infty$ and $\beta_2\to 0$, we obtain one type of fully packed 
loops (FPL) \cite{Kondev_96}. 
This model is equivalent to a dimer covering of the hexagonal lattice 
\cite{Kasteleyn_63}, which in turn can be described in the continuum limit 
by a conformal field theory of central charge $c=1$ \cite{Fisher_63,Wu_06}. 
However, no value of $J$ seems to provide these limiting values. 

2. There exists a direct mapping between the ICLAT model and the two-color 
non-overlapping Eulerian-bond model. The loops correspond to low-temperature
graphs (or domain walls) of the two copies of Ising spins. We can rewrite
the ICLAT-model partition function \reff{def_Z_ICLAT} as
\begin{eqnarray}
Z_{\rm ICLAT}(G;L_2,L'_2) &=& \sum\limits_{\{\sigma,\tau\}}
                              \prod_{e =\<ij\>\in E} 
\left[ e^{-2L'_2} \delta_{\sigma_i,\sigma_j} + 
       e^{-2L_2} \delta_{\tau_i,\tau_j} \right. \nonumber \\
   & & \qquad \qquad \qquad \left. + 
       \left(1- e^{-2L'_2} - e^{-2L_2}\right)  \delta_{\sigma_i,\sigma_j} 
       \delta_{\tau_i,\tau_j} \right] \,.
\label{def_Z_ICLAT_BIS}
\end{eqnarray} 
Indeed, the weights for each edge coincide with those normalized weights 
given in the third column of Table~\ref{table_weights_ICLAT_general}. 
One places occupied bonds on the edges of the dual graph $G^*$ as in the 
standard low-temperature expansion procedure: place an occupied bond of 
color ``A'' (resp.\/ color ``B'') on the dual edge $e^*$ if the Ising spins 
$\tau$ (resp.\/ $\sigma$) on the corresponding original edge $e=\<ij\>$ are not
equal, i.e. $\tau_i \tau_j=-1$ (resp.\/ $\sigma_i \sigma_j=-1$). 
Note that any dual edge cannot be occupied by two bonds of distinct colors, 
as the configuration $\tau_i \tau_j=\sigma_i \sigma_j=-1$ has zero weight. 
If $G$ is the triangular lattice, then $G^*$ is the hexagonal lattice, 
and loops cannot intersect. But for other lattices, loops of distinct colors 
can intersect at certain vertices.
In summary, one obtains a one-to-two correspondence between a 
two-color non-overlapping Eulerian-bond model on the dual graph $G^*$, 
and the ICLAT model on the original graph $G$. 

%
%
\section{Phase diagrams} \label{sec.phase.diagrams}

In this section we will qualitatively describe the phase diagrams for the 
ICLAT models defined on the square, triangular, hexagonal, and kagome lattice. 
We will use the analytic results obtained in the preceding section, as well
as exact results from the 8--vertex model \cite{Baxter_book}.

  Some important points will be studied using Monte Carlo (MC) simulations.
In these cases, we have used the embedding algorithm for the AT model
\cite{SS_AT} with one difference. In the ferromagnetic regime, in the 
algorithm of Ref.~\cite{SS_AT} one fixes one type of spins (either $\sigma$ or 
$\tau$) and performs a standard Swendsen-Wang cluster algorithm 
\cite{SW,Kandel_90} with the other type. In the simulation for this paper, 
we fixed one of the {\em three} types of spins ($\sigma$, $\tau$, or 
$\sigma\tau$). The fixing of the variable $\sigma\tau$ seems important in 
reducing the critical slowing down for the ICLAT model.  
For the antiferromagnetic regime, we performed the embedding using the 
Wang--Swendsen--Koteck\'y \cite{WSK} algorithm.

The Binder $Q_i$ cumulants for different observables were computed for each 
value of the parameters $(L_2,L_2')$, and from the crossing of these cumulants
for distinct values of the linear size, we extracted the critical values of
the parameters. Details will be published elsewhere. 

%
%
\subsection{Square-lattice ICLAT model} \label{sec.sq.ICLAT}

In this section we will consider the phase diagram for a square--lattice
ICLAT model. It is interesting to start with the phase diagram for the
symmetric AT model on the square lattice, as it is simpler than that of the
full AT model \cite{Wu_Lin}.

\begin{figure}[htb]
\begin{center}
\includegraphics[width=200pt]{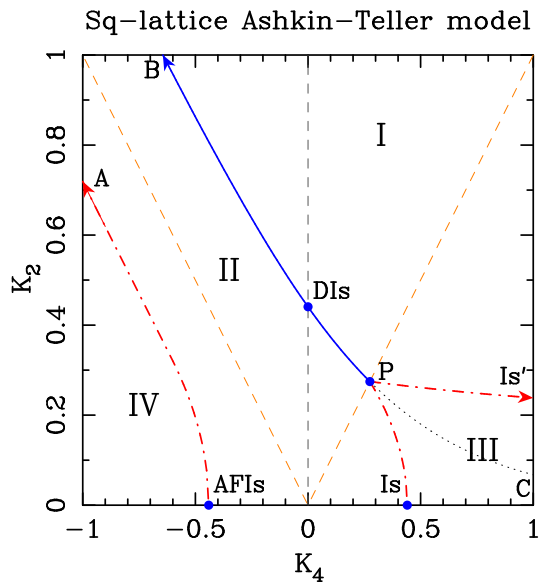}
\end{center}
\caption{\label{figure_AT_sq}
Phase diagram on the square-lattice symmetric AT model 
\reff{def_H_sAT} in the plane $(K_4,K_2)$.
The self-dual curve \reff{self_dual_curve_sAT_square} is B--DIs--P--C. 
The solid (blue) curves represent the critical part of the self-dual curve,  
the (red) dot-dashed curves represent Ising-like transition lines, 
the dashed (orange) lines represent the 4--state Potts subspace, and the
dotted curve represents the noncritical part of the self-dual curve
\reff{self_dual_curve_sAT_square}. The Roman numerals designate the different
phases of the model (see text).  
}
\end{figure}

First of all, as the square lattice is bipartite, the phase 
diagram of the symmetric AT on the square lattice should be invariant under  
the interchange $K_2\to-K_2$ [cf.\/ \reff{def_symmetries_AT_bipartite}].
Thus, we will focus on the part with $K_2>0$: See Figure~\ref{figure_AT_sq}.

There are some ``easy'' points in this phase diagram. 
\begin{itemize}
\item The line $K_4=0$ corresponds to two decoupled Ising models, so there 
      are Ising critical points at 
      $(K_4,K_2)=(0,\pm \smfrac{1}{2} \log(1+\sqrt{2}))$. 
      Point DIs in Figure~\ref{figure_AT_sq} represents the ferromagnetic one 
      ($+$ sign). 

\item The line $K_2=0$ corresponds to an Ising model in the variable 
      $\sigma\tau$. There are Ising critical points at 
      $(K_4,K_2)=(\pm \smfrac{1}{2} \log(1+\sqrt{2}),0)$. Point Is 
      in Figure~\ref{figure_AT_sq} represents the ferromagnetic one
      ($+$ sign), and point AFIs represents the antiferromagnetic one 
      ($-$ sign). 

\item The limit $K_4\to\infty$ corresponds to an Ising model with coupling
      $2K_2$; therefore, we have Ising critical points 
      at $(K_4,K_2)\to (\infty,\pm \smfrac{1}{4} \log(1+\sqrt{2}))$. The
      ferromagnetic point ($+$ sign) is represented by point Is' in 
      Figure~\ref{figure_AT_sq}. 

\item The $K_2=K_4$ subspace corresponds to the 4--state Potts model. This
      model has a ferromagnetic critical point at 
      $(K_4,K_2)=\smfrac{1}{4}\log 3$ (point P in Figure~\ref{figure_AT_sq}).
      The antiferromagnetic 4--state Potts model on the square lattice is
      disordered even at zero temperature \cite{Ferreira,SS_AF_sq}. 
\end{itemize}

%
%

The square lattice is self--dual; therefore 
the square-lattice AT model is self-dual on the curve
\reff{def_AT_selfdual}. For the symmetric AT model \reff{def_H_sAT}, 
this curve takes the form:
\be
e^{-2K_4} \;=\; \sinh(2K_2) \,.
\label{self_dual_curve_sAT_square}
\ee

The AT model on any planar graph can be mapped onto an 8--vertex model on the 
medial graph \cite{Wu_77}. In particular, the AT model on the square 
lattice can be mapped onto a {\em staggered} 8--vertex model on the
square lattice (which has not been solved in general). As a special case,
the AT model on the self--dual manifold \reff{def_AT_selfdual} maps
onto a {\em homogeneous}\/ eight--vertex model, which is exactly
solvable \cite{Baxter_book}. Finally, the   
symmetric AT model \reff{def_H_sAT} on the square lattice maps
(after a simple further transformation) onto a {\em homogeneous}\/ 6--vertex 
model. In this way, Baxter showed that the self-dual curve 
\reff{self_dual_curve_sAT_square} is critical only for  
$K_4 \le \smfrac{1}{4}\log 3$ (blue solid curve in Figure~\ref{figure_AT_sq}),
and it is noncritical for $K_4 > \smfrac{1}{4}\log 3$ (dotted curve
in Figure~\ref{figure_AT_sq}). The critical part can be described by a 
conformal field theory of central charge $c=1$ (i.e., it can be related
to the Gaussian model). Along this line the critical exponents
vary continuously, thus violating the usual notion of universality. 
 
Even though the exact solution of the symmetric square-lattice AT model 
is not known, we have evidence \cite{Ditzian,Knops_80,Kamieniarz_97} that two
critical curves emerge at the 4--state Potts model critical point P: one
goes to the Ising critical point Is, and the other one tends as $K_4\to\infty$
to the critical Ising point Is'. Finally, there is another critical curve
emerging from the Ising critical point AFIs and pointing toward 
$K_4\to-\infty$. The exact location of these three curves is still an open
problem, as well as their universality classes. However, it is generally
believed that the three curves belong to the Ising universality class.  

The phase diagram of this model shows four different phases:
\begin{enumerate}
\item[I.] This is the so-called Baxter phase \cite{Ditzian}. The spins $\sigma$
          and $\tau$ are independently ferromagnetically ordered. There are
          four extremal infinite-volume Gibbs states, one for each choice 
          for the signs of $\< \sigma\>$ and $\<\tau\>$. The sign of 
          $\<\sigma\tau\>$ is that of $\<\sigma\>\<\tau\>$. 
\item[II.] This is the paramagnetic phase, in which the three spins 
          $\sigma$, $\tau$, and $\sigma\tau$ are disordered. There is a
          unique infinite-volume Gibbs state. 
\item[III.] In this phase both $\sigma$ and $\tau$ are disordered, but 
          $\sigma\tau$ is ferromagnetically ordered. There are two 
          extremal infinite-volume Gibbs states, one for each choice of the
          sign of $\< \sigma\tau\>$. 
\item[IV.] This is the antiferromagnetic analogue of III: $\sigma$ and $\tau$ 
          are both disordered, but $\sigma\tau$ is antiferromagnetically 
          ordered. There are two extremal infinite-volume Gibbs states. 
\end{enumerate}

The critical exponents along the critical part of the self-dual curve
can be obtained by relating the AT model to the 8--vertex model or to
the Gaussian model \cite{Kadanoff_79,Nijs_79,Knops_80,Yang_87,Saleur_88}. 
We can parametrize the critical part of the self-dual curve 
\reff{self_dual_curve_sAT_square} by using the parameter $\mu$:
\be
e^{4K_4} \;=\; 1 +  2\cos\mu \,, \qquad 0\le \mu \le \frac{2\pi}{3}\,.
\label{def_mu}
\ee
This parameter $\mu$ is related to the coupling constant $g$ of the 
Gaussian model by \cite{Saleur_88}\footnote{
 Our $g$ is that of Saleur \cite{Saleur_88} and equals $2\pi$ times of
 the $K$ of Kadanoff and Brown \cite{Kadanoff_79}, and Yang \cite{Yang_87}.
}
\be
\mu \;=\; \pi \left( 1 - \frac{g}{4} \right) \,, \qquad \frac{4}{3}\le g\le 4\,.
\label{def_g}
\ee
 
Then, using renormalization--group arguments \cite{Kadanoff_79,Knops_80}
one finds that the critical exponents along the critical part of the
self-dual curve \reff{self_dual_curve_sAT_square} are given in terms of
$\mu$ as
\begin{subeqnarray}
\nu    &=& \frac{2-y}{3-2y}       \slabel{def_nu_SAT} \\
\alpha &=& \frac{2-2y}{3-2y}      \slabel{def_alpha_SAT} \\[2mm]
\beta  &=& \frac{2-y}{8(3-2y)}    \slabel{def_beta_SAT} \\
\gamma &=& \frac{7(2-y)}{4(3-2y)} \slabel{def_gamma_SAT}\\[2mm]
\beta' &=& \frac{1}{4(3-2y)}      \slabel{def_betap_SAT} \\
\gamma'&=& \frac{7-4y}{2(3-2y)}   \slabel{def_gammap_SAT}
\label{def_critical_exp_SAT}
\end{subeqnarray}
where $y$ is the renormalization-group eigenvalue related to $\mu$ via
\be
y \;=\; \frac{2\mu}{\pi} \,, \qquad 0\leq y \leq \frac{4}{3} \,,
\label{def_y}
\ee
and the magnetic exponents $\beta,\gamma$ (resp.\  $\beta',\gamma'$) 
correspond to the $\sigma$ or $\tau$ (resp.\ $\sigma\tau$) magnetization.
The value of the critical exponents \reff{def_critical_exp_SAT} vary
along the self-dual curve \reff{self_dual_curve_sAT_square}; but there are
two exponents that remain constant on this curve
\begin{subeqnarray}
\eta  &=& 2 - \frac{\gamma}{\nu} \;=\; \frac{1}{4} \slabel{def_eta_SAT}\\ 
\delta&=& 1 + \frac{\gamma}{\beta} \;=\; 15        \slabel{def_delta_SAT}
\label{def_critical_exp_SAT_bis}
\end{subeqnarray}
Finally, it is worth commenting that the AT domain walls (considered 
as extended curves) have also critical exponents (of the watermelon type). 
This has recently been investigated by Picco and Santachiara 
\cite{Picco_10,Picco_11}, and by Ikhlef and Rajabpour \cite{Ikhlef_12}. 

%
%
\begin{figure}[htb]
\begin{center}
\includegraphics[width=200pt]{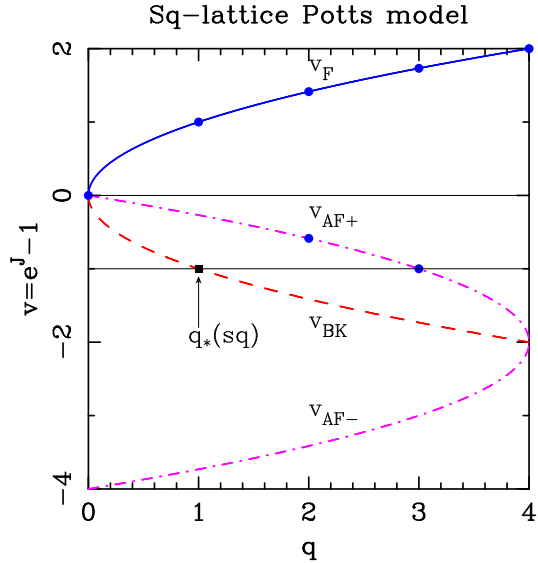}
\end{center}
\caption{\label{figure_Potts_sq}
Phase diagram on the square-lattice $q$--state Potts model  
in the plane $(q,v)$, where $v$ is related to the coupling constant $J$
by \reff{def_v_Potts}.
The upper solid (blue) curve $v_{\rm F}=+\sqrt{q}$ corresponds to the 
ferromagnetic critical curve. Its continuation (depicted as a dashed -- red --
curve) corresponds to the BK line $v_{\rm BK}=-\sqrt{q}$. 
The other two branches $v_{\rm AF+}=-2+\sqrt{4-q}$ and 
$v_{\rm AF-}=-2-\sqrt{4-q}$ correspond respectively to the critical 
antiferromagnetic and its dual counterpart curves. The dots $\bullet$ 
represent the critical values of the model for integer values of $q$. 
The square $\blacksquare$ represents the intersection of the BK
curve $v_{\rm BK}=-\sqrt{q}$ with the zero--temperature limit in the 
antiferromagnetic regime $v=-1$: i.e. the point $q_*({\rm sq})=1$.  
}
\end{figure}

There is a nice relation between the symmetric square-lattice AT model on
the self-dual curve \reff{self_dual_curve_sAT_square} 
and the ferromagnetic $q$-state Potts model at its critical temperature  
\cite{Baxter_78}.
Figure~\ref{figure_Potts_sq} shows the phase diagram of the $q$-state
Potts model in the plane $(q,v)$, where $v$ is given in terms of the Potts
model coupling constant $J_\text{Potts}$ by\footnote{
  We have used the same letter as for the temperature-like variable in the 
  HMBW model \reff{def_v_HMBW}. It should be clear from the context which 
  $v$ we are referring to.
} 
\be
v \;=\; e^{J_\text{Potts}} -1 \,.
\label{def_v_Potts}
\ee
There are four curves where the exact free energy is known 
\cite{Baxter_82,Baxter_book}: 
\begin{subeqnarray}
v &=& \pm \sqrt{q} \slabel{def_curve_sq1} \\
v &=& -2 \pm \sqrt{4-q} \slabel{def_curve_sq2} 
\label{def_curves_sq}
\end{subeqnarray} 
The curve (\ref{def_curve_sq1}${}_+$) $v_{\rm F}(q)=\sqrt{q}$ 
is known to correspond to the ferromagnetic critical point for this model, 
and it is self-dual. 
Its analytic continuation into the Berker--Kadanoff (BK) phase 
\cite{Saleur_90,Saleur_91,Jacobsen_06} corresponds to the BK curve 
(\ref{def_curve_sq1}${}_-$) $v_{\rm BK}(q)=-\sqrt{q}$, and it is also self-dual.
Curve (\ref{def_curve_sq2}${}_+$) $v_{\rm AF+}(q)=-2+\sqrt{4-q}$ is expected 
to give the  critical curve for the antiferromagnetic model; in particular, 
it gives the exactly known values of the critical temperature for $q=2,3$. 
This curve is  dual to the curve (\ref{def_curve_sq2}${}_-$) 
$v_{\rm AF-}(q)=-2-\sqrt{4-q}$. The curves $v_{\rm AF\pm}$ bound the BK phase. 
The renormalization group flow is repulsive close to this
boundary, and it is attracted by the curve $v_{\rm BK}$ lying in between.  
This curve intersects the line $v=-1$ (corresponding to the zero-temperature
Potts model in the antiferromagnetic regime) at a point $q_*({\rm sq})=1$.

The above mentioned relation between these two models stems from the
fact that both models can be written as six-vertex models \cite{Baxter_book}.
Thus, we can use $q$ to parametrize the critical part of the self-dual
curve \reff{self_dual_curve_sAT_square}. Indeed, this $q$ should be 
interpreted as an {\em effective} number of states.\footnote{
   Please note that this not imply that the AT and Potts models are
   equivalent. The AT model is untwisted, so that (unlike the
   Potts model) the central charge is c = 1 throughout the critical regime.
} 
The result is
\be
\sqrt{q} \;=\; 2\cos \mu \,, \quad q\in [0,4] \,,
\label{def_q}
\ee
where $\mu$ is related to the AT couplings via 
\reff{self_dual_curve_sAT_square}/\reff{def_mu}.

The critical 4-state Potts model corresponds to $\mu=0$, $y=0$, $g=4$, 
and $\sqrt{q}=2$ (point P in Figure~\ref{figure_AT_sq});
the two decoupled Ising models correspond to the point
$\mu=\pi/2$, $y=1$, $g=2$, and $\sqrt{q}=0$
(point DIs in Figure~\ref{figure_AT_sq}).
The part of the AT self-dual curve connecting these two models $(y\leq 1)$
is characterized by $\alpha\ge 0$, and corresponds to going from $q=4$ down to
$q=0$ along the ferromagnetic critical curve $v_{\rm F}$.   
Therefore, $0\le \mu \le \pi/2$ corresponds to the Potts--model critical curve
$v_{\rm F}$. 

If we keep moving to higher values of $\mu$ (i.e., $2\pi/3 \le \mu \le \pi/2$),
then we enter the BK branch $v_{\rm BK}$. This corresponds
to $y>1$, and $\sqrt{q}<0$. In particular, the specific heat is not
divergent in this regime (i.e., $\alpha < 0$), as the energy is {\em not}\/
a relevant operator in the BK phase. We can move down 
the curve $v_{\rm BK}$ until we enter the unphysical
regime of the Potts model: i.e., $v=-1$, which corresponds to the
zero--temperature limit of the {\em antiferromagnetic} Potts model. 
We expect that this point $q_*$ is lattice--dependent,
as universality usually does not hold in the antiferromagnetic regime.   
This limiting point along the self-dual curve \reff{self_dual_curve_sAT_square}
as $K_4\to-\infty$ (point B in Figure~\ref{figure_AT_sq}) corresponds 
to $\mu=2\pi/3$, $y=4/3$, $g=4/3$, and $\sqrt{q}=-1$. Its universality
class is characterized by the following critical exponents 
\reff{def_critical_exp_SAT}: 
\be
\nu     \;=\; 2                          \,, \quad  
\alpha  \;=\; -2          \,, \quad  
\beta   \;=\; \frac{1}{4} \,, \quad 
\beta'  \;=\; \frac{3}{4} \,, \quad  
\gamma  \;=\; \frac{7}{2} \,, \quad 
\gamma' \;=\; \frac{5}{2} \,, 
\label{def_critical_exp_sq_SAT}
\ee
and the two critical exponents $\eta$ and $\delta$ as in 
\reff{def_critical_exp_SAT_bis}.

%
%
\begin{figure}[htb]
\begin{center}
\includegraphics[width=200pt]{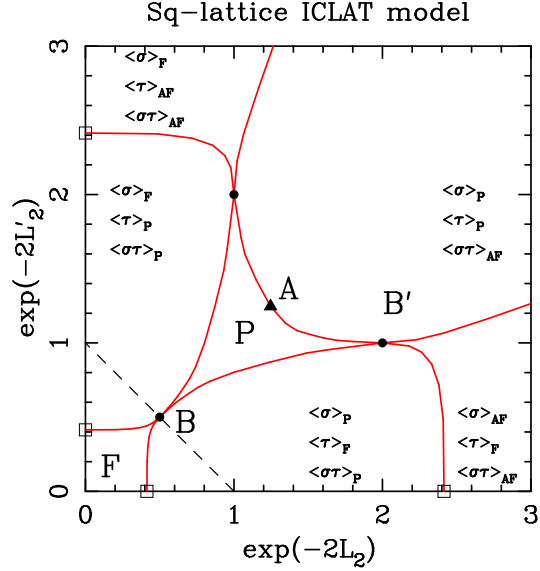}
\end{center}
\caption{\label{figure_ICLAT_sq}
Phase diagram on the square-lattice ICLAT model \reff{def_H_ICLAT} 
in the plane $(e^{-2L_2},e^{-2L'_2})$.
The (red) solid curves are Ising-like critical curves, except at the  
dots of coordinates $(1/2,1/2)$, $(1,2)$, and $(2,1)$. 
Point B of coordinates $(1/2,1/2)$ corresponds to the point where 
the self-dual curve of the symmetric AT model \reff{self_dual_curve_sAT_square}
hits this plane, and the other two points are equivalent to the former one 
due to the symmetries of this model.  All these points are depicted as 
solid circles ($\bullet$). Point A ($\blacktriangle$) shows
where the branch emerging from point AFIs in Figure~\ref{figure_AT_sq} hits
the ICLAT model. The dashed curve going through point B 
corresponds to the  union-jack HMWB model subspace \reff{rel_ICLAT_square}.  
The squares ($\Box$) on the axes correspond to the exact Ising 
critical points located at $\sqrt{2} \pm 1$. 
The distinct phases are denoted by the ordering of the distinct spins: e.g., 
$\<\sigma\>_P$, $\<\tau\>_F$, and $\<\sigma\tau\>_{AF}$ means respectively,
that the $\sigma$ spins are disordered, the $\tau$ spins are ferromagnetically 
ordered, and that the product $\sigma\tau$ is antiferromagnetically ordered.
F corresponds to the ferromagnetic phase: $\<\sigma\>_F$, $\<\tau\>_F$,
$\<\sigma\tau\>_F$; and P corresponds to the paramagnetic phase:
$\<\sigma\>_P$, $\<\tau\>_P$, $\<\sigma\tau\>_P$.
}
\end{figure}

Let us now consider the square-lattice ICLAT model. 
Its phase diagram is shown in Figure~\ref{figure_ICLAT_sq}, and 
it is symmetric under the interchange $L_2 \leftrightarrow L_2'$ 
because of \reff{def_symmetries_ICLAT}. Therefore, we will focus
on the part with $L'_2 \ge L_2$.  

First of all, we want to know where the self-dual curve 
\reff{self_dual_curve_sAT_square} hits the plane $(e^{-2L_2},e^{-2L'_2})$ 
in the limit $K_4\to-\infty$.
We can rewrite \reff{self_dual_curve_sAT_square} as
\be
e^{-2(K_2+K_4)} \;=\; \frac{1}{2} \left(1 - e^{-4K_2} \right) \to \frac{1}{2}
\,,
\label{self_dual_curve_sAT_square2}
\ee
in the limit $K_4\to-\infty$ and $K_2=K_2'\to+\infty$, with $K_2+K_4=L_2$
and $K'_2+K_4=L'_2=L_2$ held fixed. Thus, the self-dual curve 
\reff{self_dual_curve_sAT_square} hits the ICLAT model at the symmetric point
\be
\left(e^{-2L_2},e^{-2L_2'}\right) \;=\; \left(\frac{1}{2},\frac{1}{2}\right)\,.
\label{def_self_dual_sq_ICLAT}
\ee
This point is depicted as B in Figure~\ref{figure_ICLAT_sq}, and it is
equivalent to point B' of coordinates  
\be
\left(e^{-2L_2},e^{-2L'_2}\right) \;=\; (2,1) \,,
\label{def_self_dual_sq_ICLAT_bis}
\ee
because of the symmetry \reff{def_symmetries_ICLAT_bipartite}. 

\medskip

\noindent
{\bf Remark}. Ikhlef and Rajabpour \cite[Section 4]{Ikhlef_12} have consider
the square-lattice ICLAT model at the symmetric point 
\reff{self_dual_curve_sAT_square2}/\reff{def_self_dual_sq_ICLAT}. They found
that at this particular point, this ICLAT model can be mapped onto an 
integrable 19--vertex model. Furthermore, they also mapped this particular 
point of the square-lattice ICLAT model onto an integrable dilute Brauer model. 

\medskip

In the limit $L'_2\to\infty$, we have a square--lattice Ising model on the
$\sigma$ spins. Therefore, we have the following two critical points:
\be
\left(e^{-2L_2},e^{-2L_2'}\right) \;=\; \left(\sqrt{2}\pm 1,0\right)\,,
\label{def_ising_sq_ICLAT}
\ee
where the `$+$' (resp.\/ `$-$') sign corresponds to the antiferromagnetic
(resp.\/ ferromagnetic) Ising critical point. Using 
\reff{def_symmetries_ICLAT_bipartite}, we see that these two points
are  mapped to points at infinity with slopes $\sqrt{2}\mp 1$. 
All these points belong to the Ising universality class.

The point $L_2=L_2'=0$ corresponds to a zero-temperature 4--state Potts
antiferromagnet. This model is disordered on the square lattice. Therefore,
the point 
\be
\left(e^{-2L_2},e^{-2L_2'}\right) \;=\; \left(1,1\right)
\ee
belongs to the paramagnetic phase, which is indicated with the sign P
in Figure~\ref{figure_ICLAT_sq}.  

The point $L_2=L_2'\to + \infty$ corresponds to a zero-temperature point
in the ferromagnetic regime; therefore, the region around the origin
in the $(e^{-2L_2},e^{-2L_2'})$ plane corresponds to a ferromagnetically
ordered phase, denoted by F in Figure~\ref{figure_ICLAT_sq}.

We know from Section~\ref{subsec.partial.trace}, 
that the union-jack--lattice HMBW model with coupling $J$ is equivalent to an 
ICLAT model on the square lattice on the line \reff{rel_ICLAT_square}. 
This line is depicted as a (black) dashed line in Figure~\ref{figure_ICLAT_sq}. 
Indeed, the self-dual point for this HMBW model corresponds to the 
critical point \reff{def_self_dual_sq_ICLAT}.

Baxter has shown that the AT model on the square lattice can be mapped onto
a {\em staggered}\/ eight-vertex model on the square lattice \cite{Baxter_book},
with weights $(a,b,c,d)$ on one sublattice, and weights $(a,b,d,c)$ on 
the other sublattice. These weights can be written (up to some unimportant
multiplicative constant) in terms of the parameters
$L_2,L_2'$ of the square--lattice ICLAT model as follows 
(see \cite[Eqs.~(12.9.6) and~(12.9.17)]{Baxter_book}):  
\begin{subeqnarray}
a &=& \omega_0 + \omega_1 \;=\; 1+e^{-2L'_2}\\
b &=& \omega_2 - \omega_3 \;=\; e^{-2L_2}  \\
c &=& \omega_2 + \omega_3 \;=\; e^{-2L_2}  \\
d &=& \omega_0 - \omega_1 \;=\; 1-e^{-2L'_2}
\label{def_8V_weights_ICLAT}
\end{subeqnarray}
where the weights $\omega_k$ correspond to the ICLAT model, and can be read 
from the third column of Table~\ref{table_weights_ICLAT_general}.
The above staggered eight-vertex model becomes {\em homogeneous} when $c=d$.
In terms of the couplings $L_2,L'_2$, this implies that the model should
satisfy \reff{rel_ICLAT_square}. 
The vertex weights \reff{def_8V_weights_ICLAT} reduce along this line to
\begin{subeqnarray}
a &=& 2- e^{-2L_2} \\
b &=& c\;=\; d \;=\; e^{-2L_2}
\label{def_8V_weights_ICLAT_2}
\end{subeqnarray}

In order to characterize the different phases, it is useful to compute
the parameter $\Delta$:
\be
\Delta \;\equiv\; \frac{a^2 + b^2 - c^2 -d^2}{2(ab+cd)} \;=\; e^{2L_2} -1
\label{def_delta_sq}
\ee
When $0\leq e^{-2L_2} < \frac{1}{2}$, we have that $\Delta > 1$ and hence,
the eight-vertex model is ordered (it belongs to the ferroelectric Regime~I
in Baxter's notation \cite{Baxter_book}, as $a > b+c+d$).
On the other hand, when $e^{-2L_2} > \frac{1}{2}$, we have that $|\Delta|<1$,
and the system is disordered (it belongs to Baxter's Regime~III).
Thus, there is a critical point at the symmetric point 
\reff{def_self_dual_sq_ICLAT}.

In order to obtain the value of the parameter $\mu$ at this critical point,
we first need to make use of the rearrangement procedures explained
in Baxter's book \cite[Section~10.11]{Baxter_book}, so that the transformed
weights $(a_r,b_r,c_r,d_r)$ belong to Baxter's principal regime IV 
(the anti-ferroelectric phase with $c>a+b+c$ and $\Delta<-1$). 
To achieve this, we first use the duality (or weak--graph) transformation
\cite[Eq.~(10.2.5)]{Baxter_book}:
\begin{subeqnarray}
a' &=& \frac{1}{2}(a +b+c+d) \;=\; 1+e^{-2L_2} \\
b' &=& \frac{1}{2}(a +b-c-d) \;=\; 1-e^{-2L_2} \\
c' &=& \frac{1}{2}(a -b+c-d) \;=\; 1-e^{-2L_2} \\
d' &=& \frac{1}{2}(a -b-c+d) \;=\; 1-e^{-2L_2} 
\label{def_weak_graph}
\end{subeqnarray}
so that the model with weights $(a',b',c',d')$ belongs to the ferroelectric
regime I ($a'>b'+c'+d'$ with $\Delta>1$). Then, we make the change
$(a_r,b_r,c_r,d_r)=(c',d',a',b')$, so that the model now belongs to
the principal regime. 
The critical value of the parameter $\mu$ is given in terms of the weights
$(a_r,b_r,c_r,d_r)$ when $e^{-2L_2}=1/2$: 
\be
\tan \frac{\mu}{2} \;=\; \sqrt{ \frac{c_r d_r}{a_r b_r} } \;=\; \sqrt{3} \,,
\label{def_tan_mu_sq}
\ee
and hence,
\be
\mu \;=\; \frac{2\pi}{3} \,.
\label{def_mu_sq}
\ee
In this case, the critical exponents are those of the eight-vertex model for 
the same value of $\mu$ \cite[Equations~(10.12.24)/(10.12.25)]{Baxter_book}: 
\begin{subeqnarray}
\nu    &=& \frac{1}{y}  \slabel{def_nu_8V} \\
\alpha &=&  2-\frac{2}{y} \slabel{def_alpha_8V} \\[2mm]
\beta  &=& \frac{1}{8y}  \slabel{def_beta_8V} \\
\gamma &=& \frac{7}{4y}  \slabel{def_gamma_8V} \\[2mm]
\beta' &=& \frac{2-y}{4y}\slabel{def_betap_8V} \\
\gamma'&=& \frac{2+y}{2y}\slabel{def_gammap_8V}
\label{def_critical_exp_8V}
\end{subeqnarray}
where $y$ is given by \reff{def_y}.
The numerical values for these exponents at the symmetric point 
\reff{def_self_dual_sq_ICLAT} are: 
\be
\nu    \;=\; \frac{3}{4} \,, \quad  
\alpha \;=\; \frac{1}{2} \,, \quad  
\beta  \;=\; \frac{3}{32} \,, \quad 
\beta' \;=\; \frac{1}{8} \,, \quad  
\gamma \;=\; \frac{21}{16} \,, \quad 
\gamma'\;=\; \frac{5}{4} \,. 
\label{def_critical_exp_sq_ICLAT}
\ee
These results agree with the result found by Hintermann and Merlini 
\cite{Hintermann_72}:
the self-dual point \reff{def_Tc_HMBW} of the 
union-jack--lattice HMBW model is critical with critical exponent
$\alpha=1/2$.

We find that, on one hand, the value of $\mu=2\pi/3$ at the symmetric point
$e^{-2L_2}=e^{-2L'_2} = 1/2$ is the same, independently of how we
approach that point, either along the self-dual curve 
\reff{self_dual_curve_sAT_square} when $K_4\to-\infty$, or along the 
HMBW--model line \reff{rel_ICLAT_square}. However,
we get two different sets of critical exponents: if we follow the
self-dual curve \reff{self_dual_curve_sAT_square}, we should use the 
expressions \reff{def_critical_exp_SAT} obtained by renormalization-group 
arguments;
but along the line \reff{rel_ICLAT_square}, one should use instead the
eight-vertex exponents \reff{def_critical_exp_8V}. 

The exponents $1/\nu=1/2$ in Eq.~\reff{def_critical_exp_sq_SAT} and $1/\nu=4/3$ 
in Eq.~\reff{def_critical_exp_sq_ICLAT} correspond to the renormalization 
exponents along the diagonal direction (i.e. along the self-dual curve in 
the large-coupling limit) in Figure~\ref{figure_AT_sq}, and the 
perpendicular-to-diagonal direction
(i.e., along the dashed line in Figure~\ref{figure_ICLAT_sq}), respectively. 
The same behavior will be found in the triangular and hexagonal lattices
(see the next two sections).

Finally, all points (except point B and its symmetric counterparts) on the 
solid curves in Figure~\ref{figure_ICLAT_sq} 
belong to the Ising universality class. We have checked this fact numerically 
by computing via MC simulations the values of the critical exponents on 
several points on these curves.  

%
%
\subsection{Triangular-lattice ICLAT model}

The phase diagram for the symmetric AT model on the triangular lattice
is qualitatively different from that for the square lattice. The main
reason is that the triangular lattice is not bipartite; therefore, there
is no symmetry $K_2\to -K_2$. However, if we focus on the ferromagnetic
regime $K_2\ge 0$, then it is rather similar, except that phase IV does not
exist. See \cite{Lv_11} for a recent study of this model.

Another difference with respect to the same model on the square lattice 
is that the symmetric AT model on the triangular lattice is not self-dual, and 
it does not satisfy in general the star-triangle equation. However, it 
does satisfy the star-triangle equation
on a certain curve in the $(K_2,K_4)$ plane. Temperley and Ashley
\cite{Temperley_79} found that this curve is given by
\be
e^{-4K_4} \;=\; \frac{1}{2} \left( e^{4K_2} -1 \right)
\label{def_selfdual_tri}
\ee
The qualitative behavior of the symmetric triangular-lattice AT model
on this curve is similar to that for the square lattice:
\begin{itemize}
\item The model is critical only for $K_4\leq \smfrac{1}{4}\log 2$, and 
      it can be mapped onto the Gaussian model with central charge $c=1$.
\item At $K_4=K_2=\smfrac{1}{4}\log 2$, the critical curve splits 
      into two Ising-like critical curves. This point corresponds to
      the critical coupling for a ferromagnetic 4--state Potts model 
      on the triangular lattice. 
\item For $K_4 > \smfrac{1}{4}\log 2$, the curve is no longer critical.
\end{itemize}

One of the Ising-like critical curves emerging at 
$K_4=K_2=\frac{1}{4}\log 2$ goes to the Ising critical point 
at $(K_2,K_4)=(0,\smfrac{1}{4}\log 3)$. The other line goes to the Ising
critical point at $K_4\to\infty$ with $K_2=\smfrac{1}{2}\log 3$.

As the subspace $K_2=0$ corresponds to an Ising model in the variables
$\sigma\tau$, then we expect the above ferromagnetic critical point
at $K_4=\smfrac{1}{4}\log 3$, and an antiferromagnetic critical point
in the limit $K_4\to-\infty$. Thus, there is no  AFIs--A curve, like in
the phase diagram for the square lattice (see Figure~\ref{figure_AT_sq}). 

Finally, the 4--state antiferromagnetic Potts model on the triangular
lattice has a zero--temperature critical point. This point is recovered
when $K_2=K_4\to-\infty$. 

We expect that the same critical exponents will be found for the part
of the curve \reff{def_selfdual_tri} between the critical points for
the 4--state Potts and the two decoupled Ising models. 
The effective number of Potts states along the curve \reff{def_selfdual_tri}
is given by \cite[Equation~(18)]{Temperley_79}:
\be
\sqrt{q} \;=\; \sqrt{2} \, \frac{e^{2K_2}(3-e^{4K_2})}{(e^{4K_2}-1)^{3/2}} \,,
\label{def_q_tri}
\ee
and again the relation between $q$ and $\mu$ is given by \reff{def_q}.
Then, the 4--state Potts critical point is defined by $e^{4K_2}=2$; hence,
$q=4$ and $\mu=0$. The decoupled Ising models are obtained for $e^{4K_2}=3$,
and thus, $q=0$ and $\mu=\pi/2$. These values of $q$, and $\mu$ indeed
agree with those obtained for the square-lattice AT model,
as we expect universality to hold in the ferromagnetic regime.

However, when we take larger values of $\mu > \pi/2$, we enter the
BK curve $v_{\rm BK}$ for the triangular lattice \cite{Jacobsen_06}.  
This is the middle branch of the curve \cite{Baxter_78,Baxter_86,Baxter_87}
\be
v^3 + 3v^2 \;=\; q \,,
\label{def_BK_tri}
\ee
and hits the $v=-1$ line at $q_*({\rm tri})=2$. This corresponds to
$\sqrt{q}=-\sqrt{2}$, $\mu=3\pi/4$, and $y=3/2$. These values differ from 
those of the square-lattice model, and lead to a different set of critical
exponents for this point: 
\be
\nu \;=\; \infty                     \,, \quad  
\frac{\alpha}{\nu} \;=\; -2          \,, \quad  
\frac{\beta}{\nu}  \;=\; \frac{1}{8} \,, \quad 
\frac{\beta'}{\nu} \;=\; \frac{1}{2} \,, \quad  
\frac{\gamma}{\nu} \;=\; \frac{7}{4} \,, \quad 
\frac{\gamma'}{\nu}\;=\; 1 \,.
\label{def_critical_exp_tri_SAT}
\ee
Saleur \cite{Saleur_88} indicates that at this point $\mu=3\pi/4$ or $g=1$
one has a Kosterlitz-Thouless transition. In Refs.~\cite{Deng_07,Liu_11} 
the values for $\alpha/\nu$, $\gamma/\nu$, and $\beta/\nu$ in 
\reff{def_critical_exp_tri_SAT} were obtained by studying the critical 
$O(2)$ loop model on the hexagonal lattice.   
It is interesting to note that the triangular-lattice Ising $q=2$ 
antiferromagnet at zero temperature $v=-1$ is critical with $c=1$. 

\begin{figure}[htb]
\begin{center}
\includegraphics[width=200pt]{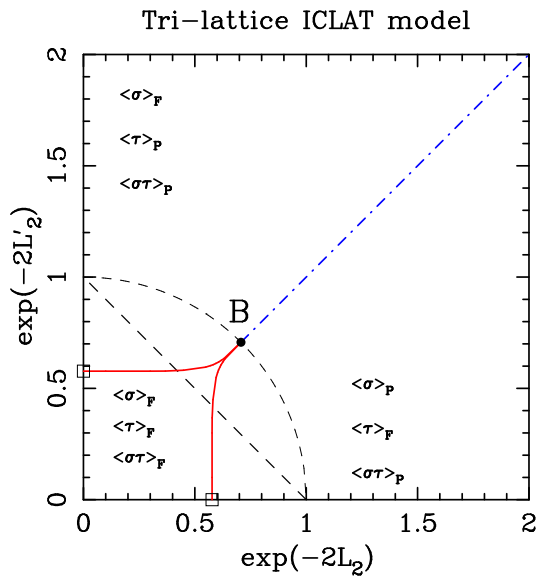}
\end{center}
\caption{\label{figure_ICLAT_tri}
Phase diagram on the triangular-lattice ICLAT model \reff{def_H_ICLAT}
in the plane $(e^{-2L_2},e^{-2L'_2})$.
Point B (depicted as a black dot) has coordinates $(1/\sqrt{2},1/\sqrt{2})$, 
and corresponds to the point where the self--dual curve of the symmetric
AT model \reff{def_selfdual_tri} hits this plane. 
The ferromagnetic Ising critical point for
$e^{-2L'_2}=0$ is located at $e^{-2L_2}=1/\sqrt{3}$. 
The dashed arc going through point B corresponds to the 
triangular-lattice HMBW model, and the dashed line joining $(0,1)$ and $(1,0)$ 
corresponds to the bisected-hexagonal-lattice HMBW model. 
The notation for curves, symbols, and phases is as in 
Figure~\ref{figure_ICLAT_sq}.
The main diagonal starting at point B (depicted as a blue dot-dashed line)
corresponds to a line of critical points belonging to the same universality
class as point $B$. 
}
\end{figure}

Let us now consider the triangular-lattice ICLAT model. Its phase
diagram is given in Figure~\ref{figure_ICLAT_tri}. Again it is symmetric
under the interchange $L_2 \leftrightarrow L'_2$, so we will focus on
the subspace $L'_2 \ge L_2$. 

First of all we would like to compute the symmetric point in the 
$(e^{-2L_2},e^{-2L'_2})$ space where the critical curve 
\reff{def_selfdual_tri} ends. This is
easily obtained from \reff{def_selfdual_tri} if we take $K_4,-K_2\to-\infty$ 
with $K_2+K_4=L_2=L'_2$ fixed. The result is
\be
\left( e^{-2L_2}, e^{-2L_2} \right) \;=\; \left( \frac{1}{\sqrt{2}},
                                                 \frac{1}{\sqrt{2}} \right)\,. 
\label{def_hit_tri}
\ee
This corresponds to point B in Figure~\ref{figure_ICLAT_tri}.

Some important particular cases are given by
\begin{itemize}

\item $e^{-2L'_2}=0$ corresponds to an Ising model on the $\sigma$ variables.
      Thus, there are phase transitions at
\begin{subeqnarray}
      e^{-2L_2} &=& \smfrac{1}{\sqrt{3}} \\
      e^{-2L_2} &=& +\infty
\end{subeqnarray}
The first solution corresponds to the ferromagnetic Ising critical point;
while the second one is the antiferromagnetic critical point at zero
temperature. 

\item The triangular-lattice HMBW model (or simply the BW model) 
      \cite{Baxter_Wu_73,Baxter_Wu_74,Baxter_74} corresponds to the circle
      \reff{rel_ICLAT_tri}. 
For $e^{-2L_2} > \smfrac{1}{\sqrt{2}}$ the model is disordered, for
    $e^{-2L_2} < \smfrac{1}{\sqrt{2}}$, the model is ordered, and finally
at  $e^{-2L_2} = \smfrac{1}{\sqrt{2}}$, the model is critical.

\item In Section~\ref{subsec.partial.trace}, we showed that the 
      bisected-hexagonal-lattice HMBW model 
      corresponds to the line \reff{rel_ICLAT_square}. 
\end{itemize}

%
%

Wu \cite{Wu_77} suggested that the triangular AT model should correspond
to a {\em uniform} 8--vertex model on the kagome lattice. The details
were worked out by Temperley and Ashley \cite{Temperley_79}.
Theorem~\ref{theo.AT_vs_mAT} tells us that the triangular-lattice
AT model with weights $\{\omega_k\}$ is equivalent to a mixed 
AT model on the triangular and hexagonal lattices with weights
$\{\widehat{\omega}_k\}$ given by \reff{def_omegahat}.
Then, Theorem~\ref{theo.mAT_ICLAT_bis} relates this mixed AT model with
couplings $\{\widehat{K}_2,\widehat{K}'_2,\widehat{K}_4\}$,
with an ICLAT model on the triangular lattice with couplings 
$\{L_2,L'_2\}$ given by \reff{def_L_ICLAT_bis}. Notice 
that we have $\omega_3 = 0$ in the triangular-lattice AT model,
so that $\widehat{\omega}_2=\widehat{\omega}_3$ in the mixed
AT model, which also means that the couplings for this model satisfy
$\widehat{K}'_2=\widehat{K}_4$. 

We also know by following Wu \cite{Wu_77} that this mixed AT model
can be related to an 8--vertex model on the medial of the triangular
lattice (i.e., a kagome lattice) with weights 
\begin{subeqnarray}
a &=& \widehat{\omega}_0 \;=\; \frac{\omega_0+\omega_1}{2} \;=\; 
                               e^{\widehat{K}_2 + 2\widehat{K}'_2}
\\[2mm]
b &=& \widehat{\omega}_3 \;=\; \frac{\omega_2-\omega_3}{2} \;=\; 
                               e^{-\widehat{K}_2}
\\[2mm]
c &=& \widehat{\omega}_2 \;=\; \frac{\omega_2+\omega_3}{2} \;=\; 
                               e^{-\widehat{K}_2}
\\[2mm]
d &=& \widehat{\omega}_1 \;=\; \frac{\omega_0-\omega_1}{2} \;=\; 
                               e^{\widehat{K}_2 - 2\widehat{K}'_2}
\label{def_weights_8V_vs_Khat}
\end{subeqnarray}
where we have taken into account that 
$\widehat{K}'_2 =\widehat{K}_4$.
Baxter \cite{Baxter_book} found that this uniform 8--vertex model on the
kagome lattice can be solved by transforming it to a square--lattice 
8-vertex model when the following condition is fulfilled\footnote{
   Notice we have followed Wu's notation \cite{Wu_77} to go from 
   spin configurations to arrow configurations, and this choice is 
   non-equivalent to that of Baxter \cite{Baxter_book} or 
   Temperley--Ashley \cite{Temperley_79}.
   We can ``translate'' Baxter's formulas to our notation by using the
   transformation $(a,b,c,d)\to (c,d,a,b)$. 
}
\be
\frac{d}{a} \;=\; \frac{c^2 - d^2}{a^2 - b^2} \,.
\label{condition_sol_kag}
\ee
When $b=c$ [cf.\/ \reff{def_weights_8V_vs_Khat}], this condition 
implies that $ad=b^2$, or equivalently, $\widehat{K}_2=0$. Therefore,
the 8--vertex model on the kagome lattice is soluble by Baxter's methods
if the weights have the form:
\begin{subeqnarray}
a &=& e^{2\widehat{K}'_2}\\[2mm]
b &=& c \;=\; 1  \\[2mm]
d &=& e^{-2\widehat{K}'_2}
\label{def_weights_8V_vs_Khat_soluble}
\end{subeqnarray}

This 8--vertex model is critical if $a=b+c+d$. This equation implies that
\be
\sinh (2\widehat{K}'_2) \;=\; 1  \quad \Leftrightarrow \quad  
e^{-2L_2} \;=\; e^{-2L'_2} \;=\; \frac{1}{\sqrt{2}}  \,.
\ee
This corresponds to the symmetric point \reff{def_hit_tri}. 

To obtain the value of $\mu$ corresponding to this point, we proceed as for
the square lattice. If $e^{-2L_2}> 1/\sqrt{2}$, the model is disordered
(it belongs to regime III). Using the weak--graph transformation
\reff{def_weak_graph}, we can get a point in regime I with weights:
\begin{subeqnarray}
a' &=&      1 + \cosh 2\widehat{K}'_2 \\[2mm] 
b' &=& c' \;=\; \sinh 2\widehat{K}'_2 \\[2mm] 
d' &=&     -1 + \cosh 2\widehat{K}'_2 
\end{subeqnarray}
If we do the transformation $(a',b',c',d')\to (c',d',a',b')=(a_r,b_r,c_r,d_r)$,
we get finally into the principal regime, with weights
\begin{subeqnarray}
a_r &=& d_r \;=\; \sinh 2\widehat{K}'_2\\[2mm]
b_r &=&      -1 + \cosh 2\widehat{K}'_2  \\[2mm]
c_r &=&       1 + \cosh 2\widehat{K}'_2
\end{subeqnarray}
Using these weights, we compute the critical value of $\mu$:
\be
\tan \frac{\mu}{2} \;=\; \sqrt{ \frac{c_r d_r}{a_r b_r} } \;=\; \sqrt{2}+1 \,,
\label{def_tan_mu_tri}
\ee
and hence,
\be
\mu \;=\; \frac{3\pi}{4} \,.
\label{def_mu_tri}
\ee
The numerical values for these exponents at the symmetric point
\reff{def_hit_tri} are:
\be
\nu    \;=\; \frac{2}{3} \,, \quad
\alpha \;=\; \frac{2}{3} \,, \quad
\beta  \;=\; \frac{1}{12} \,, \quad
\beta' \;=\; \frac{1}{12} \,, \quad
\gamma \;=\; \frac{7}{6} \,, \quad
\gamma'\;=\; \frac{7}{6} \,.
\label{def_critical_exp_tri_ICLAT}
\ee
These results agree with the result found by Baxter and Wu  
\cite{Baxter_Wu_73,Baxter_Wu_74,Baxter_74}. 

It is worth noticing that these exponents coincide with those for the
ferromagnetic 4--state Potts model $\mu=0$. In fact, these two models
are believed to belong to the same universality class \cite{Domany_78}. 
One striking difference between the BW model and the 4--state Potts model
is that the latter displays logarithmic corrections (both multiplicative
and additive) \cite{Nauenberg_80,Cardy_80,SS_97}, while the former does 
not have any of these.  

A qualitative phase diagram for this model is shown in 
Figure~\ref{figure_ICLAT_tri}. We expect, as in the square--lattice case,
that at the point B (where the self-dual curve of the symmetric AT model
hits the ICLAT plane), the critical curve splits into two curves
going to Ising--like critical points on the axis. These curves are expected to 
be Ising-like. We also know that
the line \reff{rel_ICLAT_square} corresponds to the HMBW model
on the bisected-hexagonal lattice. The phase diagram shows that
this model has {\em two}\/ critical points, contrary to what we know
about the same model on the other two triangulations (triangular and
union-jack lattices). Our numerical MC results show that their location 
is:\footnote{
   We performed the MC computation on the kagome lattice (see 
   section~\ref{sec.ICLAT.kag}). The values of the couplings for the 
   triangular lattice are the square of those obtained for the 
   kagome lattice.  
}
\begin{subeqnarray}
e^{2L_2} &=& 0.4196(15)           \,, \qquad e^{2L'_2} \;=\; 0.5803(8) \\ 
e^{2L_2} &=& 0.5803(8)\phantom{0} \,, \qquad e^{2L'_2} \;=\; 0.4196(15) 
\label{def_MC_crossings_tri}
\end{subeqnarray}
In terms of the HMBW coupling J \reff{def_L.coro}, we obtain the critical 
points 
\be
J_{1,c} \;=\; 0.5003(5)\,, \qquad J_{2,c} \;=\; 0.3857(4) \,. 
\label{def_J_critical_bh}
\ee
Indeed, we can check that these two values are mutually dual: using the
definition \reff{def_v_HMBW}, we arrive at $v_1 v_2 = 2.001(3)$.
Therefore, we suggest the following:

\begin{conjecture} \label{conj.bh_kag}
The Hintermann--Merlini--Baxter--Wu model 
\reff{def_H_HMBW}/\reff{def_Z_HMBW} on the bisected-hexagonal lattice
has two critical points 
\be
J_{1,c} \;=\; \frac{1}{2}\,, \qquad J_{2,c} \;=\; \frac{1}{2}\, 
\log\left( \frac{e+1}{e-1} \right) \;\approx\; 0.3859684164 \,.
\ee
These two points are mutually dual. 
\end{conjecture} 

Finally, the diagonal line starting at point B and depicted as a dot-dashed
line on Figure~\ref{figure_ICLAT_tri} belongs to the same universality class 
as point B. This is supported by our MC simulations along this line.  
The above statement is true for all points on this diagonal line, 
except in the limit $e^{L_2}=e^{L'_2}\to +\infty$, where we recover
a hexagonal-lattice FPL model with fugacity $n=2$. 
This FPL model is equivalent to the zero-temperature limit of the 
triangular-lattice 4-state Potts antiferromagnet \cite{Baxter_70,Moore_00}. 
Therefore, its continuum limit can be described by a conformal field theory 
of central charge $c=2$ \cite{Kondev_96}.  

%
%
\subsection{Hexagonal-lattice ICLAT model}

The phase diagram for the symmetric AT model on the hexagonal lattice
is qualitatively the same as for the square lattice, as both lattices
are bipartite. However, the AT model on the hexagonal lattice is not
self-dual; but it does satisfy the star-triangle equation
on a certain curve in the $(K_2,K_4)$ plane. Temperley and Ashley
\cite{Temperley_79} found that this curve is given by
\be
e^{-4K_4} + e^{-2K_4} \cosh 2K_2 \;=\; \sinh^2 2K_2 
\label{def_selfdual_hc}
\ee
We then find that
\begin{itemize}
\item The model is critical only for $K_4\leq \frac{1}{4}\log 5$, and
      it can be mapped onto the Gaussian model with central charge $c=1$.
\item At $K_4=K_2=\frac{1}{4}\log 5$, the critical curve splits
      into two Ising-like critical curves. This point corresponds to
      the critical coupling for a ferromagnetic 4--state Potts model
      on the hexagonal lattice.
\item For $K_4 > \frac{1}{4}\log 5$, the curve is no longer critical.
\end{itemize}

As the subspace $K_2=0$ corresponds to an Ising model in the variables
$\sigma\tau$, then we expect the above ferromagnetic critical point
at $K_4=\smfrac{1}{2}\log (2+\sqrt{3})$, and an antiferromagnetic
critical point at $K_4=-\smfrac{1}{2}\log (2+\sqrt{3})$. There is a curve 
of Ising-like critical points emerging from this point, and going to
$K_4\to-\infty$, as for the square lattice.

Finally, the 4--state antiferromagnetic Potts model on the hexagonal
lattice is disordered at all temperatures \cite{Shrock_97,Hex}, thus
none of the above critical curves cross the line $K_2=-K_4 \ge 0$.

We expect that the same critical exponents will be found for the part
of the curve \reff{def_selfdual_hc} between the critical points for
the 4--state Potts and the two decoupled Ising models.
The effective number of Potts states along the curve \reff{def_selfdual_hc}
could be obtained from the results given in Ref.~\cite{Temperley_79}:
\be
\sqrt{q} \;=\; \sqrt{1 + \xi(K_2)}\, \left[ \xi(K_2) - 2\right] \,,
\label{def_q_hc}
\ee
where $\xi(K_2)$ is given by
\be
\xi(K_2) \;=\; \frac{1}{2\tanh^2 (2K_2)} \left[ 1 +
 \sqrt{ 1 + 4\tanh^2 (2K_2) } \right] \,.
\label{def_xi_K2_hc}
\ee
Again the relation between $q$ and $\mu$ is given by \reff{def_q}.
Then, the 4--state Potts critical point is defined by $e^{2K_2}=\sqrt{5}$;
hence, $\xi=3$, $q=4$ and $\mu=0$.
The decoupled Ising models are obtained for $e^{2K_2}=2+\sqrt{3}$,
and thus, $\xi=2$, $q=0$ and $\mu=\pi/2$. These values of $q$, and $\mu$ indeed
agree with those obtained for the square-lattice AT model,
as we expect universality to hold in the ferromagnetic regime.

When we take larger values of $\mu > \pi/2$, we enter the
BK curve $v_{\rm BK}$ for the hexagonal lattice.
This is the lower branch of the curve \cite{Baxter_78,Baxter_86,Baxter_87}
\be
v^3 - 3vq \;=\; q^2 \,,
\label{def_BK_hc}
\ee
and hits the $v=-1$ line at $q_*({\rm hex})=(3-\sqrt{5})/2$.
This corresponds to
$\sqrt{q}=(1-\sqrt{5})/2$, $\mu=3\pi/5$, and $y=6/5$. This value coincides
with the one we obtain when we take the limit $K_2\to+\infty$ in
\reff{def_q_hc}, as $\lim_{K_2\to\infty}\xi(K_2)=(1+\sqrt{5})/2$.
These values lead to the following set of critical
exponents for this point:
\be
\nu    \;=\; \frac{4}{3}                \,, \quad
\alpha \;=\;-\frac{2}{3}\,, \quad
\beta  \;=\; \frac{1}{6} \,, \quad
\beta' \;=\; \frac{5}{12} \,, \quad
\gamma \;=\; \frac{7}{3} \,, \quad
\gamma'\;=\; \frac{11}{6} \,.
\label{def_critical_exp_hc_SAT}
\ee

\begin{figure}[htb]
\begin{center}
\includegraphics[width=200pt]{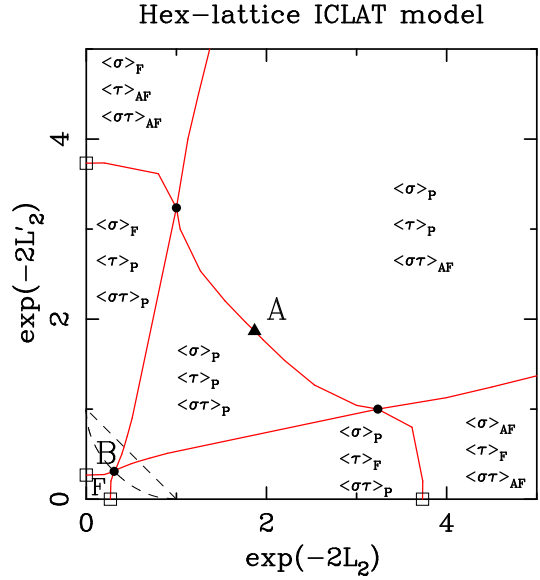}
\end{center}
\caption{\label{figure_ICLAT_hc}
Qualitative phase diagram on the hexagonal-lattice ICLAT model
\reff{def_H_ICLAT}
in the plane $(e^{-2L_2},e^{-2L'_2})$.
Point B of coordinates $((\sqrt{5}-1)/4, (\sqrt{5}-1)/4)$
corresponds to the point where
the self--dual curve of the symmetric
AT model \reff{def_selfdual_hc} hits this plane, and the other
two points are equivalent to the former one because of the symmetries
of this model.
The ferromagnetic and antiferromagnetic Ising critical points for
$e^{-2L'_2}=0$ are located at $2 \pm \sqrt{3}$, respectively.
The dashed arc going through point B corresponds to the models that are
solvable using Baxter's method. The dashed line corresponds to the
uniform HMBW model on the bisected-hexagonal lattice.
The notation for curves, symbols, and phases is as in 
Figure~\ref{figure_ICLAT_sq}.
}
\end{figure}

Let us now consider the hexagonal-lattice ICLAT model. Its phase
diagram is given in Figure~\ref{figure_ICLAT_hc}. Again it is symmetric
under the interchange $L_2 \leftrightarrow L'_2$, so we will focus on
the subspace $L'_2 \ge L_2$.

%
%

First of all we would like to compute the symmetric point in the
$(e^{-2L_2},e^{-2L'_2})$ space where the critical curve
\reff{def_selfdual_hc} ends. This is
easily obtained from \reff{def_selfdual_hc} if we take $K_4,-K_2\to-\infty$
with $K_2+K_4=L_2=L'_2$ fixed. The result is
\be
\left( e^{-2L_2}, e^{-2L_2} \right) \;=\; \left( \frac{\sqrt{5}-1}{4},
                                                 \frac{\sqrt{5}-1}{4}
\right)\,.
\label{def_hit_hc}
\ee
This corresponds to point B in Figure~\ref{figure_ICLAT_hc}, and it is
equivalent to point B' of coordinates
\be
\left(e^{-2L_2},e^{-2L'_2}\right) \;=\; (1+\sqrt{5},1) \,,
\label{def_self_dual_hc_ICLAT_bis}
\ee
because of the symmetry \reff{def_symmetries_ICLAT_bipartite}.

Some important particular cases are given by
\begin{itemize}

\item $e^{-2L'_2}=0$ corresponds to an Ising model on the $\sigma$ variables.
      Thus, there are phase transitions at
\be
e^{-2L_2} \;=\; 2\pm \sqrt{3} \,.
\ee
The `$+$' solution corresponds to the ferromagnetic Ising critical point;
while the `$-$' solution is the antiferromagnetic critical point.
Symmetry \reff{def_symmetries_ICLAT_bipartite} implies that these points
are mapped to points at infinity with slopes $2\pm \sqrt{3}$.

\item The point $(e^{-2L_2},e^{-2L'_2})=(1,1)$ corresponds to the
zero--temperature 4--state antiferromagnetic Potts model on the hexagonal
lattice. This model is known to be disordered \cite{Shrock_97,Hex}.

\item In Section~\ref{subsec.partial.trace}, we showed that the 
      bisected-hexagonal--lattice HMBW model
      corresponds to the line \reff{rel_ICLAT_square}. In agreement with
      Conjecture~\ref{conj.bh_kag}, we find two critical points
      on this line at the same values as for the triangular--lattice
      ICLAT model \reff{def_MC_crossings_tri} 
      (recall that in Section~\ref{subsec.partial.trace} we
      obtain maps from the bisected-hexagonal-lattice HMBW model
      to the ICLAT model on {\em decorated} triangular or hexagonal
      lattices).
\end{itemize}

The kagome lattice is also the medial of the hexagonal, thus we
can repeat the same argument to relate the hexagonal--lattice ICLAT
model to the kagome--lattice 8--vertex model \cite{Temperley_79},
using a mixed AT model as the intermediate step \cite{Wu_77}.
Again, Theorem~\ref{theo.AT_vs_mAT} tells us that the hexagonal--lattice
AT model with weights $\{\omega_k\}$ is equivalent to a mixed
AT model on the hexagonal and triangular lattices with weights
$\{\widehat{\omega}_k\}$ [cf.\/ \reff{def_omegahat}],
and this latter model us related to a hexagonal--lattice
ICLAT model  with couplings $\{L_2,L'_2\}$
[cf.\/ \reff{def_L_ICLAT_bis}].
Notice that we have $\omega_3 = 0$ in the hexagonal--lattice
AT model, so that $\widehat{\omega}_2=\widehat{\omega}_3$ in the
mixed AT model, which also means that the couplings for this model satisfy
$\widehat{K}'_2=\widehat{K}_4$.

The results of Wu \cite{Wu_77} ensure that this mixed AT model
can be related to an 8--vertex model on the medial of the hexagonal
lattice (i.e., a kagome lattice) with weights given by
\reff{def_weights_8V_vs_Khat}.

The only difference with respect to the triangular--lattice case is that
in this latter case, the $\sigma$ spins live on the triangular lattice,
and the $\tau$ spins on the dual hexagonal lattice; while for the
hexagonal--lattice case, the situation is reversed: the $\sigma$ spins
live on the hexagonal lattice, and the $\tau$ spins live on the dual
triangular lattice. This implies that the condition for the
corresponding kagome--lattice 8-vertex model to be soluble by
Baxter's method now reads:\footnote{
   Again, we have followed Wu's notation \cite{Wu_77} to go from
   spin configurations to arrow configurations.
   We can ``translate'' Baxter's formulas to our notation by using the
   transformation $(a,b,c,d)\to (d,c,a,b)$.
}
\be
\frac{c}{a} \;=\; \frac{d^2 - c^2}{a^2 - b^2} \,.
\label{condition_sol_kag_bis}
\ee
When $b=c$ [cf.\/ \reff{def_weights_8V_vs_Khat}], this condition
implies that the couplings $\widehat{K}_2$ and $\widehat{K}'_2$ should
satisfy the following equation:
\be
x^3 \, e^{2\widehat{K}'_2} - x^2\, e^{4\widehat{K}'_2} -
x\, e^{6\widehat{K}'_2} + 1 \;=\; 0 \,,
\label{def_sol_hc}
\ee
where
\be
x \;=\; e^{-2\widehat{K}_2} \,.
\label{def_x}
\ee
It is difficult to write the solutions of the above equation
\reff{def_sol_hc} as simple expressions in the ICLAT couplings
$L_2$ and $L_2'$. However, it is not difficult to prove
that the following points provide solutions of \reff{def_sol_hc}:
\begin{subeqnarray}
\left( e^{-2L_2}, e^{-2L'_2} \right) &=& (1,0) \\[2mm]
\left( e^{-2L_2}, e^{-2L'_2} \right) &=& \left( \frac{\sqrt{5}-1}{4},
                                          \frac{\sqrt{5}-1}{4}\right)
\end{subeqnarray}
Therefore, the symmetric point \reff{def_hit_hc} belongs to this curve
where the ICLAT model is soluble as a 8--vertex model on the square
lattice.

Furthermore, the symmetric point \reff{def_hit_hc} is the only one where
the 8--vertex model is critical. The condition for this 8--vertex model
to be critical is
\be
a \;=\; b + c + d \,,
\ee
which in terms of the couplings $\widehat{K}_2,\widehat{K}'_2$ means that
\be
\sinh2\widehat{K}'_2 \;=\; e^{-2 \widehat{K}_2 } \,.
\ee
This condition implies that
\be
e^{-2L'_2} \;=\; \tanh 2\widehat{K}_2 \;=\; e^{-2L_2} \,.
\ee
Therefore, the model is critical precisely at the symmetric point
\reff{def_hit_hc}.

To obtain the value of $\mu$ corresponding to this point, we proceed as for
the triangular lattice. By rearranging the weights appropriately so that
the model is mapped to the principal regime, we obtain for the value of $\mu$
at the critical point \reff{def_hit_hc}
\be
\tan \frac{\mu}{2} \;=\; \sqrt{\frac{5+2\sqrt{5}}{5}} \,,
\label{def_tan_mu_hc}
\ee
and hence,
\be
\mu \;=\; \frac{3\pi}{5} \,.
\label{def_mu_hc}
\ee
The numerical values for these exponents at the symmetric point
\reff{def_hit_hc} are:
\be
\nu    \;=\; \frac{5}{6} \,, \quad
\alpha \;=\; \frac{1}{3} \,, \quad
\beta  \;=\; \frac{5}{48} \,, \quad
\beta' \;=\; \frac{1}{6} \,, \quad
\gamma \;=\; \frac{35}{24} \,, \quad
\gamma'\;=\; \frac{4}{3} \,.
\label{def_critical_exp_hc_ICLAT}
\ee
Again we find that this set of critical exponents do not agree with the 
set \reff{def_critical_exp_hc_SAT} obtained by following the critical curve of 
the self-dual curve of the hexagonal-lattice symmetric AT model.  

Finally, as for the square lattice, we find numerically in our MC simulations 
that all solid curves in Figure~\ref{figure_ICLAT_hc} are Ising-like, except 
point B and its symmetric counterparts. 

%
%
\subsection{Kagome-lattice ICLAT model} \label{sec.ICLAT.kag}

The phase diagram for the symmetric AT model on the kagome lattice
is expected to be qualitatively similar to that of the same model on
the triangular lattice. The main reason is that both lattices are not
bipartite. Unfortunately, the exact location of any of the critical
curves is unknown for this case. See Ref.~\cite{Lv_11} for a recent study.

What is exactly known about the behavior of the symmetric kagome-lattice
AT model on this curve reduces to the solution of the Ising model
\cite{Kano}:

\begin{itemize}
\item For $K_2=0$, it is critical in the ferromagnetic regime at coupling
\be
K_4  \;=\;  \frac{1}{4} \log (3+2\sqrt{3}) \,.
\ee
\item In the antiferromagnetic regime it is always disordered.
\end{itemize}

Finally, the 4--state Potts model on the kagome
lattice has a ferromagnetic critical point at $K_2=K_4\approx 0.2873$,
and in the antiferromagnetic regime, it is always disordered.

\begin{figure}[htb]
\begin{center}
\includegraphics[width=200pt]{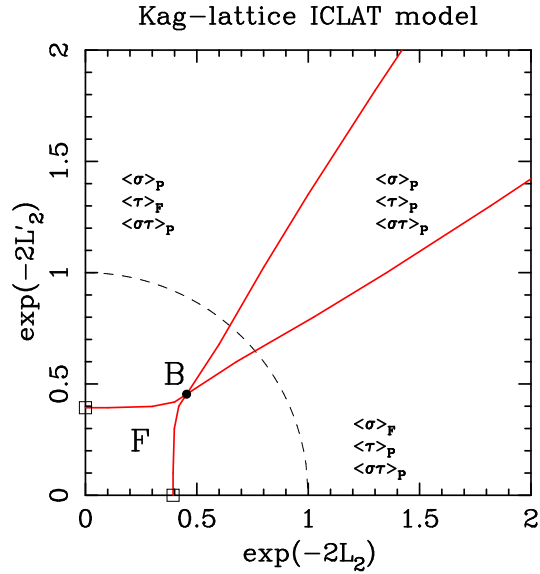}
\end{center}
\caption{\label{figure_ICLAT_kag}
Phase diagram on the kagome-lattice ICLAT model \reff{def_H_ICLAT}
in the plane $(e^{-2L_2},e^{-2L'_2})$.
Point B has coordinates $(0.45411(3), 0.45411(3))$, and
corresponds to the point where the self--dual curve of the symmetric
AT model hits this plane. The ferromagnetic Ising critical point for
$e^{-2L'_2}=0$ is located at $e^{-2L_2}=0.39360791$.
The dashed arc corresponds to the bisected-hexagonal-lattice HMBW model.
The notation for curves, symbols, and phases is as in 
Figure~\ref{figure_ICLAT_sq}.
}
\end{figure}

Let us now consider the kagome-lattice ICLAT model. Its phase
diagram is given in Figure~\ref{figure_ICLAT_kag}. Again it is symmetric
under the interchange $L_2 \leftrightarrow L'_2$, so we will focus on
the subspace $L'_2 \ge L_2$.

The position of the critical point on the symmetric subspace is not
exactly known for this model. Our numerical MC results show that
this point has coordinates $(0.45411(3), 0.45411(3))$.
This corresponds to point B in Figure~\ref{figure_ICLAT_kag}.

Some important particular cases are given by
\begin{itemize}

\item $e^{-2L'_2}=0$ corresponds to an Ising model on the $\sigma$ variables.
      Thus, there are phase transitions at
\be
      e^{-2L_2} \;=\; \smfrac{1}{\sqrt{3+2\sqrt{3}}}\,.
\ee

\item In Section~\ref{subsec.partial.trace}, we showed that the uniform
      bisected-hexagonal-lattice HMBW model
      corresponds to the circle \reff{rel_ICLAT_tri}.
      In agreement with Conjecture~\ref{conj.bh_kag}, we find that the 
      critical curve for the kagome--lattice ICLAT model crosses the above
      circle at {\em two} points:
\begin{subeqnarray}
e^{2L_2} &=& 0.6478(5) \,, \qquad e^{2L'_2} \;=\; 0.7618(3) \\
e^{2L_2} &=& 0.7618(3) \,, \qquad e^{2L'_2} \;=\; 0.6478(5)
\label{def_MC_crossings_kag}
\end{subeqnarray}
      Indeed, the values of the HMBW coupling $J$ are the same as in
      \reff{def_J_critical_bh}. 
\end{itemize}

%
%

%
%
\section{Discussion} \label{sec.discussion} 

The main goal of this paper is the study of the infinite-coupling-limit 
Ashkin--Teller model on several common lattices, namely, square, triangular,
hexagonal, and kagome. We also have considered a generalization of the 
well-known Baxter--Wu model: the Hintermann--Merlini--Baxter--Wu model
defined on any plane Eulerian triangulation. 

We have first investigated the relations between these two models. As a side
effect we have also considered the Ashkin--Teller and the mixed Ashkin--Teller
models, which are known in the literature. In particular we find that 

\begin{itemize}

\item The partition functions for the AT model on a graph $G$ \reff{def_Z_AT}
      and the mixed AT model \reff{def_Z_mAT} on $G$ and its dual $G^*$ 
      are equal (modulo some unimportant multiplicative factors): 
$$
Z_{\rm AT}(G;\omega_0,\omega_1,\omega_2,\omega_3)
\;\propto\; 
Z_{\rm mAT}(G,G^*;\widehat{\omega}_0,\widehat{\omega}_1,
                  \widehat{\omega}_2,\widehat{\omega}_3) \,,
$$
       where the weights $\widehat{\omega}_k$ are given in terms of the
       weights $\omega_k$ by \reff{def_omegahat}. (See 
       Theorem~\ref{theo.AT_vs_mAT}.) 
       As a side effect, we recover the duality relation for the AT model 
       \reff{def_Z_AT_vs Z_mAT_theo}. (See Theorem~\ref{theo.duality_AT}.)

\item The partition function for the ICLAT model on a graph $G$ 
      \reff{def_Z_ICLAT} and the partition function of the mixed AT model
      \reff{def_Z_mAT} on $G$ and $G^*$ are again equal (modulo 
      some unimportant multiplicative factors):
$$
 Z_{\rm mAT}(G,G^*;\widehat{\omega}_0,\widehat{\omega}_1,
                   \widehat{\omega}_2,\widehat{\omega}_2)
 \;\propto\; Z_{\rm ICLAT}(G;L_2,L'_2)
$$
       where the weights couplings $L_2,L'_2$ are given
       in terms of the couplings $\widehat{K}_2,\widehat{K}'_2$ (or the 
       weights $\widehat{\omega}_k$) by \reff{def_L_ICLAT_bis}. 
       (See Theorem~\ref{theo.mAT_ICLAT}.) 

\item Using the partial-trace transformation (Lemma~\ref{lemma.partial.trace})
      we obtain that the HMBW model on certain lattices is a subspace of the 
      ICLAT on other lattices. In particular Corollary~\ref{coro.partial.trace}
      implies: 
      \begin{itemize}
        \item The triangular-lattice HMBW is equivalent to a triangular-lattice
              ICLAT model on the subspace \reff{rel_ICLAT_tri}. 
        \item The bisected-hexagonal-lattice HMBW is equivalent to a 
              kagome-lattice ICLAT model on the subspace \reff{rel_ICLAT_tri}. 
        \item The union-jack-lattice HMBW is equivalent to a 
              square-lattice ICLAT model on the subspace 
              \reff{rel_ICLAT_square}.      
        \item The bisected-hexagonal-lattice HMBW is equivalent to a
              hexagonal-lattice ICLAT model on the subspace
              \reff{rel_ICLAT_square}.
        \item The bisected-hexagonal-lattice HMBW is equivalent to a
              triangular-lattice ICLAT model on the subspace
              \reff{rel_ICLAT_square}.
      \end{itemize} 

\item We have found in Section~\ref{sec.BW.2CNEB} that the partition 
      function for the BW model (i.e., the
      triangular-lattice HMBW model \reff{def_Z_HMBW}) can be written as
      a partition function of a 2--color--non-overlapping Eulerian bond 
      model on a hexagonal lattice with weights \reff{def_weights_On}.  

\end{itemize}

The second important result is the computation of the phase diagram of the ICLAT
model on the square, triangular, hexagonal, and kagome lattices. In the 
first three cases, we know the exact point (labeled B in the figures) 
where the self-dual 
curve of the corresponding symmetric AT model hits the ICLAT model in the limit
of infinite couplings. Using the expression for the critical exponents of the
symmetric AT along this self-dual curve, we are able to make predictions about 
the critical exponents at B.  
On the other hand, in the ICLAT plane, we can also use the relation between
the AT model (and its infinite-coupling limit) and the 8-vertex model. From
this relation another set of critical exponents arise. In the case of the 
square and triangular lattice, these predictions agree with the results of
Hintermann--Merlini \cite{Hintermann_72} and Baxter--Wu 
\cite{Baxter_Wu_73,Baxter_Wu_74}.
It is interesting to note that these two sets of critical exponents do 
{\em not} coincide. 

Finally, the exact location of the self-dual curve of the kagome-lattice AT 
model is not known. In this case, we have obtained a high-precision Monte Carlo 
estimate for the positions of point B and of the two points where the 
critical curves cross the bisected-hexagonal HMBW subspace. 
From these latter values, we have determined the two critical points of the 
bisected-hexagonal-lattice HMBW  model (see Conjecture~\ref{conj.bh_kag}). 

%
%
\appendix 
\section{Decimation transformation} 
\label{appendix.decimation}

%
%
\begin{figure}[htb]
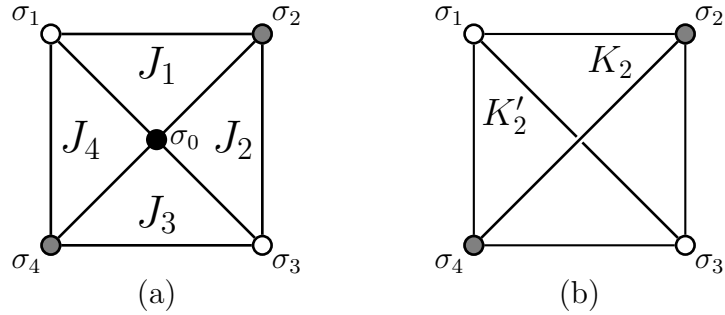

 \centering 
 \psset{xunit=40pt}
 \psset{yunit=40pt}
 \pspicture(-0.5,-0.7)(6.5,2.5)
 \rput{0}(0,0){
  %
  %
  \psline[linecolor=black,linewidth=1pt](0,0)(2,0)(2,2)(0,2)(0,0)
  \psline[linecolor=black,linewidth=1pt](0,0)(2,2)
  \psline[linecolor=black,linewidth=1pt](2,0)(0,2)
  \pscircle*[linecolor=gray] (0,0){4pt}
  \pscircle[linewidth=1pt]   (0,0){4pt}
  \pscircle*[linecolor=gray] (2,2){4pt}
  \pscircle[linewidth=1pt]   (2,2){4pt}
  \pscircle*[linecolor=white](2,0){4pt}
  \pscircle[linewidth=1pt]   (2,0){4pt}
  \pscircle*[linecolor=white](0,2){4pt}
  \pscircle[linewidth=1pt]   (0,2){4pt}
  \pscircle*[linecolor=black](1,1){4pt}
  \uput[90] (1,1.4){\Large $J_1$}
  \uput[270](1,0.6){\Large $J_3$}
  \uput[ 0] (1.4,1){\Large $J_2$}
  \uput[180](0.6,1){\Large $J_4$}
  \uput[225](0,0){$\sigma_4$}
  \uput[45] (2,2){$\sigma_2$}
  \uput[315](2,0){$\sigma_3$}
  \uput[135](0,2){$\sigma_1$}
  \uput[0]  (1,1){$\sigma_0$}
  \uput[270](1.0,-0.2){(a)}
 }

 \rput{0}(4,0){
  %
  %
  \psline[linecolor=black](0,0)(2,0)(2,2)(0,2)(0,0)
  \psline[linecolor=black,linewidth=1pt](2,0)(0,2)
  \psline[linecolor=black,linewidth=1pt,border=1pt](0,0)(2,2)
  \pscircle*[linecolor=gray] (0,0){4pt}
  \pscircle[linewidth=1pt]   (0,0){4pt}
  \pscircle*[linecolor=gray] (2,2){4pt}
  \pscircle[linewidth=1pt]   (2,2){4pt}
  \pscircle*[linecolor=white](2,0){4pt}
  \pscircle[linewidth=1pt]   (2,0){4pt}
  \pscircle*[linecolor=white](0,2){4pt}
  \pscircle[linewidth=1pt]   (0,2){4pt}
  \uput[225](0,0){$\sigma_4$}
  \uput[45] (2,2){$\sigma_2$}
  \uput[315](2,0){$\sigma_3$}
  \uput[135](0,2){$\sigma_1$}
  \uput[135](1.6,1.5){\large $K_2$}
  \uput[225](0.6,1.5){\large $K'_2$}
  \uput[270](1.0,-0.2){(b)}
 }
\endpspicture
\caption{\label{fig_map_sqa_decimation}
(a) The HMBW model with generic 3-spin interaction
    $J_1 \sigma_0\sigma_1\sigma_2 + J_2 \sigma_0\sigma_2\sigma_3 
       +J_3 \sigma_0\sigma_3\sigma_4 + J_4 \sigma_0\sigma_4\sigma_1$. 
(b) The AT model with 2-spin interaction
    $K_2 \sigma_2\sigma_4 + K_2' \sigma_1\sigma_3$, and 4-spin interaction  
    $K_4 \sigma_1\sigma_2\sigma_3\sigma_4$ obtained from (a)
    after summing over the values of $\sigma_0=\pm 1$. 
}
\end{figure}

In this appendix we will consider the decimation transformation, that can
be considered as a generalization of the $k=2$ case of the partial-trace
transformation dealt with in Section~\ref{subsec.partial.trace}.

Let us suppose we have a general (i.e., face-dependent coupling) 
HMBW model defined on an Eulerian plane triangulation $G=(V,E)$. Then
the vertex set has a tripartition $V=V_1\cup V_2 \cup V_3$. Let us further 
suppose that the vertices belonging to the sublattice $V_3$ have all degree 
$\Delta_3=4$. 
Then, this general HMBM model can be transformed into a mixed AT model 
by performing the sum over all spins living on $V_3$. 
(See Figure~\ref{fig_map_sqa_decimation}.)  
\begin{eqnarray}
 & & \sum_{\sigma_0=\pm 1} 
\exp\left[\sigma_0\left(J_1 \sigma_1\sigma_2 + J_2 \sigma_2\sigma_3 
       +J_3 \sigma_3\sigma_4 + J_4 \sigma_4\sigma_1\right)\right]  \nonumber\\ 
 & & \quad\quad\quad\quad\quad
  \;=\; 2 \exp\left[ K_0 + K_2 \sigma_2\sigma_4 + K_2' \sigma_1\sigma_3 + 
          K_4 \sigma_1\sigma_2\sigma_3\sigma_4\right] \,.
\end{eqnarray}
Inspecting all the possible configurations for the spins $\sigma_i$ with
$i=1,2,3,4$, one has the following 4 independent equations
\begin{subeqnarray}
\omega_0 &=& \exp (K_0+K_2+K'_2+K_4) \;=\; \cosh (J_1+J_2+J_3+J_4) \\
\omega_1 &=& \exp (K_0-K_2+K'_2-K_4) \;=\; \cosh (J_1+J_2-J_3-J_4) \\
\omega_2 &=& \exp (K_0+K_2-K'_2-K_4) \;=\; \cosh (J_1-J_2-J_3+J_4) \\
\omega_3 &=& \exp (K_0-K_2-K'_2+K_4) \;=\; \cosh (J_1-J_2+J_3-J_4) 
\label{def_omega_decimation}
\end{subeqnarray}
These equations have the solution:
\begin{subeqnarray} 
e^{4K_0}  &=&  
      \cosh (J_1+J_2+J_3+J_4) \cosh (J_1+J_2-J_3-J_4) \nonumber \\
 & &  \quad \times \cosh (J_1-J_2-J_3+J_4)\cosh (J_1-J_2+J_3-J_4)  \\[2mm]
e^{4K_2} &=& 
     \frac{\cosh (J_1+J_2+J_3+J_4) \cosh (J_1-J_2-J_3+J_4)}
          {\cosh (J_1+J_2-J_3-J_4) \cosh (J_1-J_2+J_3-J_4) }  \\[2mm]
e^{4K'_2} &=& 
     \frac{\cosh (J_1+J_2+J_3+J_4) \cosh (J_1+J_2-J_3-J_4)}
          {\cosh (J_1-J_2-J_3+J_4) \cosh (J_1-J_2+J_3-J_4) }  \\[2mm]
e^{4K_4} &=& 
     \frac{\cosh (J_1+J_2+J_3+J_4) \cosh (J_1-J_2+J_3-J_4)}
          {\cosh (J_1+J_2-J_3-J_4) \cosh (J_1-J_2-J_3+J_4) }  
\label{def_couplings_decimation}
\end{subeqnarray}
Notice that the 2--spin terms couple spins placed on the diagonal; not 
nearest-neighbor ones. Therefore, we obtain a mixed AT model on 
the graph $G_1=(V_1,E_1)$ [defined by the vertices in $V_1$], and its
dual  $G_1^*=(V_2,E_1^*)$ [defined by the vertices in $V_2$]: 
\begin{subeqnarray}
Z_{\rm HMBW}(G;\bm{J}) &=& 2^{|V_3|} \,  
Z_{\rm mAT}(G_1,G_1^*;\bm{K}_2,\bm{K}_2',\bm{K}_4) \\[2mm]
&=& 2^{|V_3|} \, 
Z_{\rm mAT}(G_1,G_1^*;\bm{\omega}_0,\bm{\omega}_1,\bm{\omega}_2,\bm{\omega}_3) 
\label{def_Z_Is4s1_OK}
\end{subeqnarray} 
where the relation between the mixed AT couplings and the original 
HMBW couplings is given in \reff{def_couplings_decimation},
and the weights $\bm{\omega}_k$ are given in \reff{def_omega_decimation}.
To emphasize that the couplings are face- or edge-dependent, we use boldface
letters. The relation between general HMBW, AT, and mixed AT models can be 
generalized easily from the mappings discussed in Section~\ref{sec.mappings}.
Indeed, we can interchange $V_1$ and $V_2$ in the above discussion.
 
We can rewrite \reff{def_Z_Is4s1_OK} in terms of the partition
function of an AT model on $G_1^*$ by using \reff{def_relations_ATs}:
\be
Z_{\rm HMBW}(G;\bm{J}) \;=\; 2^{|V_3|-|V_2|+1}\,
Z_{\rm AT}(G_1^*;\widehat{\bm{\omega}}_0,\widehat{\bm{\omega}}_2,
                 \widehat{\bm{\omega}}_1,\widehat{\bm{\omega}}_3)\,,
\label{def_Z_Is4s1_OK2}
\ee
where the new weights are given by  
\begin{subeqnarray}
\widehat{\omega}_{0} &=& \omega_0 + \omega_2 \;=\; 2\cosh(J_1+J_2)
                                                    \cosh(J_3+J_4) \\[2mm] 
\widehat{\omega}_{1} &=& \omega_0 - \omega_2 \;=\; 2\sinh(J_1+J_2)
                                                    \sinh(J_3+J_4) \\[2mm] 
\widehat{\omega}_{2} &=& \omega_1 + \omega_3 \;=\; 2\cosh(J_1-J_2)
                                                    \cosh(J_3-J_4) \\[2mm] 
\widehat{\omega}_{3} &=& \omega_1 + \omega_3 \;=\; 2\sinh(J_1-J_2)
                                                    \sinh(J_3-J_4)
\label{def_omega_hat_decimation}
\end{subeqnarray}

The most interesting case corresponds to a homogeneous HMBW model 
\reff{def_H_HMBW}/\reff{def_Z_HMBW} with coupling $J$. Then, 
\reff{def_Z_Is4s1_OK2} reduces to a ICLAT model:
\begin{subeqnarray}
Z_{\rm HMBW}(G;J) &=& 2^{|V_3|-|V_2|+1}\,
Z_{\rm AT}(G_1^*;2\cosh^2(2J),2\sinh^2(2J),2,0) \\
 &=& Z_{\rm ICLAT}(G_1^*;L_2,L_2') 
\label{def_Z_Is4s1_OK3}
\end{subeqnarray}
where the couplings $L_2,L_2'$ are given by:
\begin{subeqnarray}
e^{-2L_2} &=& \frac{1}{\cosh^2 (2J)}
\\[2mm]
e^{-2L'_2}&=& \tanh^2(2J) 
\label{def_L_decimation}
\end{subeqnarray}
Indeed, the uniform HMBW model on an Eulerian planar triangulation $G$
maps into the line \reff{rel_ICLAT_square} of the ICLAT model on the
graph $G_1^*$.    

If we apply this transformation to the union-jack lattice we get case (b) of 
Section~\ref{subsec.partial.trace}; while if we apply it to the 
bisected-hexagonal lattice, we obtain cases (c) and (d) of 
Section~\ref{subsec.partial.trace}.

%
%
\section*{Acknowledgments}

We wish to warmly thank Tim Garoni, Alan Sokal, and Andrea Sportiello for 
their collaboration at early stages of this work. 

J.S. is grateful to the kind hospitality of the Department of Physics 
of New York University, and the Hefei National Laboratory for Physical 
Sciences at Microscale and Department of Modern Physics of the University 
of Science and Technology of China, where part of this work was done.  

The research of Y.H. and Y.D. is supported by 
National Nature Science Foundation of
China under grants No.~11275185 and 10975127, and the Chinese Academy of 
Sciences.
The work of J.L.J. was supported by the Agence Nationale de la Recherche (grant 
ANR-10-BLAN-0414: DIME), and the Institut Universitaire de France.
The research of J.S. was supported in part by Spanish MEC grants
FPA2009-08785 and MTM2011-24097 and by U.S.\ National Science
Foundation grant PHY--0424082.


\begin{thebibliography}{99}

\bibitem{Onsager_44} L. Onsager, Phys. Rev. {\bf 65}, 117 (1944).

\bibitem{Hintermann_72} A. Hintermann and D. Merlini, Phys. Lett. {\bf 41A},
         208 (1972). 

\bibitem{Baxter_book} R.J. Baxter, {\em Exactly Solved Models in Statistical
        Mechanics}\/ (Academic Press, London--New York, 1982).

\bibitem{Baxter_Wu_73} R.J. Baxter and F.Y. Wu, Phys. Rev. Lett. {\bf 31},
        1294 (1973).

\bibitem{Baxter_Wu_74} R.J. Baxter and F.Y. Wu, Aust. J. Phys. {\bf 27},
        357 (1974).

\bibitem{Domany_78} E. Domany and E.K. Riedel, J. Appl. Phys. {\bf 49},
        1315 (1978).

\bibitem{Potts_52} R.B. Potts, Proc. Camb. Phil. Soc. {\bf 48}, 106 (1952). 

\bibitem{Nauenberg_80} M. Nauenberg and D.J. Scalapino, Phys. Rev. Lett.
          {\bf 44}, 837 (1980).

\bibitem{Cardy_80} J.L. Cardy, M. Nauenberg and D.J. Scalapino, Phys. Rev. B
         {\bf 22}, 2560 (1980).

\bibitem{SS_97} J. Salas and A.D. Sokal, J. Stat. Phys. {\bf 88}, 567 (1997),
        [arXiv:hep-lat/9607030]. 

\bibitem{Shchur_10} L.N. Shchur and W. Janke, Nucl. Phys. B {\bf 840}, 491
        (2010), [arXiv:1007.1838].

\bibitem{Merlini_72} D. Merlini and C. Gruber, J. Math. Phys. {\bf 13}, 1814
        (1972).

\bibitem{Gruber_77} C. Gruber, A. Hintermann, and D. Merlini, {\em Group
        Analysis of Classical Lattice Systems}. Lecture Notes in Physics,
        Vol.~60 (Springer-Verlag, Berlin--Heidelberg--New York, 1977).

\bibitem{Ashkin_Teller_43} J. Ashkin and E. Teller, Phys. Rev. {\bf 64}. 178
        (1943).

\bibitem{Fan_72b} C. Fan, Phys. Lett. {\bf 39A}, 136 (1972). 

\bibitem{Diestel} R. Diestel, {\em  Graph Theory}\/ (Springer--Verlag, 
        Heildelberg, New York, 2005).

\bibitem{Tsai_West} M.--T. Tsai and D.B. West, Ars Math. Contemp. {\bf 4},
        73 (2011).  

\bibitem{Deng_11} Y. Deng. Y. Huang, J.L. Jacobsen, J.Salas, and A.D. Sokal,
        Rev. Phys. Lett. {\bf 107},  150601 (2011) [5 pages], 
        [arXiv:1108.1743].

\bibitem{tilings} B. Gr\"unbaum and G.C. Shepard, {\em Tilings and Patterns}\/ 
        (W.H. Freeman and Company, New York, 1987). 

\bibitem{Wu_77} F.Y. Wu, J. Math. Phys. {\bf 18}, 611 (1977). 

\bibitem{Ikhlef_12} Y. Ikhlef and M.A. Rajabpour, J. Stat. Mech., P01012, 
        (2012) [arXiv:1111.3197].

\bibitem{Fan_72} C. Fan, Phys. Rev. B {\bf 6}, 902 (1972). 

\bibitem{Wu_Wang_76} F.Y. Wu and Y.K. Wang, J. Math. Phys. {\bf 17}, 4439
        (1976). 

\bibitem{Nienhuis_82} B. Nienhuis, Phys. Rev. Lett. {\bf 49}, 1062 (1982).

\bibitem{Kondev_96} J. Kondev, J. de Gier, and B. Nienhuis, J. Phys. A 
        {\bf 29},  6489 (1996) [cond-mat/9603170]. 

\bibitem{Kasteleyn_63}  P.W. Kasteleyn, J. Math. Phys. 4, 287 (1963).

\bibitem{Fisher_63} M.E. Fisher and J. Stephenson, Phys. Rev. {\bf 132}, 
        1411 (1963).
 
\bibitem{Wu_06} F.Y. Wu, Int. J. Mod. Phys. B {\bf 20}, 5357 (2006),
        [arXiv:cond-mat/0303251].

\bibitem{SS_AT} J. Salas and A.D. Sokal, J. Stat. Phys. {\bf 85}, 297 (1996),
        [arXiv:hep-lat/9511022].

\bibitem{SW} R.H. Swendsen and J.S. Wang, Phys. Rev. Lett. {\bf 58}, 2 (1987).

\bibitem{Kandel_90} D. Kandel, R. Ben-Av, and E. Domany, Phys. Rev. Lett. 
         {\bf 65}, 941 (1990).

\bibitem{WSK} J.S. Wang, R.H. Swendsen, and R.Koteck\'y, Phys. Rev. Lett. 
        {\bf 63}, 109 (1989). 

\bibitem{Wu_Lin} F.Y. Wu and K.Y. Lin, J. Phys. C {\bf 7}, L181 (1974).

\bibitem{Ferreira} S.J. Ferreira and A.D. Sokal, J. Stat. Phys. {\bf 96}, 461
        (1999), [arXiv:cond-mat/9811345]. 

\bibitem{SS_AF_sq}  J. Salas and A.D. Sokal, J. Stat. Phys. {\bf 92}, 729
        (1998), [arXiv:cond-mat/9801079].
 
\bibitem{Ditzian} R.V. Ditzian, J.R. Banavar, G.S. Grest, and L.P. Kadanoff,
         Phys. Rev. B {\bf 22}, 2542 (1980).

\bibitem{Knops_80} H.J.F. Knops, Ann. Phys. (N.Y.) {\bf 128}, 448 (1980).

\bibitem{Kamieniarz_97} G. Kamieniarz, P. Kozlowski, and R. Dekeysen,
        Phys. Rev. E {\bf 55}, 3724 (1997). 

\bibitem{Kadanoff_79} L.P. Kadanoff and A.C. Brown, Ann. Phys. (N.Y.) 
        {\bf 121}, 318 (1979).

\bibitem{Nijs_79} M.P.M. den Nijs, J. Phys. A {\bf 12}, 1857 (1979).

\bibitem{Yang_87} S.--K. Yang, Nucl. Phys. B {\bf 285}, 183 (1987).

\bibitem{Saleur_88} H. Saleur, J. Stat. Phys. {\bf 50}, 475 (1988).

\bibitem{Picco_10} M. Picco and  R. Santachiara, J. Stat. Mech., P07027 (2010),
        [arXiv:1005.0493].
 
\bibitem{Picco_11} M. Picco and  R. Santachiara, Critical interfaces and 
        duality in the Ashkin--Teller model  
        [arXiv:1011.1159].
 
\bibitem{Baxter_78} R.J. Baxter, H.N.V. Temperley, and S.E. Ashley, 
        Proc. Royal Soc. London A {\bf 358}, 535 (1978).  

\bibitem{Baxter_82} R.J. Baxter, Proc. Royal. Soc. London A {\bf 383}, 
        43 (1982).  

\bibitem{Saleur_90} H. Saleur, Comm. Math. Phys. {\bf 132}, 657 (1990).

\bibitem{Saleur_91} H. Saleur, Nucl. Phys. B {\bf 360}, 219 (1991).

\bibitem{Jacobsen_06} J.L. Jacobsen and H. Saleur, Nucl. Phys. B {\bf 743}, 
        207 (2006) [arXiv:cond-mat/0512058].

\bibitem{Lv_11} J.-P. Lv, Y. Deng, and Q.-H. Chen, Phys. Rev E {\bf 84},
        021125 (2011), [arXiv:1009.3172]. 

\bibitem{Temperley_79} H.N.V. Temperley and S.E. Ashley, Proc. Royal Soc. 
        London A {\bf 365}, 371 (1979).
 
\bibitem{Baxter_86} R.J. Baxter, J. Phys. A {\bf 19}, 2821 (1986).

\bibitem{Baxter_87} R.J. Baxter, J. Phys. A {\bf 20}, 5241 (1987).

\bibitem{Deng_07} Y. Deng, T.M. Garoni, W. Guo, H.W.J. Bl\"ote, and A.D. Sokal,
         Phys. Rev. Lett. {\bf 98}, 120601 (2007), [arXiv:cond-mat/0608447].

\bibitem{Liu_11} Q. Liu, Y. Deng, and T.M. Garoni, Nucl. Phys. B {\bf 846},
         283 (2011), [arXiv:1011.1980].

\bibitem{Baxter_74} R.J. Baxter, Aust. J. Phys. {\bf 27},
        369 (1974).

\bibitem{Baxter_70} R.J. Baxter,  J. Math. Phys. {\bf 11}, 784 (1970).

\bibitem{Moore_00} C. Moore and M. E. J. Newman, J. Stat. Phys. {\bf 99}, 
         629 (2000), [arXiv:cond-mat/9902295]. 

\bibitem{Shrock_97} R. Shrock R and S.-H. Tsai, J. Phys. A {\bf 30}, 495 
        (1997), [arXiv:cond-mat/9608095].

\bibitem{Hex} J. Salas, J. Phys. A {\bf 31}, 5969 (1998), 
        [arXiv:cond-mat/9802145].

\bibitem{Kano} K. Kano and S. Naya, Progr. Theor. Phys. {\bf 10}, 158 (1953).  

\end{thebibliography}
\end{document}